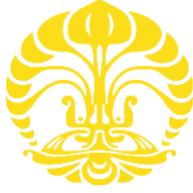

**UNIVERSITAS INDONESIA**

**PERANCANGAN *USER INTERFACE* SIAK-NG DENGAN METODE *DESIGN THINKING* UNTUK MENDUKUNG INTEGRASI SISTEM**

**SKRIPSI**

**NAILA ZAAFIRA**

**1906354791**

**FAKULTAS TEKNIK**

**PROGRAM STUDI TEKNIK INDUSTRI**

**DEPOK**

**JUNI 2023**

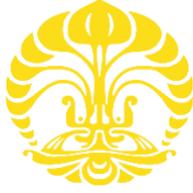

**UNIVERSITAS INDONESIA**

**PERANCANGAN *USER INTERFACE* SIAK-NG DENGAN METODE *DESIGN THINKING* UNTUK MENDUKUNG INTEGRASI SISTEM**

**SKRIPSI**

**Diajukan sebagai salah satu syarat memperoleh gelar Sarjana Teknik**

**NAILA ZAAFIRA**
**1906354791**

**FAKULTAS TEKNIK**
**PROGRAM STUDI TEKNIK INDUSTRI**
**DEPOK**
**JUNI 2023**

**HALAMAN PERNYATAAN ORISINALITAS**

**Skripsi ini adalah hasil karya saya sendiri,**

**dan semua sumber baik yang dikutip maupun**

**dirujuk telah saya nyatakan benar.**

**Nama**           **: Naila Zaafira**

**NPM**           **: 1906354791**

**Tanda Tangan**   **:** 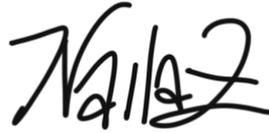

**Tanggal**        **:** 2 Juni 2023

# HALAMAN PERSETUJUAN SIDANG

Skripsi dengan judul:

## Perancangan User Interface SIAK-NG Berbasis Design Thinking untuk Mendukung Integrasi Sistem

Dibuat untuk melengkapi sebagian persyaratan untuk menjadi Sarjana Teknik pada Program Studi Teknik Industri Departemen Teknik Industri Fakultas Teknik Universitas Indonesia, dan disetujui untuk diajukan dalam sidang ujian skripsi.

Depok, 6 Juni 2023

Pembimbing Skripsi,

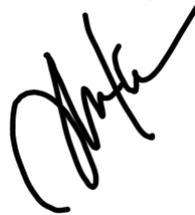

Dr. Maya Arlini Puspasari, S.T., M.T.
NIP/NUP 198803222012122003

Skripsi ini diajukan oleh

| | | |
|---|---|---|
| Nama | : | Naila Zaafira |
| NPM | : | 19063547913 |
| Program Studi | : | Teknik Industri |
| Judul Skripsi | : | Perancangan User Interface SIAK-NG dengan Metode Design Thinking untuk Mendukung Integrasi Sistem |

**Telah berhasil dipertahankan di hadapan Dewan Penguji dan diterima sebagai bagian persyaratan yang diperlukan untuk memperoleh gelar Sarjana Teknik pada Program Studi Teknik Industri, Fakultas Teknik, Universitas Indonesia**

### DEWAN PENGUJI

Pembimbing  : Dr. Maya Arlini Puspasari, S.T., M.T.   ( 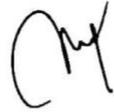 )

Penguji 1    : Dr.-Ing. Amalia Suzianti S.T., M.Sc   ( 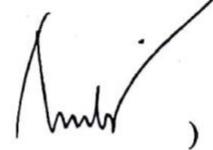 )

Penguji 2    : Rheinanda Kaniaswari, MT   ( 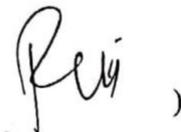 )

Ditetapkan di : Depok
Tanggal     : 22 Juni 2023

## KATA PENGANTAR

Puji dan syukur Penulis panjatkan kepada Tuhan Yang Maha Esa yang telah melimpahkan berkah dan rahmat kepada penulis. Karena atas rahmat dan karunia-Nya penulis dapat menyelesaikan skripsi yang berjudul "Perancangan *User Interface* SIAK-NG dengan Metode *Design Thinking* untuk Mendukung Integrasi Sistem". Pembuatan skripsi ini dilakukan sebagai salah satu syarat dalam menyelesaikan mata kuliah skripsi untuk mendapatkan gelar Sarjana Teknik di Departemen Teknik Industri, Fakultas Teknik, Universitas Indonesia.

Dalam menyusun skripsi ini tentu penulis mendapati beberapa rintangan dan cobaan. Pada proses pembuatan skripsi ini penulis menghadapi beberapa kesulitan, tetapi bukanlah hal yang mustahil untuk penulis menyelesaikan skripsi ini karena adanya bantuan, bimbingan, serta dukungan dari berbagai pihak dan orang terkasih bagi penulis. Untuk itu, penulis ingin mengucapkan terima kasih kepada:

(1) Tuhan yang Maha Esa yang telah memberikan begitu banyak berkah, kesempatan, dan bimbingan bagi penulis. Yang memberikan penulis kehidupan yang penuh arti dan keberuntungan, dan tidak pernah meninggalkan penulis dalam masa-masa kelam.

(2) Ibu Dr. Maya Arlini Puspasari, S.T., M.T., selaku dosen pembimbing skripsi dan akademik yang telah menyediakan waktu, pikiran, tenaga, dukungan, motivasi dan doa serta memberikan bimbingan dan masukan dalam penyelesaian penulisan penelitian ini dan selama masa perkuliahan.

(3) Ibu Dr.-Ing. Amalia Suzianti S.T., M.Sc dan Ibu Rheinanda Kaniaswari, MT selaku dosen penguji yang telah memberikan kritik dan saran yang membangun dan membantu penulis di penelitian ini.

(4) Seluruh dosen Teknik Industri UI yang telah memberikan banyak pembelajaran dan ilmu kepada penulis selama masa perkuliahan.

(5) Pihak Direktorat Sistem & Teknologi Informasi yang telah memberi saya kesempatan untuk melakukan penelitian ini.

(6) Seluruh responden yang telah bersedia meluangkan waktunya untuk memberikan masukan selama proses pengerjaan penelitian ini.

(7) Ibu penulis yang terus memberikan semangat, perhatian, doa, dan cinta yang tidak henti-hentinya kepada penulis.

(8) Ibu penulis yang membantu penulis bangkit dari titik-titik rendah dalam hidup.

(9) Ibu penulis sebagai inspirasi dan dukungan emosional dalam hidup penulis, yang telah menjadikan penulis orang yang lebih baik dari hari ke hari, dan memberikan begitu banyak pengorbanan yang tidak akan bisa penulis gantikan sepenuhnya.

(10) Bapak penulis yang memberikan kestabilan dan sumber dukungan bagi penulis.

(11) Keluarga penulis yang memberikan dukungan, perhatian, dan banyak kenangan indah dalam hidup penulis.

(12) Alexey Gavryushin yang telah menemani penulis di berbagai negara saat penulis jauh dari rumah, serta telah meluangkan waktu di tengah kesibukan pribadi untuk memberi wawasan dari perspektif teknis dengan membantu membangun kode untuk website ini.

(13) Serta seluruh pihak yang telah membantu penulis selama penyusunan skripsi sebagai tugas akhir ini

Dengan ini, penulis ingin mengungkapkan rasa terima kasih kepada semua pihak yang telah turut serta dalam proses ini. Meskipun tidak mungkin untuk menyebutkan satu per satu, semoga Tuhan yang Maha Esa membalas segala kebaikan yang telah diberikan oleh mereka yang telah membantu. Penulis juga ingin meminta maaf jika terdapat kesalahan dalam penggunaan kata yang mungkin kurang berkenan. Penulis sadar bahwa penelitian ini masih belum mencapai kesempurnaan. Oleh karena itu, segala kritik dan saran yang dapat membantu perbaikan dan pengembangan penelitian ini di masa depan akan sangat dihargai. Semoga skripsi ini dapat memberikan manfaat bagi para pembaca dan dapat menjadi landasan untuk penelitian selanjutnya.

Depok, 2 Juni 2023

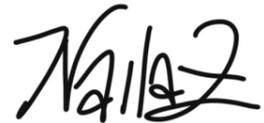

Naila Zaafira

**Universitas Indonesia**

**HALAMAN PERNYATAAN PERSETUJUAN PUBLIKASI TUGAS AKHIR**

**UNTUK KEPENTINGAN AKADEMIS**

Sebagai sivitas akademik Universitas Indonesia, saya yang bertanda tangan di bawah ini:

Nama                   : Naila Zaafira

NPM                   : 1906354791

Program Studi         : Teknik Industri

Fakultas               :Teknik

Jenis karya            : Skripsi

Demi pengembangan ilmu pengetahuan, menyetujui untuk memberikan kepada Universitas Indonesia **Hak Bebas Royalti Noneksklusif (*Non-exclusive Royalty-Free Right*)** atas karya ilmiah saya yang berjudul : **Perancangan *User Interface* SIAK-NG dengan Metode *Design Thinking* untuk Mendukung Integrasi Sistem** beserta perangkat yang ada (jika diperlukan). Dengan Hak Bebas Royalti Noneksklusif ini Universitas Indonesia berhak menyimpan, mengalihmedia/format-kan, mengelola dalam bentuk pangkalan data (*database*), merawat, dan memublikasikan tugas akhir saya selama tetap mencantumkan nama saya sebagai penulis/pencipta dan sebagai pemilik Hak Cipta.

Demikian pernyataan ini saya buat dengan sebenarnya.

                             Dibuat di        : Depok

                             Pada tanggal    : 21 Juni 2023

                                  Yang menyatakan

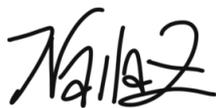

                                  (Naila Zaafira)

# ABSTRAK

| | |
|---|---|
| Nama | : Naila Zaafira |
| Program Studi | : Teknik Industri |
| Judul | : Perancangan *User Interface* SIAK-NG dengan Metode *Design Thinking* untuk Mendukung Integrasi Sistem |
| Pembimbing | : Dr. Maya Arlini Puspasari, S.T.,M.T.,MBA |

Penelitian ini dilakukan untuk meningkatkan antarmuka pengguna SIAK-NG (Sistem Informasi Akademik Next Generation), website portal akademik milik Universitas Indonesia melalui pendekatan *design thinking*. Meskipun SIAK-NG telah ada selama beberapa dekade dan mengalami beberapa perbaikan, belum ada upaya khusus untuk mengevaluasi kualitas antarmuka pengguna. Dalam rangka perombakan SIAK-NG sesuai dengan masterplan Universitas Indonesia 2019-2024, penelitian ini bertujuan untuk memberikan wawasan yang diperlukan. Penelitian ini berfokus pada perancangan ulang antarmuka website SIAK-NG untuk mengatasi keluhan dan kesulitan pengguna, terutama terkait dengan desain antarmuka yang belum memuaskan. Pendekatan *Design Thinking* digunakan untuk menghasilkan solusi yang sesuai dengan kebutuhan mahasiswa aktif Universitas Indonesia sebagai pengguna utama. Melalui metode seperti *storyboarding, empathy map, usability testing,* dan *tools* lainnya, penulis akan merancang rekomendasi yang sesuai dengan kebutuhan pengguna.

Kata Kunci: *Design Thinking*, Portal Akademik, Portal Universitas, *User Interface, User Persona, Usability Testing*

**Universitas Indonesia**

# ABSTRACT

Name : Naila Zaafira

Study Program : Industrial Engineering

Title : SIAK-NG User Interface Design with Design Thinking Method to Support System Integration

Advisor : Dr. Maya Arlini Puspasari, S.T.,M.T.,MBA

This research aims to improve the user interface of SIAK-NG (Next Generation Academic Information System), the academic portal website of the University of Indonesia through a design thinking approach. Despite being in existence for several decades and undergoing multiple improvements, there has been no specific effort to evaluate the quality of the user interface. In line with the revamping of SIAK-NG according to the University of Indonesia 2019-2024 master plan, this study aims to provide the necessary insights. The research focuses on redesigning the website interface of SIAK-NG to address user complaints and difficulties, particularly related to the unsatisfactory interface design. The Design Thinking approach is employed to generate solutions that meet the needs of active University of Indonesia students who are the primary users. Through methods such as storyboarding, empathy mapping, usability testing, and others, the author will design recommendations that align with user requirements.

Keywords: Design Thinking, University Portal, Academic Portal, User Persona, User Interface, Usability Testing

# DAFTAR ISI

**BAB 5**

**Universitas Indonesia**

# DAFTAR TABEL

# DAFTAR GAMBAR

**Universitas Indonesia**

# BAB 1
# PENDAHULUAN

## 1.1. Latar Belakang

SIAK-NG (Sistem Informasi Akademik Next Generation) merupakan salah satu sistem yang sangat penting di Universitas Indonesia (UI) dan digunakan oleh seluruh sivitas UI di seluruh fakultas. Dalam manual resmi SIAK-NG yang diterbitkan oleh Direktorat Sistem dan Teknologi Informasi (DSTI) UI, SIAK-NG didefinisikan sebagai aplikasi berbasis web yang digunakan untuk mendukung kegiatan akademik di UI. Aplikasi ini bersifat *online* dan dapat diakses oleh mahasiswa melalui internet dari mana saja dan kapan saja. SIAK-NG juga mengintegrasikan proses bisnis dari semua fakultas di UI sehingga memudahkan pemantauan proses akademik.

Universitas Indonesia memiliki visi dan misi yang bertujuan untuk menciptakan iklim yang mendukung kegiatan akademik yang lancar. Visi UI adalah menjadi universitas yang unggul, mandiri, dan bertaraf internasional dalam bidang pendidikan, penelitian, dan pengabdian kepada masyarakat. Misi UI mencakup empat poin, yaitu memberikan pendidikan berkualitas, melakukan penelitian yang inovatif, menyebarkan hasil penelitian kepada masyarakat, dan berperan aktif dalam pengembangan masyarakat (Pengembangan dan Pelayanan Sistem Informasi Universitas Indonesia, 2008)

Fungsi utama SIAK-NG adalah sebagai platform informasi tentang segala hal yang berhubungan dengan kegiatan belajar mengajar di UI. Mulai dari pendaftaran kelas hingga pengajuan skripsi dan cuti. SIAK-NG memiliki beberapa kegunaan bagi mahasiswa, antara lain:

1. Memudahkan pemantauan nilai-nilai dan kegiatan akademis secara *online*: Melalui SIAK-NG, mahasiswa dapat dengan mudah memeriksa nilai-nilai yang telah diperoleh dan mengikuti perkembangan akademik mereka secara *online*. Ini memungkinkan mahasiswa untuk mengetahui pencapaian akademik dan mengidentifikasi area yang perlu diperbaiki.

2. Memudahkan registrasi secara *online*: SIAK-NG memungkinkan mahasiswa untuk mendaftar mata kuliah secara *online*. Dengan fitur ini, proses registrasi dapat dilakukan dengan lebih efisien dan mengurangi kemungkinan kesalahan dalam pengisian data.

3. Melihat jadwal kuliah dan jadwal ujian: Mahasiswa dapat mengakses jadwal kuliah dan jadwal ujian melalui SIAK-NG. Ini memudahkan mahasiswa untuk merencanakan jadwal dengan baik dan menghindari bentrok antara mata kuliah atau ujian.

4. Mengisi IRS (Isian Rencana Studi) secara *online*: SIAK-NG memungkinkan mahasiswa untuk mengisi IRS atau melakukan registrasi kelas secara *online*.

5. Menambah atau membatalkan mata kuliah pada semester yang sedang berjalan: SIAK-NG memungkinkan mahasiswa untuk menambah atau membatalkan mata kuliah pada semester yang sedang berjalan. Fitur ini memberikan fleksibilitas kepada mahasiswa dalam mengatur rencana studi mereka

Dalam konteks ini, dapat terlihat betapa penting dan krusialnya SIAK- NG bagi keberlangsungan akademik di UI. Menurut DSTI, SIAK-NG merupakan salah satu website yang paling penting di UI, dengan *Service Level Agreement* (SLA) yang memiliki target tingkat pelayanan lebih besar dari 80%.

Oleh karena itu, pada akhir 2019, Majelis Wali Amanat Universitas Indonesia (MWA UI) merilis dokumen yang berisi rencana strategis Universitas Indonesia untuk periode 2020-2024. Salah satu agenda yang tercantum dalam dokumen tersebut adalah merombak total SIAK-NG guna meningkatkan performanya. Rencana ini termasuk dalam Rencana Pembangunan Jangka Panjang (RPJP) Universitas Indonesia 2019-2024, yang bertujuan untuk membantu UI mencapai visi jangka panjangnya menjadi salah satu universitas terbaik di Asia Tenggara.

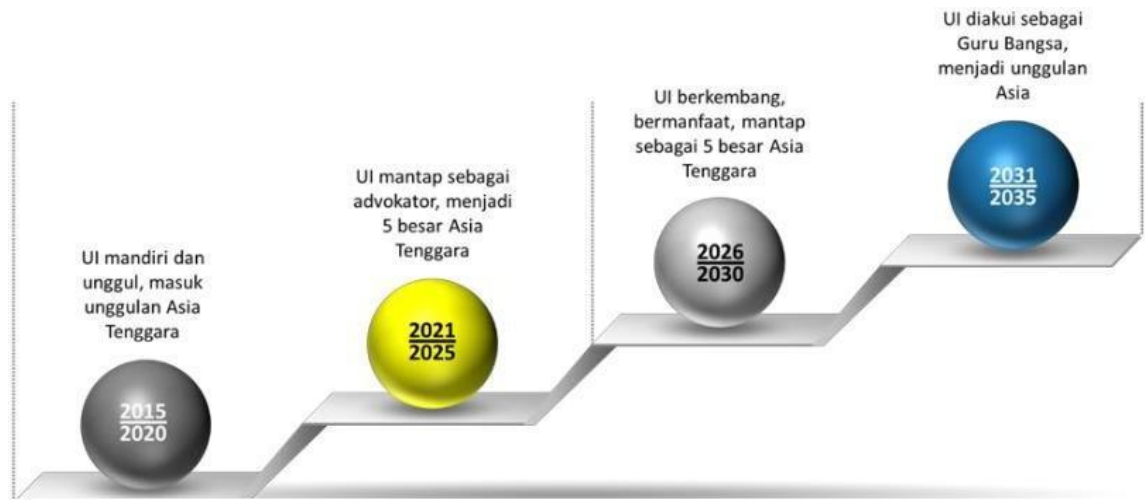

**Gambar 1. 1 Rencana Jangka Panjang Universitas Indonesia Tahun 2015-2035**

Sumber: Majelis Wali Amanat Universitas Indonesia, 2019

Pada pertengahan 2023, website SIAK-NG memasuki tahap awal pengembangan dan survei. Menurut hasil interview dengan DSTI, dalam sejarahnya, SIAK-NG belum dirancang dengan mempertimbangkan antarmuka pengguna (UI/UX) karena keterbatasan sumber daya selama tahap pengembangannya. Namun, DSTI, sebagai direktorat yang bertanggung jawab atas SIAK-NG, berupaya menerima masukan dari para pemangku kepentingan, termasuk pengguna, untuk meningkatkan website tersebut. Pemahaman tentang UI/UX menjadi aspek yang ingin ditingkatkan oleh DSTI, sebagaimana didasarkan pada dua studi pendahuluan yang dilakukan.

Studi pertama dilakukan dalam bentuk kuesioner yang menjangkau mahasiswa UI dari berbagai tahun dan mengumpulkan 35 responden untuk melihat pendapat dari pengguna lama dan baru. Hasil studi ini menunjukkan bahwa semua mahasiswa telah menggunakan SIAK-NG, yang membuktikan pentingnya SIAK-NG di Universitas Indonesia. Mayoritas merasa bahwa SIAK-NG memiliki peran signifikan dalam merepresentasikan Universitas Indonesia.

**Apakah Anda merasa desain dan rupa SIAK-NG memiliki peran signifikan dalam representasi image Universitas Indonesia?**

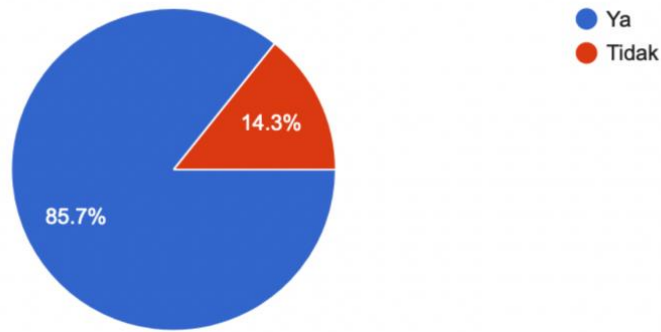

**Apakah menurut Anda desain SIAK-NG saat ini mencerminkan Universitas Indonesia sebagai universitas modern, terdepan, dan berteknologi yang unggul jika bersaing dengan universitas-universitas luar negeri?**

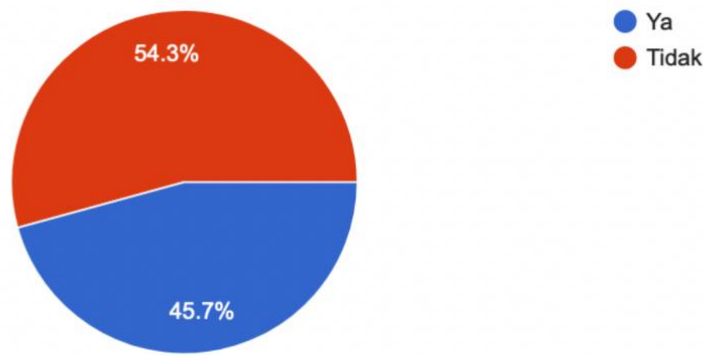

**Gambar 1. 2 Persentase responden pendapat terhadap representasi SIAK-NG pada Universitas Indonesia.**

Studi ini juga menunjukkan bahwa 86% pengguna merasa SIAK-NG membantu mereka dalam mengorganisir segala hal terkait dengan studi mereka, dan hal ini penting untuk menjaga pendidikan yang lancar

**Apakah Anda pernah mengakses SIAK-NG?**

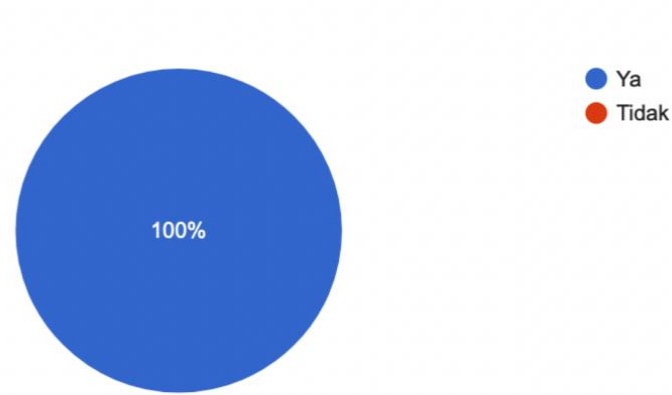

**Apakah SIAK-NG yang mudah dinavigasi dalam mencari informasi (kelas, profil dosen, database, dsb.) akan sangat membantu Anda dalam mengorganisir kegiatan akademik Anda di Universitas Indonesia?**

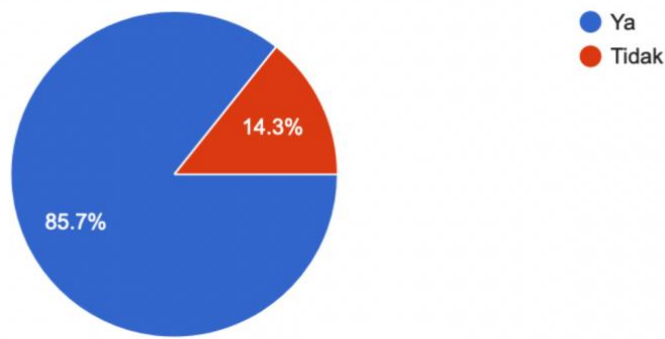

**Gambar 1. 3 Persentase responden yang pernah mengakses dan mendukung SIAK-NG yang intuitif.**

Sayangnya, hampir separuh pengguna mengalami kesulitan dalam mengoperasikan website karena *user interface*-nya kurang intuitif dan membutuhkan waktu untuk mempelajarinya. Meskipun separuh pengguna bersedia mengabaikan kekurangan tersebut, hanya 43% pengguna SIAK-NG yang tidak puas dengan *user interface*-nya. Namun, lebih dari 74% pengguna berharap adanya perbaikan pada SIAK-NG karena mereka percaya bahwa *user interface* yang lebih baik dapat meningkatkan kepuasan pengguna secara keseluruhan, meskipun dengan kekurangan teknis.

**Apakah di saat pertama Anda mengakses SIAK-NG Anda dengan mudahnya langsung dapat bernavigasi dan mengerti di mana harus mencari informasi yang Anda butuhkan?**

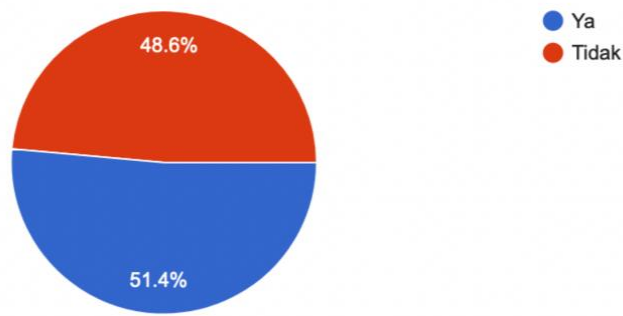

**Apakah Anda puas dengan estetika desain dan rupa (user interface) SIAK-NG?**

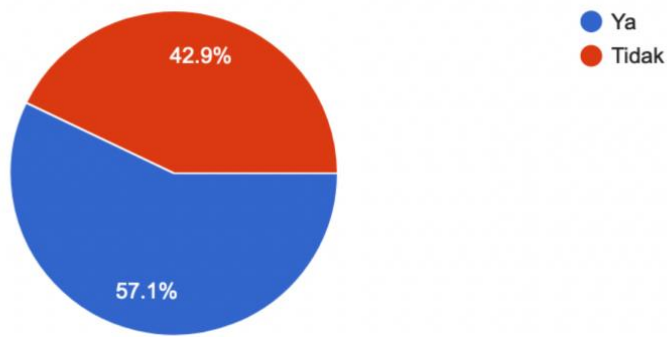

**Apakah user interface SIAK-NG yang efisien dapat secara signifikan membantu menutupi kekurangan teknis website SIAK-NG?**

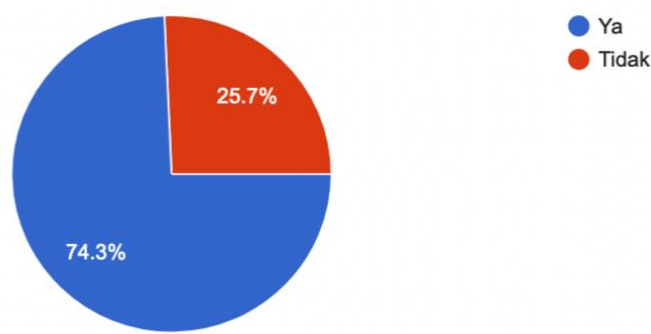

**Gambar 1. 4 Persentase user-satisfaction responden terhadap UI SIAK-NG**

Studi kedua dilakukan dengan membandingkan SIAK-NG dengan portal akademik dari universitas-universitas terkemuka, baik lokal maupun internasional. Dua portal yang digunakan adalah KU Loket oleh KU Leuven dan Simaster oleh Universitas Gadjah Mada. Sekilas, terlihat bahwa SIAK-NG tertinggal dalam hal *user interface*. Dua universitas lain menggunakan desain yang lebih baru yang didukung oleh CSS, sedangkan SIAK-NG sebagian besar menggunakan desain yang didukung oleh *legacy* HTML. Menurut interview dengan DSTI, hal ini disebabkan akibat kurangnya departemen *front-end* yang khusus ditugaskan untuk mengembangkan *user interface*. Alhasil, pengembang SIAK-NG hanya fokus pada peningkatan fungsionalitas dan mengabaikan *user interface*, sehingga versi awal SIAK-NG (JUWITA) digunakan sebagai *template* dasar untuk SIAK-NG saat ini, menjelaskan mengapa banyak fitur *legacy* yang masih ada.

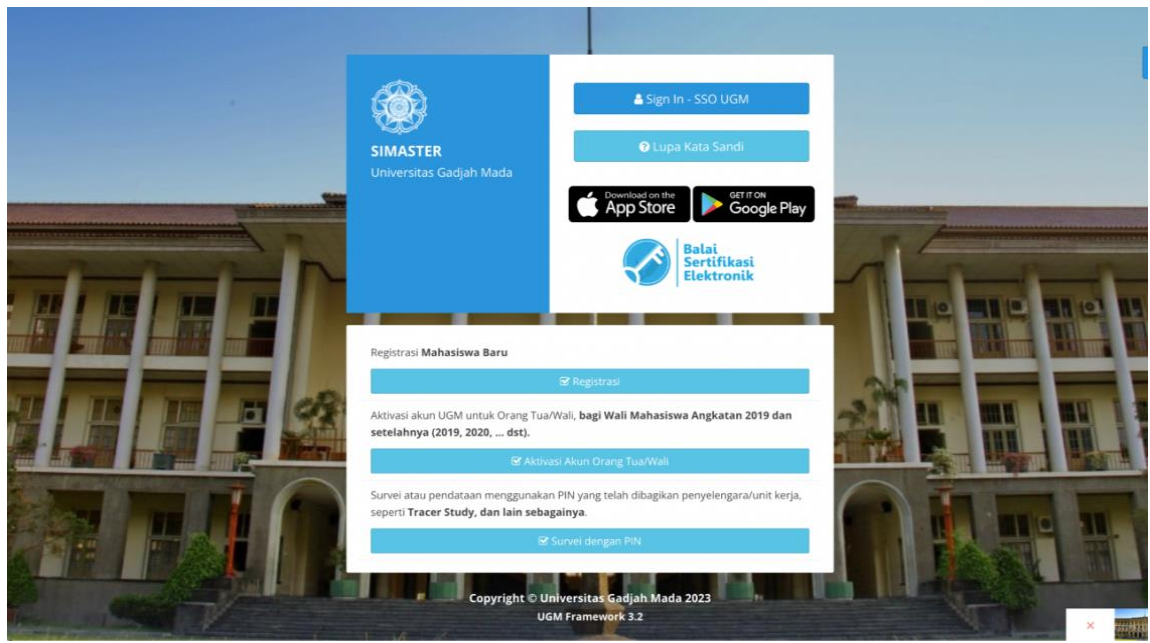

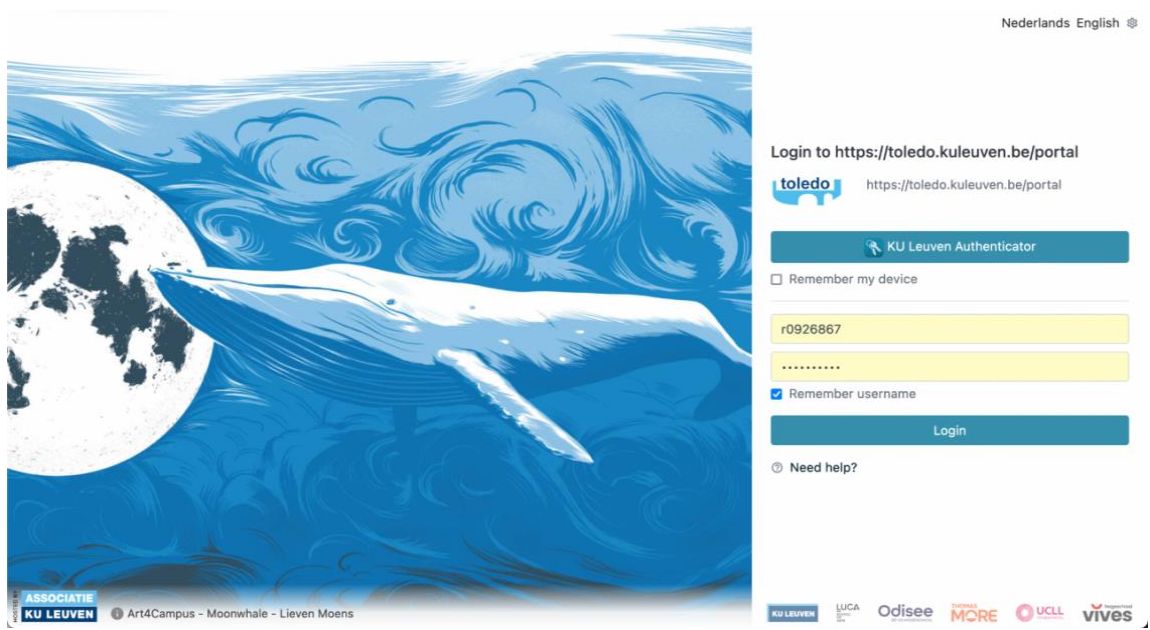

**Gambar 1. 5 Tampilan Antarmuka Halaman Login Simaster (Universitas Gadjah Mada) dan KU Loket (KU Leuven)**

Berdasarkan hal ini, dapat dikatakan bahwa SIAK-NG berada pada tingkat terendah (absent) dalam model kematangan UX, yang mengevaluasi besarnya perhatian yang diberikan pada user interface sebuah website (Pernice, 2021). Model kematangan ini terdiri dari 6 tingkat, yaitu:

1. *Absent*: UX diabaikan atau tidak ada.

2. *Limited*: Pertimbangan terhadap UX yang kurang, atau dilakukan dengan asal-asalan.

3. *Emergent*: Pertimbangan terhadap UX yang cukup, tetapi dilakukan secara tidak konsisten dan tidak efisien.

4. *Structured*: Organisasi memiliki metodologi terkait UX yang semi-sistematis dan tersebar luas, tetapi dengan tingkat efektivitas dan efisiensi yang kurang terstandar.

5. *Integrated*: Pertimbangan terhadap UX yang komprehensif, efektif, dan meluas.

6. *User-driven*: Dedikasi terhadap UX di semua tingkatan membuat hasil yang komprehensif dan hasil desain yang mempertimbangkan perilaku pengguna dengan baik.

Untuk bisa maju ke tingkat berikutnya, Nielsen merekomendasikan membantu pengembang menyadari pentingnya *user interface*, tidak sebagai desain yang bersifat dekoratif, tapi sebagai bagian dari fungsionalitas sebuah website. Hal ini, bersama dengan permintaan resmi dari RPJP untuk mengakui kebutuhan pengguna, termasuk *user interface*, minimal dapat membantu SIAK-NG mencapai tingkat 2 dalam model kematangan UX.

Masalah lain yang dapat dilihat dengan mudah dari observasi sekilas adalah perbedaan dalam hal area interaktif. Dibandingkan dengan tombol di KU Loket, ukuran layout tombol di SIAK-NG jauh lebih kecil dengan ketinggian 21 pixel. Ini merupakan pilihan desain yang mengkhawatirkan untuk *menu bar* utama, karena menurut WCAG 2.1, ukuran target minimum untuk klik harus 44×44 pixel (WCAG, 2018).

Penting untuk mengingat bahwa dengan berkembangnya e-learning, peran portal akademik seperti SIAK-NG menjadi lebih signifikan, karena aspek yang penting dalam proses pembelajaran online meliputi *platform user interfac*e, penyampaian informasi tepat waktu, dan desain (Borrellia, 2021).

*Platform-platform* perlu direncanakan ulang dengan memperhatikan perilaku mahasiswa, karena sebuah *platform* yang tidak bisa diandalkan akan menimpakan beban lebih kepada baik mahasiswa maupun guru dengan ekspektasi untuk mengadaptasi *user-behavior* untuk mengakomodir kekurangan *platform* (Ofosu-Asare, et al., 2019)

Sistem portal harus dirancang untuk menyerupai sistem-sistem konvensional pendidikan di dunia nyata agar bisa memberikan kebutuhan pengguna (Tateo, 2021). Kesuksesan tidak dapat tercapai jika tidak ada kesamaan; bahkan, seringkali masalah-masalah Desain *User Interface* Pengguna untuk E-Learning (UIDEL) berasal dari hal tersebut (Cho et al., 2009). *E-learning* yang baik harus dapat mendukung penyederhanaan pembelajaran agar pengguna dapat melihat, mendengar, berdiskusi, mengalami, dan mengajar orang lain.

Oleh karena itu, melalui penelitian ini, penulis berharap dapat melakukan studi khusus yang ditujukan untuk mengevaluasi masalah desain *user interface* yang ada pada website SIAK-NG saat ini, serta memberikan solusi desain yang lebih baik. Harapannya, ini dapat memberikan masukan yang bermanfaat dan konstruktif untuk pengembang dan pengelola SIAK-NG dalam upaya meningkatkan pengalaman pengguna secara keseluruhan.

## 1.2. Diagram Keterkaitan Masalah

Diagram keterkaitan masalah merupakan bagian dari serangkaian *tools* yang dikenal sebagai tujuh alat kontrol mutu baru (*7 new QC tools*) (Doggett, 2005). Secara khusus, diagram keterkaitan masalah berguna untuk mengidentifikasi hubungan sebab dan akibat di antara masalah yang rumit atau multivariat dengan hasil yang diinginkan. Penulis membangun diagram keterkaitan masalah untuk menggambarkan hubungan antara masalah-masalah dan solusi-solusi yang mungkin diperlukan, berdasarkan kesulitan yang dihadapi di bawah ini (Gambar 1.6).

Berikut ini adalah deskripsi identifikasi sebab dan akibat berbagai aspek penelitian ini dengan menggunakan diagram keterkaitan masalah.

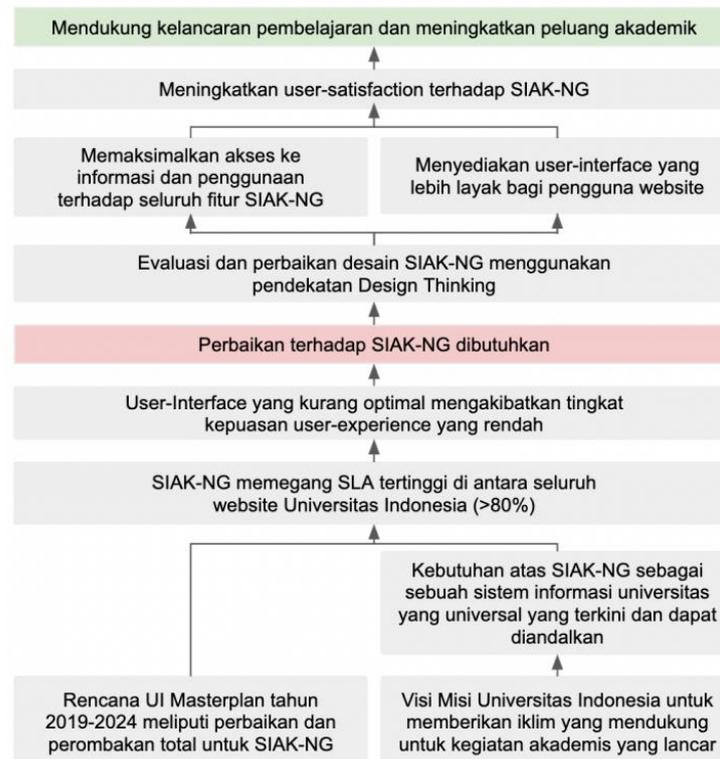

**Gambar 1. 6 Diagram Keterkaitan Masalah**

## 1.3. Rumusan Masalah

Berdasarkan konteks yang telah diuraikan sebelumnya, terdapat rumusan masalah yang merupakan pertanyaan utama yang perlu dijawab dalam penelitian ini yaitu "Bagaimana merancang ulang *user interface* SIAK-NG dengan pendekatan *design thinking*?"

## 1.4. Tujuan Penelitian

Berdasarkan perumusan masalah yang telah dijelaskan sebelumnya, tujuan dari penelitian ini adalah merancang ulang *user interface* SIAK-NG menggunakan pendekatan berpikir desain (*design thinking*).

## 1.5. Batasan Penelitian

Dalam menjalankan penelitian ini, terdapat lingkup atau batasan masalah yang ditetapkan untuk memastikan hasil penelitian sesuai dengan tujuan yang telah dijelaskan sebelumnya. Berikut adalah batasan-batasan masalah yang telah ditetapkan:

1. Objek yang akan dilakukan penelitian adalah website SIAK-NG yang dikembangkan oleh DSTI UI.

2. Penelitian ini akan menggunakan pendekatan *design thinking*.

3. Penelitian ini dilakukan dari Februari-Juni 2023.

4. Responden yang digunakan pada penelitian ini adalah pengguna SIAK-NG dengan role "Mahasiswa" dan masih menjadi mahasiswa aktif selama linimasa penelitian ini.

5. Responden yang didapat walaupun merepresentasikan seluruh rumpun di Universitas Indonesia cenderung lebih representatif terhadap rumpun ilmu teknik dan kesehatan.

6. Penelitian ini dibatasi dengan perancangan *prototype* sebagai solusi.

## 1.6. Metodologi Penelitian

Metodologi penelitian ini akan disusun menjadi lima bagian, dimulai dengan pendahuluan, dilanjutkan dengan tinjauan pustaka, pengumpulan data, pengolahan data, dan kesimpulan.

1. Pendahuluan

   Pada tahap ini, peneliti menentukan topik penelitian berdasarkan materi yang dipelajari selama program sarjana dan diskusi dengan pembimbing skripsi. Pendahuluan berfungsi untuk melatarbelakangi topik dan menuntun penelitian ke akar masalah. Penelitian awal dilakukan untuk memverifikasi lebih lanjut tentang masalah tersebut. Diagram keterkaitan masalah juga dibuat untuk membantu memahami hubungan masalah-masalah tersebut. Pendahuluan juga akan mencakup rumusan masalah, tujuan masalah, dan lingkup masalah sehingga hasil penelitian akan sesuai dengan tujuan peneliti.

2. Tinjauan Pustaka

   Setelah bab sebelumnya, bab dua melakukan tinjauan pustaka yang memberikan pembaca konteks, teori dasar, dan *tools* yang relevan dengan proyek penelitian. Teori-teori utama yang diperlukan untuk mendukung penelitian ini meliputi

*Design Thinking, User Experience, User Interface, Storyboarding, Usability Testing,* dan *PSSUQ*

3. Pengumpulan dan Pengolahan Data

Pada tahap ini, data akan dikumpulkan dan diproses. Pengumpulan data dan pengolahan data akan mengikuti pendekatan *design thinking*, yang dimulai dengan fase *empathize* dan berakhir dengan fase *define*. Fase *empathize* akan dilakukan dengan menyebarkan survei kepada responden. Data yang terkumpul kemudian akan dikembangkan menjadi demografi pengguna untuk pengelompokan persona dalam fase *empathize*. Hasil persona pengguna dalam fase *empathize* kemudian akan diproses sebagai karakteristik partisipan pada fase *define*. Fase *define* sendiri akan dilakukan melalui *in-depth interview* untuk menemukan fungsi-fungsi yang paling penting dari SIAK-NG, serta *empathy map*. Data yang didapat akan menjadi dasar untuk fase berikutnya: fase *ideate*. Fase ini menggunakan *storyboarding* sebagai *tools* untuk menggambarkan dan memahami *user journey* dengan lebih baik dan idealnya menemukan kebutuhan pengguna yang tidak terungkap langsung pada kuesioner maupun *interview*. Setelah fase *ideate* adalah fase *prototype*, di mana *low-fidelity wireframe* dan *rapid prototype* (*mock-up* yang dapat diklik) akan dirancang berdasarkan analisis *storyboarding* dan *benchmarking* situs web lainnya. *Rapid prototype* tersebut kemudian akan diuji pada fase terakhir, yaitu *testing*. Hal ini akan dibantu dengan *usability testing* dan survei PSSUQ untuk melihat apakah website SIAK-NG yang dimodifikasi mampu meningkatkan user experience sesuai kebutuhan pengguna.

4. Analisis Data

Bab ini menganalisis hasil setiap data yang dikumpulkan pada setiap fase metodologi *design thinking*. Analisis pada fase *empathize* akan dilakukan pada *user persona*, sedangkan pada fase *define*, *user persona* akan dikembangkan menjadi *empathy map* untuk mengetahui *pain point* dan *gain point* dari setiap persona. Selanjutnya pada fase *ideate*, ide-ide dikumpulkan, divalidasi, dan dieksplorasi berdasarkan skenario. Ide-ide tersebut, pain point, dan *gain point* dari setiap persona akan menjadi referensi dalam pembuatan fase *prototype*. Pada tahap terakhir yaitu fase *testing*, akan dilakukan analisis perbedaan antara performa *user interface* pengguna yang lama dan baru dan perasaan pengguna melalui PSSUQ untuk memverifikasi kesuksesan penelitian.

5. Kesimpulan dan Saran

   Pada fase akhir ini, penulis mensintesis temuan dari keseluruhan penelitian. Penulis juga akan menjelaskan rekomendasi untuk penelitian masa depan.

## 1.7. Sistematika Penulisan

Penelitian ini disusun secara sistematis menjadi lima bagian, yaitu pendahuluan, studi pustaka, pengumpulan dan pengolahan data, analisis dan pembahasan, serta kesimpulan dan saran.

Bab 1 merupakan pendahuluan yang menjelaskan tahap awal penelitian, termasuk latar belakang, diagram keterkaitan masalah, rumusan masalah, tujuan penelitian, batasan penelitian, metodologi penelitian, dan sistematika penulisan.

Bab 2 merupakan studi pustaka yang membahas teori-teori yang menjadi landasan penelitian.

Bab 3 membahas tentang pengumpulan dan pengolahan data. Penjelasan dan deskripsi terkait data yang digunakan dalam penelitian serta pengolahan data yang telah diperoleh dilakukan pada bab ini. Metode pengumpulan data meliputi penggunaan kuesioner, pengelompokan data, usability testing, dan PSSUQ yang menjadi acuan untuk merancang desain user interface sebagai solusi dari penelitian ini.

Bab 4 menjelaskan hasil dan analisis dari penelitian ini. Analisis dilakukan berdasarkan kelima tahapan *design thinking*, yaitu *empathize, define, ideate, prototype,* dan *testing.* Hasil yang diungkapkan meliputi desain *user interface* berdasarkan *storyboarding* dan wawancara dengan responden, serta hasil pengujian menggunakan metrik kinerja dan PSSUQ.

Bab 5 merupakan penutup yang berisi kesimpulan dari hasil penelitian dan saran untuk penelitian selanjutnya.

## BAB 2

## TINJAUAN PUSTAKA

### 2.1. User Interface

Berdasarkan Interaction Design Foundation (2023), desain *user interface* (UI) atau antarmuka pengguna adalah proses yang digunakan oleh desainer untuk membangun tampilan user interface dalam perangkat lunak atau perangkat komputer. Desainer bertujuan untuk menciptakan *user interface* yang mudah digunakan dan menyenangkan bagi pengguna karena kualitas *user interface* ikut mempengaruhi performa fungsionalitas sistem. Tujuan dari desain *user interface* adalah mengembangkan interaksi yang sederhana, efisien, dan ramah pengguna, dengan keseimbangan yang baik antara kegunaan dan estetika yang menarik bagi pengguna. Dalam studi ini, penulis telah memilih 4 kaidah yang akan dijadikan dasar desain *user interface:*

### 2.1.1. Hukum-Hukum UX (Yablonski, 2023)

Berdasarkan Interaction Design Foundation (2023), desain *user interface* (UI) atau antarmuka pengguna adalah proses yang digunakan oleh desainer untuk mendesain *user interface* dalam perangkat lunak atau perangkat komputer. Hukum-hukum ini dipilih untuk mengatasi masalah yang mungkin muncul dari berbagai pendekatan, guna memastikan adanya perspektif yang luas selama pengembangan *user interface* tersebut.

21 hukum tersebut diklasifikasikan ke dalam 4 kategori yaitu heuristik, prinsip, gestalt, dan bias kognitif. Berikut adalah beberapa hukum tersebut:

    1. Hukum Jakob (*Jakob's Law*): Pengguna lebih memilih situs web yang didesain sesuai dengan situs lain yang pernah mereka gunakan. Desain yang familiar membantu meningkatkan kemudahan mempelajari sistem tersebut.

    2. Efek Estetika-Kebergunaan (*Aesthetic-Usability Effect*): Pengguna cenderung menganggap situs web yang memiliki tampilan menarik secara visual lebih mudah digunakan.

    3. Ambang Doherty (*Doherty Threshold*): Interaksi antara manusia dan komputer optimal ketika respon dari sistem dapat diberikan dalam waktu kurang dari 400 ms, untuk menjaga *attention span* pengguna.

4. Hukum Fitts (*Fitts's Law*): Desain lebih fungsional jika elemen interaktif mudah dijangkau oleh pengguna, baik dalam hal jarak maupun ukuran target.

5. Hukum Hick (*Hick's Law*): Waktu yang dibutuhkan oleh pengguna untuk membuat keputusan meningat seiring dengan jumlah pilihan yang ada. Mengurangi jumlah pilihan dapat meningkatkan efisiensi.

6. Hukum Miller (*Miller's Law*): Kapasitas memori manusia terbatas, sehingga membatasi jumlah informasi yang disajikan pada satu waktu dapat meningkatkan pemahaman dan retensi.

7. Efek Gradien Tujuan (*Goal-gradient effect*): Pengguna cenderung semakin termotivasi jika mereka tahu persis seberapa jauh progress yang sudah mereka lakukan untuk sampai ke tujuan akhir.

8. Hukum Wilayah Umum (*Law of Common Region*): Pengguna cenderung mengelompokkan elemen-elemen yang berada dalam wilayah yang sama sebagai satu kesatuan.

9. Hukum Proximity (*Law of Proximity*): Elemen yang berdekatan cenderung dianggap terkait oleh pengguna.

10. Hukum Keterhubungan Seragam (*Law of Uniform Connectedness*): Elemen yang terhubung dengan suatu bentuk atau garis cenderung dianggap terkait disbanding elemen yang tidak diikat secara visual.

11. Hukum Prägnanz *(Law of Prägnanz)*: Pengguna cenderung menyederhanakan bentuk-bentuk kompleks menjadi bentuk- bentuk yang sederhana dan lebih teratur.

12. Hukum Kesamaan (*Law of Similarity*): Pengguna cenderung mengelompokkan elemen yang memiliki atribut serupa, seperti bentuk, ukuran, atau warna.

13. Prinsip Occam (*Occam's Razor*): Ketika ada beberapa penjelasan atau solusi yang memungkinkan, yang sederhana cenderung menjadi yang terbaik.

14. Prinsip Pareto (*Pareto Principle*): Sebagian besar hasil diperoleh dari sebagian kecil dari upaya yang dilakukan. Fokus pada sedikit hal yang memberikan dampak paling signifikan.

15. Hukum Parkinson (*Parkinson's Law*): Durasi yang dibutuhkan untuk menyelesaikan suatu pekerjaan akan melambat untuk mengisi waktu yang disediakan, oleh karena itu penting untuk memberikan batasan waktu.

16. Aturan Puncak-Akhir (*Peak-End Rule*): Pengguna cenderung mengingat pengalaman pada klimaks dan pada akhir. Menciptakan momen positif di klimaks dan akhir dapat meningkatkan kesan keseluruhan.

17. Hukum Postel (*Postel's Law*): Saat berinteraksi dengan sistem eksternal atau sistem yang tidak bisa kita kontrol, kita harus bisa fleksibel dan mengakomodir, namun sistem yang dapat kita control harus dipertahankan untuk tetap sesuai standar.

18. Efek Posisi Seri (*Serial Position Effect*): Pengguna cenderung lebih mudah mengingat informasi di awal dan akhir dari suatu urutan. Memperhatikan penempatan informasi penting di awal dan akhir dapat meningkatkan pemahaman dan ingatan pengguna.

19. Hukum Tesler (*Tesler's Law*): Terdapat batasan di mana sistem tidak dapat disimplifikasi lebih lanjut. Jika terus dipaksakan untuk disimplifikasi lebih lanjut, fungsionalitas akan berkurang.

20. Efek Von Restorff (*Von Restorff Effect*): Elemen yang menonjol atau berbeda dengan sekitarnya lebih mudah diperhatikan dan diingat oleh pengguna.

21. Efek Zeigarnik (*Zeigarnik Effect*): Pengguna cenderung lebih mudah mengingat tugas yang belum selesai daripada yang sudah selesai. Mengingatkan pengguna tentang tugas yang belum selesai dapat mendukung interaksi lebih lanjut oleh pengguna.

## 2.1.2.  Five Principles of Visual Design

Dalam studi oleh Gordon (2020), ditemukan bahwa saat manusia melihat sebuah elemen visual, biasanya orang tersebut dapat langsung menilai apakah elemen tersebut indah atau tidak. Namun, hanya sedikit yang dapat menjelaskan persisnya mengapa. Grafik atau desain yang menggunakan prinsip-prinsip desain visual yang baik dapat meningkatkan *traffic* dan kegunaan. Prinsip desain visual menjelaskan bagaimana mengkombinasikan komponen desain secara efektif, seperti garis, bentuk, warna, dan *grid,* dapat menghasilkan visualisasi yang seimbang.

Menerapkan prinsip-prinsip ini dapat membantu desainer mengorganisir aspek visual, seperti elemen-elemen pada menu, sehingga pengguna dapat dengan cepat memahami informasi di situs tersebut. Desain tidak bersifat hanya dekoratif, tapi juga menuntun bagaimana pengguna akan berinteraksi dengan situs tersebut dengan manipulasi penempatan elemen-elemen secara strategis, seperti:

1. Skala

Skala digunakan dengan cara membuat elemen-elemen yang paling penting dalam desain menjadi lebih besar dibandingkan dengan yang kurang penting. Teori di balik prinsip ini cukup sederhana: ketika sesuatu memiliki ukuran yang besar, akan mendapatkan perhatian lebih banyak dan lebih terlihat.

2. Hirarki Visual

Hirarki visual berkaitan dengan mengarahkan mata pada halaman sehingga fokusnya tertuju pada elemen-elemen desain yang berbeda sesuai dengan tingkat kepentingannya. Variasi dalam skala, nilai, warna, jarak, penempatan, dan berbagai sinyal lainnya dapat digunakan untuk menetapkan hierarki visual.

3. Keseimbangan

Keseimbangan terjadi ketika ada distribusi yang seimbang (tetapi tidak terlalu simetris) dari visual di kedua sisi sumbu yang melalui tengah layar. Sumbu ini biasanya vertikal, tetapi juga dapat horizontal.

4. Kontras

Kontras mengacu pada rangkaian elemen yang secara visual berbeda, efeknya adalah menekankan perbedaan yang terdapat antara kedua elemen tersebut. Dengan kata lain, kontras memperkuat perbedaan yang signifikan (misalnya, dalam ukuran atau warna) antara dua item (atau antara dua set objek). Kontras adalah pendekatan desain yang banyak digunakan untuk membuat elemen-elemen tertentu dalam desain menonjol bagi pengguna.

5. Hukum Gestalt

Prinsip-prinsip Gestalt menjelaskan bagaimana mata manusia dapat mengelompokkan dan mensimplifikasi banyak elemen yang ramai dan rumit menjadi satu elemen berdasarkan kesamaan karakteristik mereka. Hukum ini berkaitan dengan Law of Prägnanz. Gordon (2020) juga mencatat bahwa ketika prinsip-prinsip skala, hierarki visual, keseimbangan, kontras, dan Gestalt diterapkan dengan efektif, tidak hanya menghasilkan desain yang menarik tetapi juga meningkatkan kegunaan.

### 2.1.3.  *Atomic Design*

*Atomic design* merupakan suatu metode pendekatan desain *user interface* yang terdiri dari lima tahap, yang bekerja bersama untuk menciptakan sistem desain *user interface* dengan cara yang lebih terencana dan berhirarki (Frost, 2016). Kelima tahap dalam atomic design meliputi:

1. Atom
2. Molekul
3. Organisme
4. Template
5. Halaman (*page*)

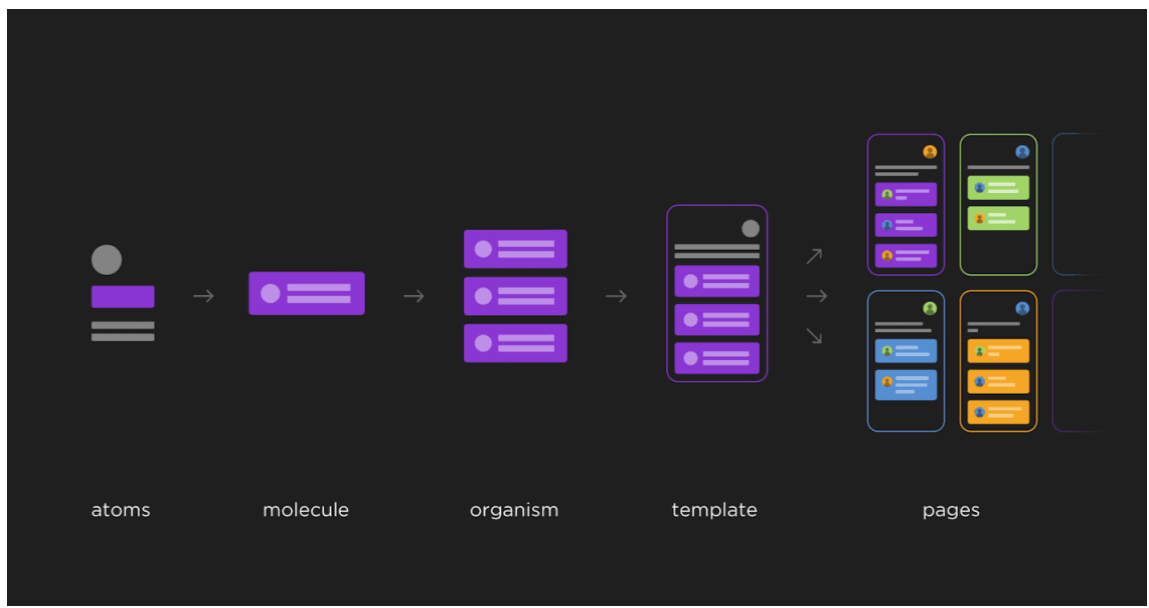

**Gambar 2. 1 Pengembangan *User Interface* dengan sistem *Atomic Design***

Sumber: Signal. (2020)

Proses *atomic design* tidak bersifat linear, melainkan hanyalah suatu *mental model* yang membantu pengembang memandang *user interface* sebagai sebuah kesatuan agar bagian-bagian dari user interface yang berbeda dapat tetap mempertahankan keunikannya namun secara keseluruhan tetap saling terhubung sebagai 1 sistem terintegrasi.

Salah satu keuntungan utama yang diberikan oleh *atomic design* adalah kemampuannya untuk dengan cepat beralih antara konsep abstrak menjadi konkret, di mana dalam pendekatan *atomic design*, desain dilihat tidak sebagai desain namun sebagai bagian dari fungsi, yaitu sesuatu yang mendikte bagaimana sebuah fitur akan dipakai oleh penggunanya. Keuntungan lainnya adalah pemisahan yang jelas antara struktur dan konten. Melalui *atomic design*, ada pemisahan yang jelas antara *template* dengan *page* Terakhir, *atomic design* bersifat universal dalam pemakaiannya, dan sistem ini dapat diterapkan pada berbagai jenis *user interface*, tidak hanya terbatas pada yang berbasis web saja.

*Atomic design* mendorong penggunaan s*pace* maksimum namun tetap menjaga pentingnya penggunaan *white space* dengan membantu memberikan ruang lebih untuk font yang lebih besar namun tetap mencegah *layout* untuk tidak terlihat berantakan. Sistem *atomic design* juga mendorong pengembang untuk memiliki elemen yang konsisten untuk seluruh website yang memperkuat penggunaan Hukum Jakob dengan mengulangi variasi desain elemen. Hal ini membantu pengguna mengidentifikasi desain elemen yang mirip sehingga menciptakan website yang lebih intuitif.

### 2.1.4. Data-Ink Ratio

Konsep *Data-Ink Ratio* dimaksud sebagai *framework* yang dipakai dalam segala hal yang berhubungan dengan data atau *dashboard*, di mana pengembang diajak untuk selalu mempertimbangkan apakah semua elemen dalam grafik tersebut relevan dengan data. Menurut Edward Tufte (2018), sebuah fitur yang fungsi utamanya adalah untuk menyampaikan dan berbagi data, data harus ditaruh sebagai prioritas no 1, dan segala elemen yang tidak berhubungan harus dihilangkan. Dalam bukunya yang berjudul "*The Visual Display of Quantitative Information*", menyebutkan dua prinsip penghapusan untuk mencapai rasio *data-ink* yang lebih baik:

1.   Menghapus selain data-ink

Elemen visual dan estetika yang tidak menambah informasi atau mengkomunikasikan data harus dipertimbangkan untuk dihapus. Kategori ini mencakup garis bantu, warna tanpa makna atau tujuan, efek 3D, dan lain sebagainya.

2.   Menghapus data-ink yang redundan

Dalam pembuatan sebuah grafik, sering terdapat kecenderunga untuk mencoba memuat lebih banyak informasi daripada yang diperlukan atau menambahkan elemen-elemen tambahan. Hal ini dilakukan banyak orang dengan harapan data yang lebih komplit berarti lebih baik, padahal kenyataanya data yang baik adalah data yang berguna bagi pengguna. Prinsip ini menjelaskan untuk tidak menambahkan variabel agar terkesan lengkap, yang pada akhirnya variabel tersebut tidak dimengerti apa fungsinya oleh pengguna. Prinsip ini menyatakan bahwa informasi tambahan yang tidak perlu harus dihapus. Kategori ini mencakup legenda yang tidak perlu, label, informasi berlebih yang tidak berhubungan dengan pesan grafik, dan lain-lain.

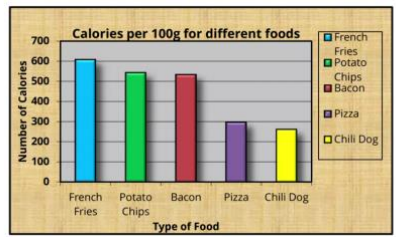

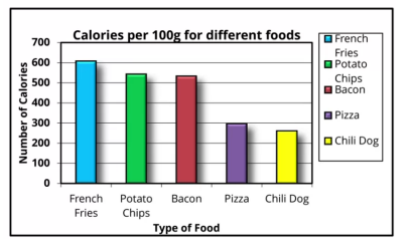

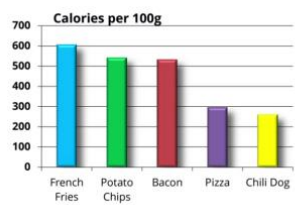

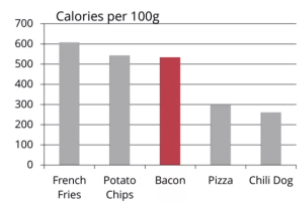

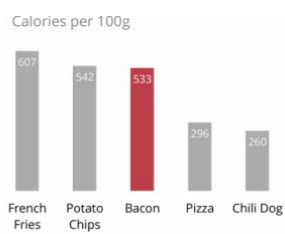

**Gambar 2. 2 Aplikasi *Data-Ink Ratio***

Sumber: Raj (2020)

*Data-Ink Ratio* juga berkaitan dengan prinsip *Call of Action color*, dimana penggunaan warna dikhususkan bagi elemen yang didesain untuk menarik perhatian pengguna agar berinteraksi dengan elemen tersebut, seperti tombol.

**2.2. User Experience**

*User Experience (UX)* dijelaskan oleh Interaction Design Foundation (2022) sebagai desain yang memberikan *product experience* yang bermakna dan terhubung. Sebuah *user interface* yang baik dapat dicapai dengan kemudahan bagi pengguna untuk menggunakan elemen-elemen interaktif pada website. Menggabungkan desain yang berorientasi pada pengguna dengan fungsionalitas yang praktis merupakan bagian dari proses integrasi desain UX yang baik, yang dapat memberikan kepuasan yang baik bagi pengguna.

**2.3. Usability**

*Usability* adalah atribut kualitas yang menilai seberapa mudah *user interface* digunakan. Istilah *usability* juga mengacu pada metode untuk meningkatkan kenyamanan penggunaan selama proses desain. *Usability* berasal dari kata "*usable*", yang diartikan sebagai "dapat digunakan". Definisi dari *usability* yang baik adalah sistem yang dapat menghilangkan atau meminimalkan kegagalan ketika pengguna melakukan berbagai tugas dan memberikan keuntungan dan kepuasan kepada pengguna. *Usability* didefinisikan sebagai sejauh mana pengguna tertentu dapat menggunakan produk untuk mencapai tujuan dengan efektif, efisien, dan puas (ISO 9241-11, 2018).

Menurut Teori Nielsen, *usability* adalah aspek kualitas yang dapat ditingkatkan selama proses desain untuk membuat produk lebih *user-friendly* (Nielsen, 2003). Terdapat lima atribut kualitas *usability* menurut Nielsen (2003) yaitu:

1. Kemudahan dipelajari (*Learnability*)

*Learnability* adalah atribut *usability* yang mendasar, karena sebagian besar sistem harus mudah dipelajari dan hal ini berdampak pada kesan pertama pengguna terhadap sistem. Pengguna harus dapat belajar menggunakan suatu sistem dengan cepat dan mudah.

2. Efisiensi (*Efficiency*)

Efisiensi adalah seberapa cepat pengguna dapat menyelesaikan aktivitas setelah belajar menggunakan suatu sistem.

3. Kemampuan Mengingat (*Memorability*)

Kemampuan mengingat berkaitan dengan orang-orang yang sudah terbiasa menggunakan suatu sistem, tetapi mengalami beberapa kesulitan dalam menggunakannya atau hanya menggunakan sistem tersebut secara jarang. Kemampuan mengingat adalah metrik yang menilai seberapa efektif orang dapat mengingat berbagai fungsi setelah mempelajarinya.

4. Kepuasan (*Satisfaction*)

Istilah "kepuasan dengan suatu sistem" mengacu pada seberapa menyenangkan penggunaan sistem tersebut. Hal ini berdampak pada motivasi pengguna dan, sebagai hasilnya, efektivitas penggunaan sistem tersebut.

5. Kesalahan (*Error*)

Ketika pengguna mengalami kesalahan dalam menyelesaikan tugas, hal itu dapat digambarkan sebagai sistem yang tidak menghasilkan hasil yang diinginkan.

## 2.4. *Design Thinking*

Terdapat berbagai pengertian mengenai *design thinking* menurut para ahli di bidangnya. Salah satu pandangan yang dikemukakan oleh K. Tschimmel (2012) menyebutkan bahwa *design thinking* adalah suatu alur berpikir yang mengarahkan pada transformasi, evolusi, dan inovasi. Pandangan lain datang dari K. Gurusamy et al. (2016), yang menjelaskan bahwa *design thinking* berperan sebagai katalisator dalam proses tersebut.

Proses *design thinking* dapat dilakukan dengan langkah-langkah tertentu, salah satunya adalah dengan mengeksplorasi masalah dan solusi melalui proses konvergensi yang melibatkan penggabungan dan pemilihan, sebagaimana diungkapkan oleh Plattner et al. (2013). Dalam hal ini, *design thinking* menjadi pendekatan inovatif yang berfokus pada manusia (*human-centered*) dan menggunakan *tools* untuk mengintegrasikan kebutuhan pengguna, potensi teknologi, serta persyaratan untuk mencapai kesuksesan, sebagaimana disampaikan oleh Tim Brown (2009).

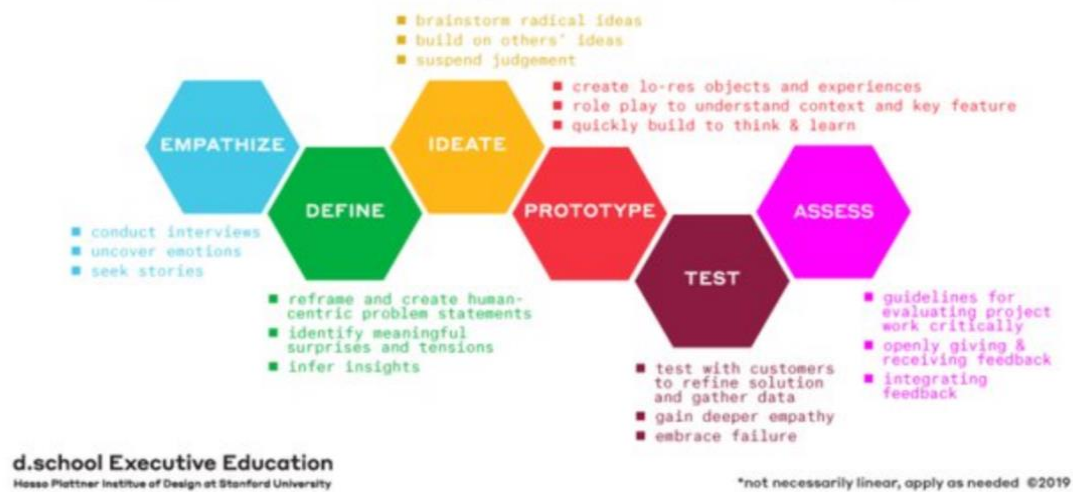

**Gambar 2. 3 Diagram Proses *Design Thinking* oleh Institut Desain Hasso-Plattner di Stanford.**

Sumber: Pulido, Mar. (2021)

Terdapat 5 tahap dari *design thinking* yang menjadi kerangka berpikir ketika mengembangkan sebuah produk atau jasa, di antaranya adalah *empathize, define, ideate, prototype,* dan *testing.*

### 2.4.1. *Empathize*

*Empathize* merupakan tahap pertama dari proses *design thinking*. Tujuan dari tahap pertama ini adalah untuk memperoleh pengertian dari permasalahan yang ingin diatasi. Empati adalah hal yang krusial dalam proses desain berbasis *human-centered* dan memungkinkan desainer mengesampingkan asumsi pribadi untuk mengerti keinginan dan kebutuhan pengguna yang sebenarnya (Interaction Design Foundation, 2021).

### 2.4.1.1. Kuesioner

Menurut McLeod (2018), kuesioner adalah alat penelitian yang terdiri dari serangkaian pertanyaan yang dirancang untuk mengumpulkan data dari responden. Kuesioner dapat berbentuk kertas atau dilakukan secara online, dan terdiri dari serangkaian pertanyaan yang harus dijawab oleh responden (Adams & Cox, 2008).

Penggunaan kuesioner memungkinkan pengumpulan data kuantitatif yang terstandarisasi untuk memastikan konsistensi dan koherensi data yang akan dianalisis. Penting bagi kuesioner untuk memiliki tujuan yang jelas terkait dengan tujuan penelitian dan mempertimbangkan penggunaan hasil data sejak awal. Selain itu, mengingat keterbatasan peneliti, kuesioner cocok digunakan karena sumber daya terbatas dan biaya yang relatif murah namun dengan keunggulan dapat mendapatkan sampel data yang besar sehingga representatif secara statistik.

Ada beberapa jenis kuesioner yang dapat digunakan dalam merancang kuesioner menurut Ndukwu (2019), antara lain:

1. Pertanyaan terbuka, yaitu gaya pertanyaan yang memungkinkan responden bereaksi lebih bebas dan terbuka.

2. Pertanyaan pilihan ganda, yaitu pertanyaan dengan beberapa kemungkinan jawaban yang dapat dipilih satu atau lebih.

3. Pertanyaan dikotomis, yaitu pertanyaan dengan hanya dua pilihan, seperti "benar" atau "salah."

4. Pertanyaan berskala, yaitu pertanyaan dengan beberapa pilihan jawaban dalam bentuk skala berdasarkan preferensi responden. Skala penilaian, skala Likert, skala diferensial semantik, dan pertanyaan visual adalah beberapa bentuk pertanyaan berskala.

5. Mensubstitusi teks dengan diagram visual.

### 2.4.1.2. *Clustering*

Dalam penelitian ini, digunakan dua metode *clustering*: *hierarchical clustering* menggunakan metode Ward dan *K-modes clustering*. *Hierarchical clustering* mengklasifikasikan data ke dalam kelompok-kelompok yang relatif homogen di dalam dirinya sendiri dan relatif heterogen antara satu sama lain (Landau & Chis Ster, 2010). Metode *clustering* Ward digunakan untuk mengklasifikasikan responden dan menciptakan grup baru dengan meminimalisir varian (Everitt et al., 2011).

*Cluster analysis* sendiri digunakan untuk menemukan pola dalam data pengguna yang tidak terlihat secara kasat mata, berdasarkan persamaan karakteristik (variabel) yang tersedia (Brickey et al., 2010). Setelah menentukan jumlah kelompok yang ada menggunakan *hierarchical clustering*, dilakukan *K-modes clustering*. *K-Modes clustering* adalah metode yang sejalan dengan *K-means* namun lebih cocok untuk clustering data kategorikal (Chatuverdi, 2001). *K-modes* digunakan karena memiliki sifat nonparametrik yang cocok untuk mengklasifikasi data kategorikal, menghindari kebutuhan penentuan jarak, dan mengoptimasi matriks yang serasi (Bonthu, 2021). Dari hasil clustering menggunakan *K-modes*, akan ditemukan sifat-sifat untuk masing-masing kelompok data.

### 2.4.1.3. User Persona

*User Persona* adalah karakter fiksi yang dibuat berdasarkan penelitian untuk mewakili berbagai jenis pengguna yang mungkin menggunakan layanan, produk, situs, atau merek dengan cara yang sama (Interaction Design Foundation, 2021; Olsen, 2015). Membuat narasi singkat untuk persona membantu dalam mengembangkan cerita yang menggambarkan interaksi persona dengan produk (Quesenbery & Brooks, 2010). *User persona* membantu penelitian dalam memfokuskan penelitian terhadap target pengguna aplikasi utama dan memahami kebutuhan, pengalaman, perilaku, dan tujuan pengguna (Harley, 2015; Interaction Design Foundation, 2022).

Komponen penting dalam pembuatan persona meliputi *cluster persona*, nama (boleh nama fiksi), demografi (usia, jenis kelamin, pendidikan atau pekerjaan), foto yang mewakili persona grup, pekerjaan atau tanggung jawab utama, tujuan yang ingin dicapai, kekhawatiran atau frustrasi, serta komentar penting menurut mereka (Harley, 2015).

Proses pembuatan Persona melibatkan langkah-langkah berikut:
1. Menentukan tipe pengguna yang penting dan berpengaruh dalam *domain* produk.
2. Memproses data dengan mengumpulkan informasi tentang pengguna dan produk melalui data mentah, kemudian menentukan hubungan antara keduanya menggunakan pendekatan statistik seperti pengelompokan atau analisis *cluster*.
3. Identifikasi dan perancangan kerangka, termasuk evaluasi data yang diproses, verifikasi kategori pengguna, dan identifikasi sub-kategori pengguna.
4. Mengurutkan prioritas dengan mengevaluasi kebutuhan bisnis atau organisasi pemilik sistem terhadap hasil data yang diperoleh dari *framework*.

5. Pembangunan persona dengan cara mengisi detail dan karakteristik yang kurang untuk memberikan persona kepribadian dan konteks.

6. Validasi dan konfirmasi persona untuk memastikan kesesuaian dengan data yang dikumpulkan (Pruitt & Adlin, 2006).

Dengan menggunakan user persona dalam penelitian skripsi ini, akan memudahkan pemahaman kebutuhan dan keluhan target pengguna SIAK-NG sehingga solusi yang tepat dapat diimplementasikan dalam pembuatan *prototype*.

### 2.4.2.  Define

Tahap *define* adalah ketika peneliti merumuskan pernyataan masalah dan menjelaskan kebutuhan serta masalah yang dialami oleh pengguna. Semua data dan informasi yang dikumpulkan selama tahap *empathize* diperiksa dan digabungkan untuk menentukan masalah mendasar yang diidentifikasi. *Tools* yang digunakan dalam penelitian ini pada tahap *define* adalah *empathy map* yang diperoleh dari *in-depth interview*.

### 2.4.2.1. In-depth Interview

*In-depth interview* atau wawancara mendalam adalah metode penelitian kualitatif yang melibatkan melakukan wawancara intensif dengan sejumlah kecil responden untuk mengeksplorasi pandangan mereka tentang ide, program, atau situasi tertentu (Boyce & Neale, 2006). Berdasarkan hal tersebut, *in-depth interview* dengan pengguna digunakan dalam penelitian ini untuk mempelajari perasaan, tindakan, motivasi, tujuan, nilai, tanggapan, keluhan, dan hambatan yang mereka alami saat menggunakan produk yang sedang diteliti. Pelaksanaan *in-depth interview* akan didukung oleh *empathy map* yang sesuai dengan penelitian ini. Pertanyaan-pertanyaan dalam *in-depth interview* juga dirancang agar bersifat terbuka untuk membantu dalam pembuatan *empathy map*.

### 2.4.2.2. Empathy Mapping

*Empathy mapping* merupakan sebuah *tools* yang digunakan untuk memahami pengguna dengan melihat dari sudut pandang mereka, baik melalui observasi maupun wawancara, dengan menggabungkan data untuk menghasilkan hasil yang tak terduga (Stanford d.school, 2010). Persona dapat digunakan bersama dengan *empathy map* sebagai tindak lanjut untuk memberikan gambaran tentang bagaimana pengalaman persona akan digunakan dalam fase selanjutnya. Tujuannya adalah memberikan rangsangan visual bagi pengguna untuk merenungkan dan menjelajahi perspektif, pengaruh, kebutuhan, emosi, keinginan, dan ketakutan mereka terkait dengan kerangka proyek (Tschimmel, 2012).

Selain itu, menurut Bratsberg (2012), *empathy mapping* dapat membantu memahami bagaimana perubahan kecil dalam desain dapat memberikan dampak besar bagi pengguna. Menurut Gray (2017), empathy map terdiri dari: "lihat", "rasakan", "dengar", "pikirkan", "katakan", "lakukan", "*pain points*", dan "*gain points*" dari pengguna. Namun, berdasarkan perubahan dari Gibbons (2018), empathy mapping terdiri dari empat bidang, yaitu "Say", "Feel", "Do", dan "Think", dan keempat bidang ini harus ditemukan secara keseluruhan saat membuat empathy map. Gambar 2.4 menunjukkan template empathy map yang terbaru.

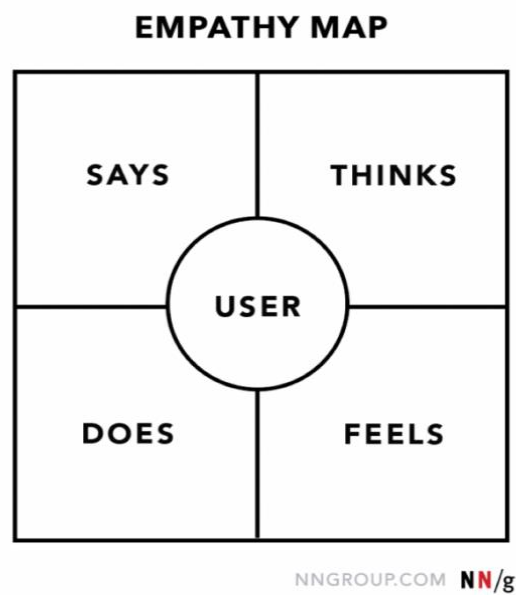

**Gambar 2. 4 Empathy Map**

Sumber: Gibbons (2018)

### 2.4.3. *Ideate*

Menurut Interaction Design Foundation (2022), *ideate* adalah proses di mana desainer berpikir secara bersama-sama dalam kelompok untuk menghasilkan ide-dan digunakan untuk menghasilkan sebanyak mungkin ide untuk mengatasi masalah. Tujuan dari *ideate* adalah menghasilkan sejumlah besar ide yang kemudian dapat dipilih menjadi ide-ide terbaik, paling praktis, atau paling orisinal untuk menginspirasi solusi desain dan produk yang baru dan lebih baik. Berbagai tools dapat digunakan, namun pendekatan *storyboarding* akan digunakan dalam fase ideate ini.

### 2.4.3.1. Storyboarding

*Storyboard* merupakan alat komunikasi yang menyampaikan sebuah cerita melalui gambar-gambar yang ditampilkan dalam urutan panel yang memetakan peristiwa-peristiwa utama cerita secara kronologis (Krause, 2018). Dalam dunia UI/UX, kita menggunakan *storyboard* untuk memberikan konteks tambahan kepada tim dan stakeholder. Dengan menggunakan gambar, cerita dapat dengan cepat dipahami dengan sekali pandang dan mudah diingat.

Terdapat tiga elemen umum dalam sebuah *storyboard* yaitu skenario, visual, dan keterangan.

1. Skenario

*Storyboard* didasarkan pada suatu skenario atau cerita pengguna. Deskripsi singkat dari skenario juga disertakan. Deskripsi skenario atau cerita tersebut cukup jelas sehingga anggota tim atau stakeholder dapat memahami apa yang digambarkan sebelum melihat visualnya. Sebagai contoh: Pembeli perlu mengisi ulang persediaan kantor.

2. Visual

Setiap langkah dalam skenario tersebut direpresentasikan secara visual dalam urutan tertentu. Langkah-langkah tersebut dapat berupa sketsa, ilustrasi, atau foto. Bergantung pada tujuan *storyboard* dan audiensnya, gambar-gambar ini dapat berupa gambaran cepat dengan tingkat kejelasan rendah atau artefak dengan tingkat kejelasan tinggi. Gambar-gambar tersebut mencakup detail-detail yang relevan dengan cerita, seperti tampilan lingkungan pengguna, balon ucapan dengan kutipan dari pengguna, atau sketsa dari layar yang digunakan pengguna.

3. Keterangan

Setiap visual memiliki keterangan yang sesuai. Keterangan tersebut menggambarkan tindakan pengguna, lingkungan, kondisi emosional, perangkat yang digunakan, dan sebagainya. Karena gambar merupakan konten utama dalam *storyboard*, keterangan-keterangan tersebut singkat dan umumnya tidak melebihi dua poin penting.

*Storyboard* dapat menjadi *tools* yang baik untuk tahap *ideate* dalam proses *Design Thinking*. *Storyboard* dapat menggambarkan ide tentang bagaimana seorang pengguna mungkin menggunakan fitur dan membantu memvisualisasikan pengalaman yang mungkin akan terjadi. Penggunaan *storyboard* dalam *prototype* UI website menawarkan beberapa keunggulan dibandingkan dengan teknik SCAMPER.

1. Representasi visual: *Storyboard* menyediakan representasi visual user interface dan interaksi, memungkinkan desainer untuk mempresentasikan konsep desain secara konkret dan intuitif

2. Desain berpusat pada pengguna: *Storyboard* membantu desainer untuk berempati dengan pengguna dan mempertimbangkan kebutuhan, tujuan, dan masalah yang dialami pengguna dalam *user journey*. Hal ini memastikan antarmuka website dirancang dengan memperhatikan pengguna, sehingga menghasilkan pengalaman yang lebih ramah pengguna (Cooper, 2007).

### 2.4.4. Prototype

Tahap keempat dari *design thinking* adalah *prototype*. *Prototype* berfungsi untuk memberikan solusi terhadap masalah yang dikemas sebagai sebuah model. Pada tahap ini, ide dan solusi yang diperoleh dari tahap ideasi akan dipilih, dikembangkan, dan diwujudkan menjadi produk yang dapat diuji sebagai solusi potensial (Henriksen, Richardson, & Mehta, 2017). *Prototype* memiliki fungsi untuk mengetes ide desain sehingga dapat dilihat bagaimana kondisi ide desain dapat direalisasikan sebagai objek fisik atau solusi yang memungkinkan.

*Prototyping* terdiri dari tiga prosedur atau tahapan, yaitu pembuatan *prototype*, *review*, dan penyempurnaan. *Review* melibatkan pendistribusian *prototype* kepada pengguna atau *stakeholder* dan mengumpulkan umpan balik mengenai apa yang berfungsi dan perlu diperbaiki berdasarkan umpan balik pengguna. Penyempurnaan mengidentifikasi bagian-bagian yang perlu diperbaiki berdasarkan *review*. *Prototype* dapat berupa sketsa kertas dengan tingkat kejelasan rendah hingga sesuatu yang memungkinkan pengguna untuk menjelajahi beberapa konten hingga situs yang dapat diklik (tingkat kejelasan tinggi). Metode yang digunakan pada tahap ini meliputi prototype dengan *low-fidelity wireframe* dan *high-fidelity rapid prototype* (Usability.gov, n.d.).

### 2.4.4.1. Wireframe Low-Fidelity

*Wireframing* adalah teknik untuk menetapkan struktur dan alur solusi desain potensial dengan menggambar gambaran produk interaktif. Wireframe ini didasarkan pada kebutuhan pengguna dan bisnis (Interaction Design Foundation, n.d.).

Wireframe dibuat sebelum desain dilakukan sehingga desainer dapat fokus pada layout tanpa terganggu oleh aspek visual seperti warna, bayangan, dan batas (Kaniaswari, 2018). Tujuan dari wireframe adalah untuk bereksperimen menemukan layout yang baik tanpa terganggu dengan elemen desain (Interaction Design Foundation, n.d.).

Menurut Lynch (2008), wireframe harus memiliki elemen-elemen berikut:

1. Logo
2. Breadcrumb
3. Header, termasuk judul
4. Konten utama
5. Sistem navigasi
6. Kolom pencarian
7. Kontak
8. Footer

Menurut Mendoza (2013), ada beberapa faktor yang perlu dipertimbangkan sebelum membuat wireframe:

1. Ukuran layar
2. *Platform*
3. Interaksi
4. Aplikasi vs. situs web
5. *Aspect Ratio*

2.4.4.2. **Rapid Prototype**

Menurut American Graphic Institute (2021), *rapid prototype* adalah proses iteratif untuk menciptakan UI atau UX dari situs web atau aplikasi. *Rapid prototyping* memungkinkan calon konsumen dan stakeholder lainnya untuk menguji UX atau UI sebelum dimasukkan ke dalam produksi skala penuh karena memungkinkan calon konsumen dan pemangku kepentingan untuk memberikan masukan sejak awal, sehingga melakukan perbaikan dan perubahan bisa lebih murah sehingga proses iterasi dapat lebih eksploratif. Pengembangan jangka panjang menjadi lebih efisien karena membutuhkan penyesuaian dan pembaruan yang minimal. Dalam penelitian ini penulis menggunakan Figma untuk pembuatan *rapid prototyping*.

### 2.4.5. *Testing*

Tahap kelima dalam proses *Design Thinking* adalah pengujian (*testing*). *Testing* dilakukan menggunakan hasil *prototype*, dan jika dilakukan dengan benar, dapat mengungkapkan banyak informasi tentang pengguna serta peluang untuk meningkatkan *prototype* (Dam and Siang, 2021). Tujuan dari tahap ini adalah mendapatkan umpan balik dan memvalidasi kesuksesan desainer dalam mengembangkan data menjadi sebuah output yang dapat menyelesaikan keseluruhan masalah, serta menjadi kesempatan bagi desainer untuk lebih memahami kebutuhan dan keinginan pengguna target; dan kesempatan bagi desainer untuk gagal, belajar, dan meningkatkan solusi terbaik (Seidel and Fixson, 2013). Penulis sebagai desainer akan melakukan pemeriksaan, wawancara, survei, atau meminta umpan balik atau saran dari pengguna setelah menguji *prototype* dengan pengguna (Henriksen et al., 2017).

Tahap ini dilakukan dengan menggunakan *usability testing* untuk mengukur metrik performa dan dilanjutkan dengan survei PSSUQ. Selain itu, metrik performa pada website SIAK-NG akan dibandingkan sebelum dan sesudah didesain ulang untuk melihat perbedaannya.

### 2.4.5.1. *Usability Testing*

*Usability Testing* adalah kegiatan yang mengevaluasi seberapa mudah suatu produk atau desain dapat digunakan dengan memantau sekelompok pengguna saat mereka mencoba menyelesaikan tugas-tugas (Interaction Design Foundation, 2021). Manfaat utama *usability testing* adalah mengidentifikasi masalah ketergunaan pada desain sejak dini, sehingga masalah yang ada dapat diatasi sebelum desain diterapkan atau diproduksi secara luas.

Menurut Dumas dan Redish (1999), *usability testing* memiliki lima tujuan:
1. Memberikan tugas kepada pengguna untuk diselesaikan dengan sistem.
2. Melibatkan pengguna akhir dalam pengujian sistem.
3. Memungkinkan penguji untuk merekam dan memantau interaksi pengguna dengan sistem.
4. Memungkinkan penguji untuk mengevaluasi data yang terkumpul dan membuat modifikasi yang relevan.
5. Meningkatkan *usability* sistem.

*Usability testing* melibatkan tiga tahap utama: perancangan dan persiapan materi pengujian, pengujian, analisis temuan pengujian, dan pelaporan hasil.
Selain itu, menurut Nielsen (2014), ada tiga langkah yang harus diambil saat melakukan *usability testing*:
1. Buat tugas se-realistis mungkin.
2. Buat tugas mudah diselesaikan.
3. Tidak diperbolehkan memberikan instruksi atau menjelaskan langkah-langkah yang harus diambil untuk mendapatkan hasil ketergunaan yang sebenarnya.

Dalam usability testing, digunakan juga metrik performa untuk menilai performa aplikasi berdasarkan hasil usability testing kepada responden (Nielsen, 2001). Metrik performa membantu dalam mengambil keputusan tentang apa yang perlu diperbaiki dalam aplikasi dan memprioritaskan masalah-masalah yang perlu diselesaikan (Albert & Tullis, 2013). Terdapat enam metrik dasar, yaitu tingkat keberhasilan tugas (*task success*), waktu penyelesaian tugas (*time on task*), jumlah kesalahan (*errors*), efisiensi (*efficiency*), dan tingkat pembelajaran (*learnability*) (Albert & Tullis, 2013).

*2.4.5.2. Survei PSSUQ*

Penggunaan kuesioner PSSUQ menjadi metode yang umum digunakan untuk mengevaluasi kepuasan pengguna terhadap situs web, program, sistem, atau produk pada akhir penelitian. PSSUQ terdiri dari 16 item pertanyaan yang dikembangkan oleh IBM sebagai *tools* penelitian untuk *usability testing* (Sauro & Lewis, 2012). Skala skor pada kuesioner menggunakan skala Likert yang terdiri dari 7 poin, di mana nilai yang lebih rendah menunjukkan tingkat kepuasan yang lebih tinggi.

Dalam analisis PSSUQ, 16 pertanyaan tersebut dapat dikelompokkan ke dalam empat kategori PSSUQ, dan menghasilkan empat jenis skor:

1. Nilai rata-rata keseluruhan (Overall): Menghitung rata-rata dari pertanyaan 15 yang menunjukkan kepuasan keseluruhan terhadap aplikasi yang dinilai.

2. Kegunaan sistem (SYSUSE): Merupakan rata-rata nilai dari pertanyaan 1-6 yang mengindikasikan kepuasan pengguna terhadap kegunaan aplikasi, yaitu kemampuan pengguna dalam menyelesaikan tugas dengan sukses dan tepat.

3. Kualitas informasi (INFOQUAL): Merupakan rata-rata nilai dari pertanyaan 7-12 yang menggambarkan tingkat kepuasan pengguna dalam menemukan informasi yang dibutuhkan dalam aplikasi.

4. Kualitas antarmuka (INTERQUAL): Merupakan rata-rata nilai dari pertanyaan 13-15 yang mencerminkan tingkat kepuasan pengguna terhadap tampilan antarmuka aplikasi.

5. Rata-rata total (Total Average): Merupakan rata-rata nilai dari pertanyaan 1-15 yang mencakup penilaian secara menyeluruh terhadap tampilan antarmuka aplikasi, termasuk pertanyaan tentang SYSUSE, INFOQUAL, dan INTERQUAL.

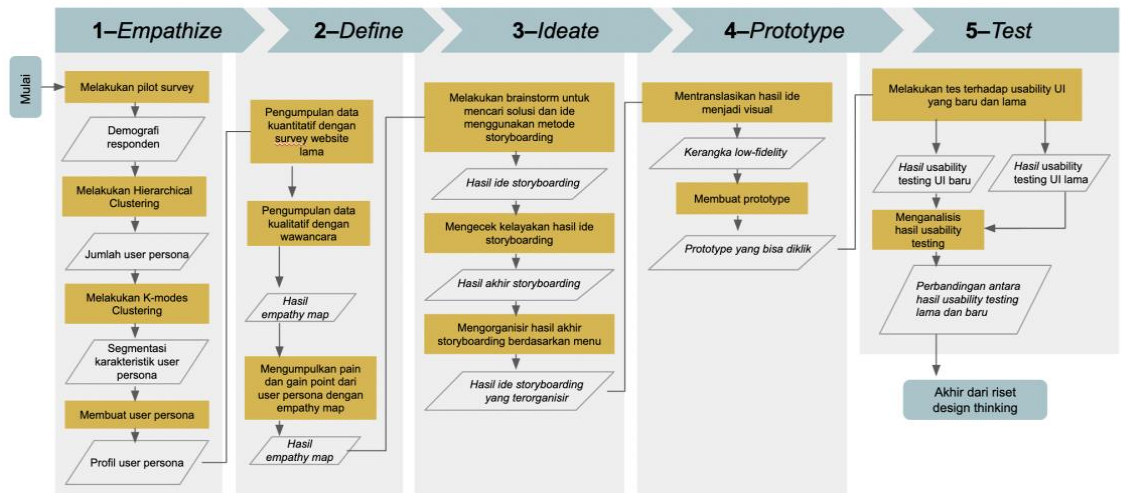

**Gambar 2. 5 Metodologi Tahapan Penelitian**

# BAB 3

# PENGUMPULAN DAN PENGOLAHAN DATA

Pada bab 3 ini, penulis akan memaparkan data yang telah dikumpulkan serta menjelaskan tahap pengumpulan dan pengolahan data tersebut. Bab ini juga akan mendemonstrasikan penggunaan seluruh metode dan *tools* yang digunakan dalam setiap tahap *design thinking*.

## 3.1. SIAK-NG

Aplikasi Sistem Informasi Akademik Next Generation (SIAK-NG) merupakan suatu sistem teknologi informasi yang telah diimplementasikan di lingkungan akademik Universitas Indonesia. Aplikasi berbasis web ini memiliki peran penting dalam menunjang kegiatan akademik bagi para mahasiswa. Dengan keunggulan utamanya yang berbasis online, SIAK-NG memberikan fleksibilitas akses kepada pengguna mahasiswa yang terhubung dengan jaringan internet. Dengan demikian, aplikasi ini dapat diakses secara bebas, baik dalam hal tempat maupun waktu.

Tujuan utama dari dibentuknya SIAK-NG adalah untuk mengintegrasikan proses bisnis yang terdapat di semua fakultas di Universitas Indonesia. Hal tersebut diharapkan dapat memberikan kemudahan bagi pihak pengelola dalam melakukan pemantauan dan pengelolaan data akademik. Aplikasi ini memiliki empat kategori pengguna utama, yaitu penyelenggara pendidikan (sub bagian akademik), mahasiswa, dosen, dan direktorat pendidikan. Setiap pengguna juga dapat memiliki lebih dari satu peran dalam program studi yang berbeda cukup dengan menggunakan satu akun.

Salah satu kegunaan dari SIAK-NG adalah memberikan kemudahan bagi *user-role* mahasiswa.

Dengan fungsi-fungsi SIAK-NG, mahasiswa dapat dengan mudah memantau nilai-nilai dan kegiatan akademik secara online, menghindari kerumitan dan pembatasan yang sering terjadi dalam proses konvensional. Proses registrasi pun menjadi lebih praktis melalui sistem online yang disediakan. Mahasiswa dapat dengan cepat melihat jadwal kuliah dan jadwal ujian mereka tanpa perlu mencari informasi secara terpisah. Pengisian IRS juga dapat dilakukan dengan mudah melalui platform yang telah tersedia. Selain itu, mahasiswa juga memiliki fleksibilitas untuk menambah atau membatalkan mata kuliah yang diambil dalam semester yang sedang berjalan hanya melalui aplikasi ini. (Pengembangan dan Pelayanan Sistem Informasi Universitas Indonesia, 2008)

## 3.2. Jenis dan Metode Pengumpulan Data

Penelitian ini dilakukan menggunakan pengumpulan data primer. Langkah pertama dalam pengumpulan data adalah melakukan survei awal untuk mengevaluasi situasi terkini dari website SIAK-NG. Survei awal menggunakan kuesioner kualitatif yang ditujukan kepada mahasiswa aktif Universitas Indonesia dari berbagai angkatan untuk memperoleh pendapat pengguna lama dan baru yang representatif. Setelah didapatkan dasar untuk melakukan penelitian dengan hasil survei awal, dilakukan survei sebenarnya menggunakan metode yang sama namun dengan populasi yang lebih besar.

Selanjutnya, penelitian dilanjutkan dengan melakukan in-depth interview yang juga bersifat kualitatif. Dalam *in-depth interview*, peneliti mengajukan pertanyaan kepada responden mengenai pengalaman mereka dengan website saat ini, termasuk hambatan yang dihadapi, tantangan yang dihadapi, kebutuhan yang belum terpenuhi, dan pertanyaan lain yang dirancang untuk mengungkapkan kebutuhan dan pendapat mengenai website. Diskusi tersebut menjadi dasar untuk proses *storyboard*. Setelah data dari proses ideate diolah dalam proses *prototyping*, ide-ide dari *storyboard* yang telah dimanifestasikan dalam bentuk *prototype* kemudian didiskusikan dengan pengguna untuk memverifikasi tingkat kepuasan dan mendapatkan umpan balik sebelum dilakukan ke tahap *testing*.

Pada akhir penelitian, tahap testing dilakukan menggunakan *usability testing* dan PSSUQ untuk website lama dan baru sehingga hasilnya dapat dibandingkan dengan performa UI SIAK-NG yang lama.

### 3.3. Pengumpulan Data

#### 3.3.1. Kuesioner

Survei digunakan untuk mengetahui demografi responden, situasi mereka, keluhan, saran, dan kebutuhan mereka, guna membantu mendefinisikan situasi dan kebutuhan saat ini untuk website SIAK-NG.

Menurut Marshall et al. (2013), kuesioner kualitatif minimal harus mencapai 20 responden. Survei menggunakan kuesioner kualitatif yang ditujukan kepada mahasiswa aktif Universitas Indonesia dari seluruh rumpun ilmu (sosial, teknik, dan kesehatan) untuk memperoleh pendapat pengguna yang representatif dan dapat diterapkan dalam skala universitas. Kuesioner ini didistribusikan secara online melalui Google Form dan mencapai 199 responden. Ada beberapa jenis pertanyaan yang diajukan dalam kuesioner antara lain:

#### 3.3.1.1. Data diri responden

Bagian ini berisikan nama, jenis kelamin, umur, fakultas, dan departemen. Adapun hasil pengumpulan data dari kuesioner adalah sebagai berikut:

1. Umur

Data yang terkumpul terdiri dari responden dengan umur antara 17- 21 tahun dengan range data yang representatif terhadap mahasiswa S1 Universitas Indonesia dari setiap angkatan. Mayoritas responden berumur 18 Tahun sebesar 42.2%, diikuti dengan responden dengan umur 19 tahun sebesar 41.7%, dan di urutan ketiga yaitu responden berumur 20 tahun sebesar 11.6%.

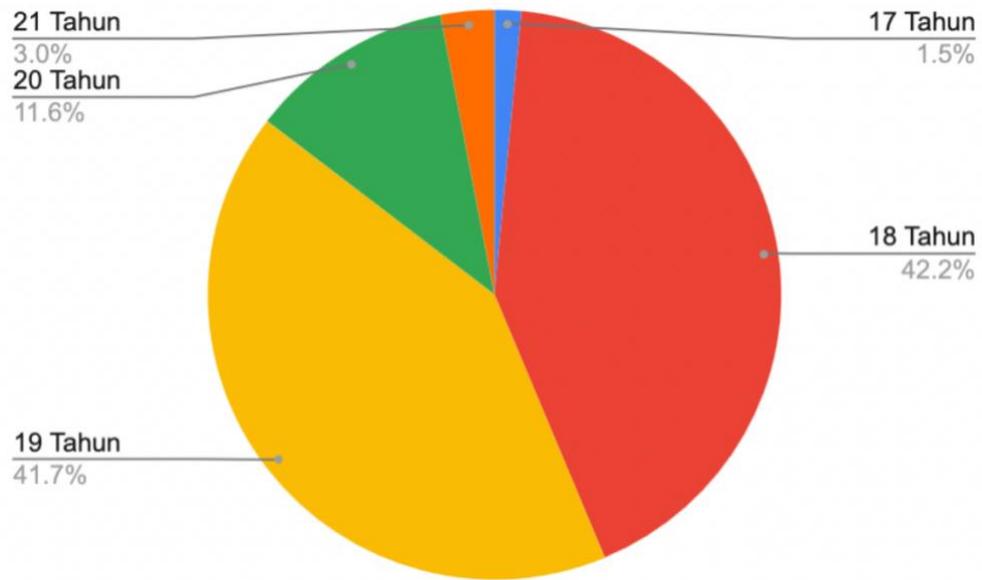

**Gambar 3.1 Pie Chart Umur Responden**

2. Kelamin

Kedua jenis kelamin diwakili dengan cukup seimbang. Mayoritas responden memiliki jenis kelamin perempuan sebesar 57,85%, sedangkan jumlah responden laki-laki mencapai 42,2%.

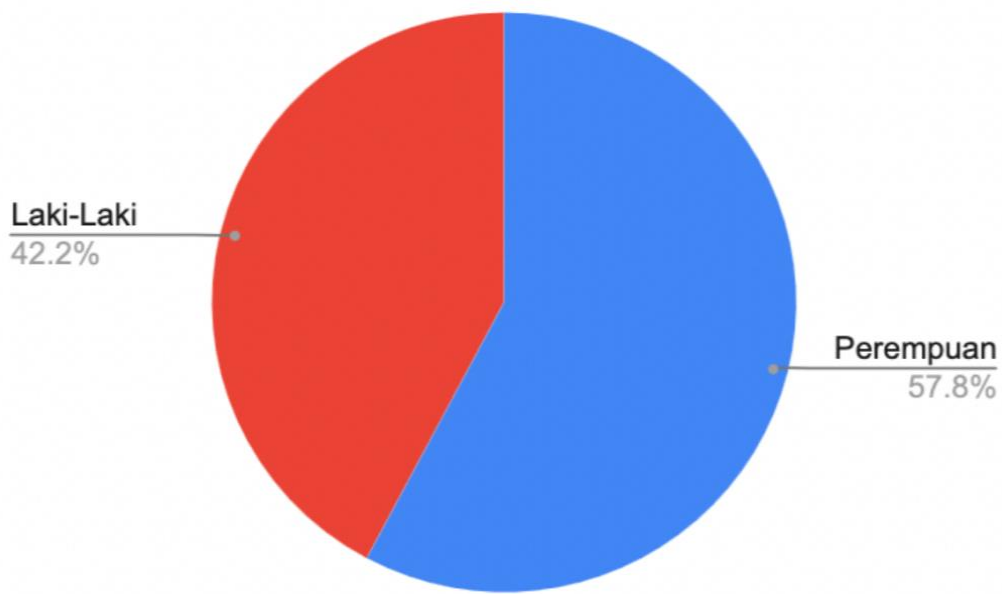

**Gambar 3.2 Pie Chart Umur Responden**

3. Fakultas

Seluruh rumpun ilmu di Universitas Indonesia berhasil direpresentasikan, walau jumlah responden terbanyak berasal dari rumpun ilmu sains & teknologi sebesar 48.2%, diikuti dengan rumun ilmu kesehatan sebesar 21.6%, dan sosial & humaniora sebesar 30.2%.

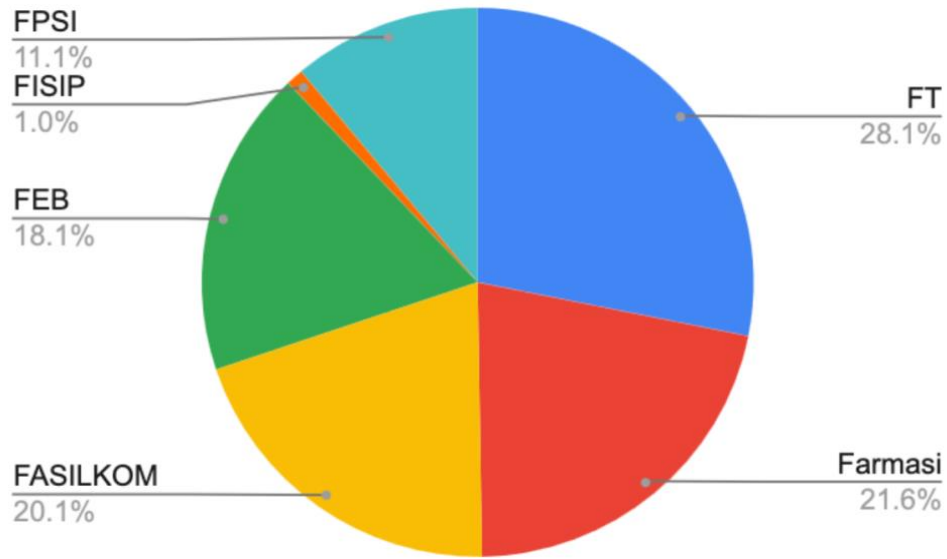

**Gambar 3.3 Pie Chart Fakultas Responden**

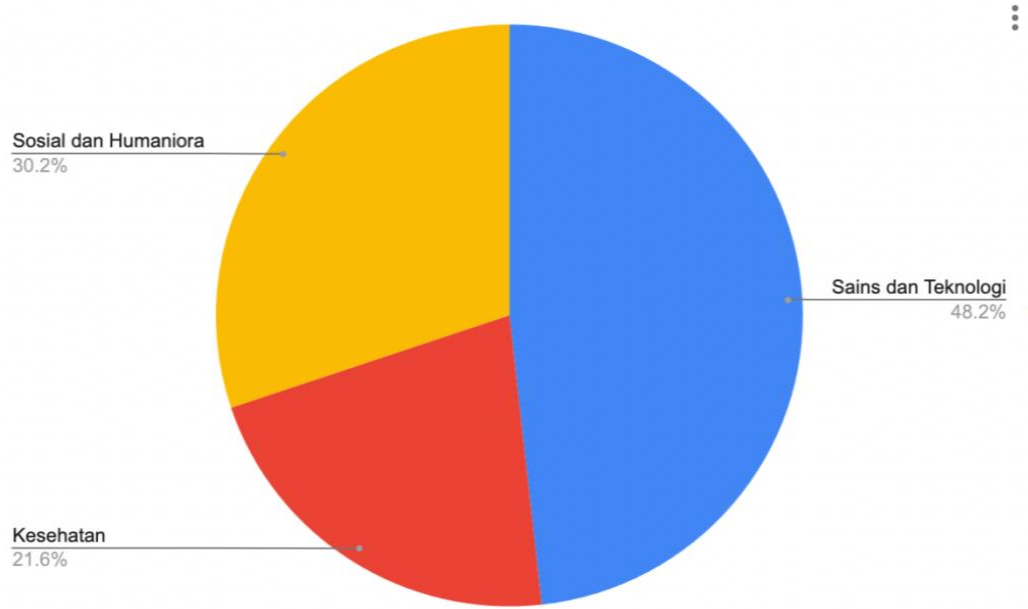

**Gambar 3.4 Pie Chart Rumpun Ilmu Responden**

4. Fakultas

Populasi responden terbesar datang dari Fakultas Farmasi sebesar 21.6%, diikuti dengan Fakultas Ilmu Komputer sebesar 20.1%, dan di urutan ketiga yaitu Departemen Teknik Kimia sebesar 15.1%. Data departemen responden menunjukkan variasi yang cukup tinggi yakni 11 departemen yang masing masing berisi populasi yang tidak sendiri sehingga cukup representatif untuk mencapai inklusivitas.

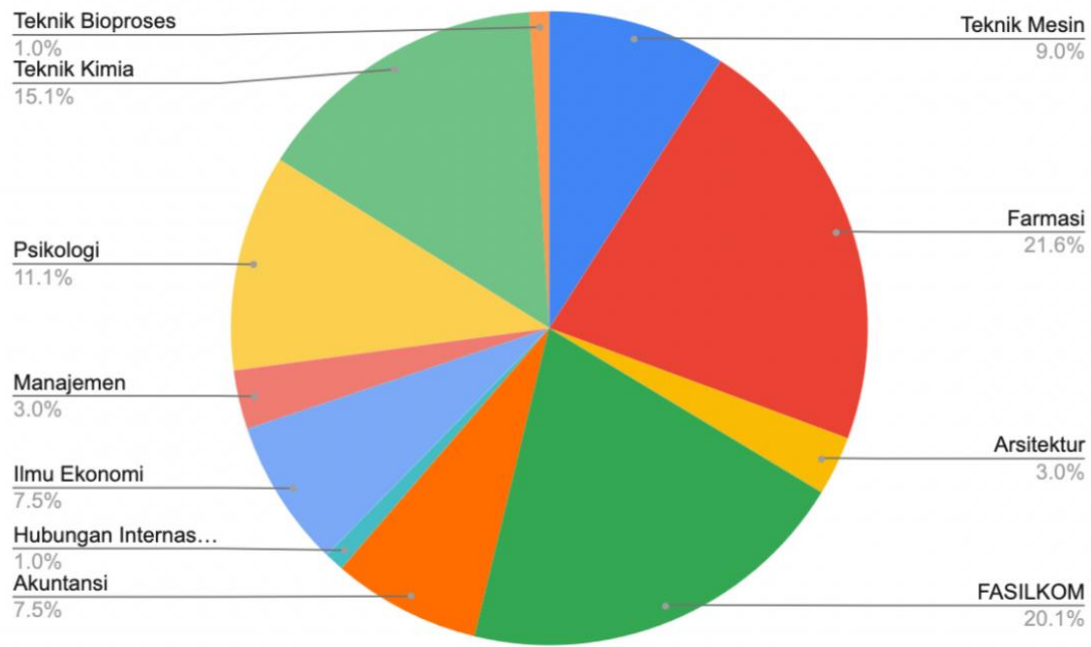

**Gambar 3.5 Pie Chart Departemen Responden**

### 3.3.1.2. Pendapat Responden

Kategori kedua dari kuesioner terdiri dari 3 pertanyaan yang digunakan untuk mengevaluasi kebutuhan (*support*) dan urgensi untuk perbaikan UI SIAK-NG dari sudut pandang pengguna

1. Pendapat responden terhadap pembaharuan SIAK-NG

Pertanyaan ini diberikan untuk mencari tahu demografi dukungan terhadap pembaharuan SIAK-NG. Jawaban dapat dipilih dari 5 pilihan antara lain:

- Tidak boleh, sudah sempurna
- Sedikit keberatan
- Tidak berpengaruh bagi saya
- Sedikit mendukung
- Mendukung

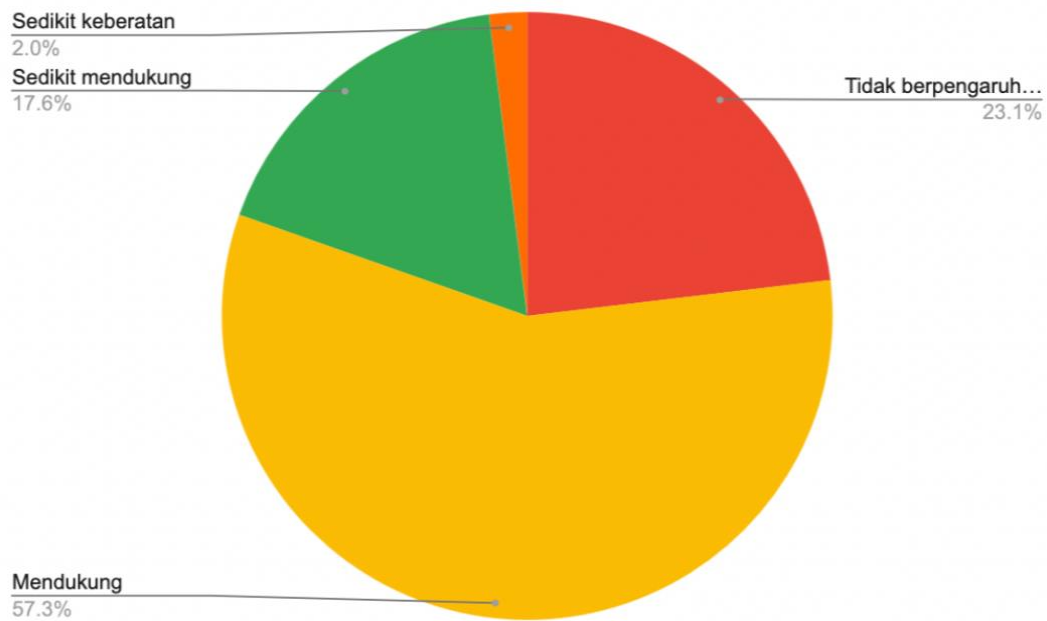

**Gambar 3.6 Pie Chart Dukungan Responden terhadap Pembaharuan SIAK-NG**

Hasil menunjukkan bahwa jumlah yang besar hampir 75% responden memberikan sambutan dan dukungan positif terhadap perbaikan UI SIAK-NG, dan dimana proporsi lebih besar (57.3%) sangat mendukung hal tersebut. Tidak terdapat responden yang memberikan penolakan kuat, yang menunjukkan bahwa SIAK-NG saat ini kurang sukses dalam membuat pengguna merasa terikat. Selanjutnya, hanya 2% yang menunjukkan rasa keberatan terhadap pembaharuan SIAK-NG. Untuk menanggulangi populasi kecil yang sedikit keberatan, penting untuk menerapkan prinsip Jakob's Law agar transisi ke website baru familiar dan tidak membutuhkan adaptasi yang signifikan.

2. Kapan pembaharuan SIAK-NG layaknya dilakukan

Pertanyaan ini diberikan untuk mencari tahu urgensi terhadap pembaharuan SIAK-NG. Jawaban dapat dipilih dari 3 pilihan antara lain:

- Tidak mendukung
- Secepatnya
- Nanti saja (>4 tahun lagi)

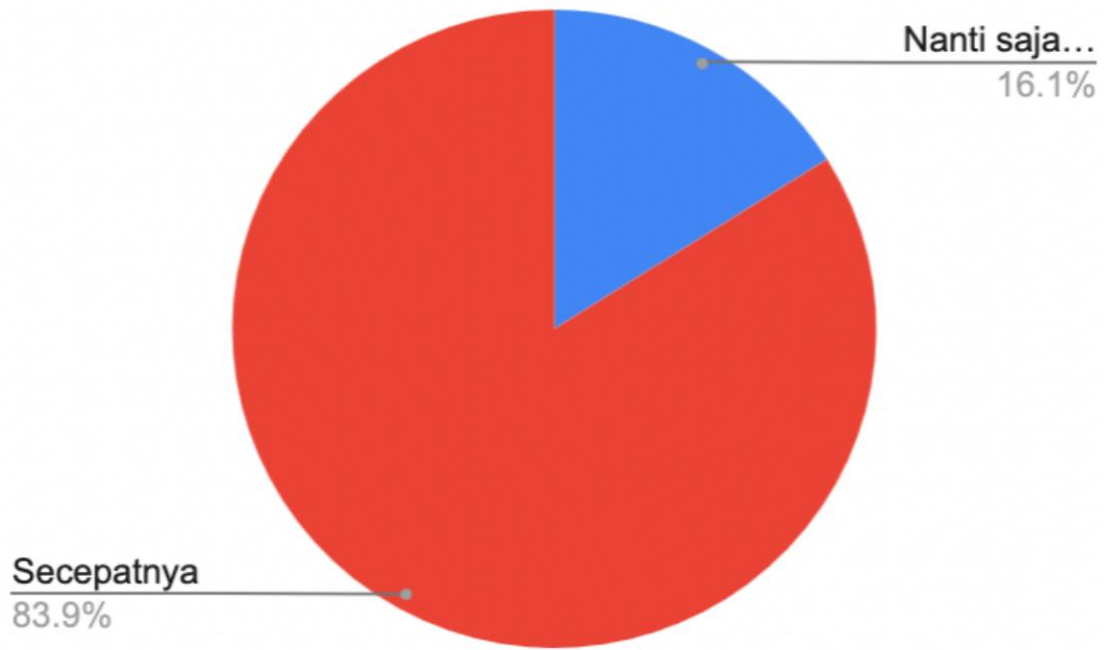

**Gambar 3.7 Pie chart Urgensi terhadap Pembaharuan SIAK-NG**

Hasil menunjukkan bahwa jumlah yang besar sebesar 83.9% mendukung perubahan secepatnya, yang memvalidasi adanya urgensi atas perbaikan UI SIAK-NG.

3. Pengaruh SIAK-NG terhadap produktivitas dan kenyaman perkuliahan

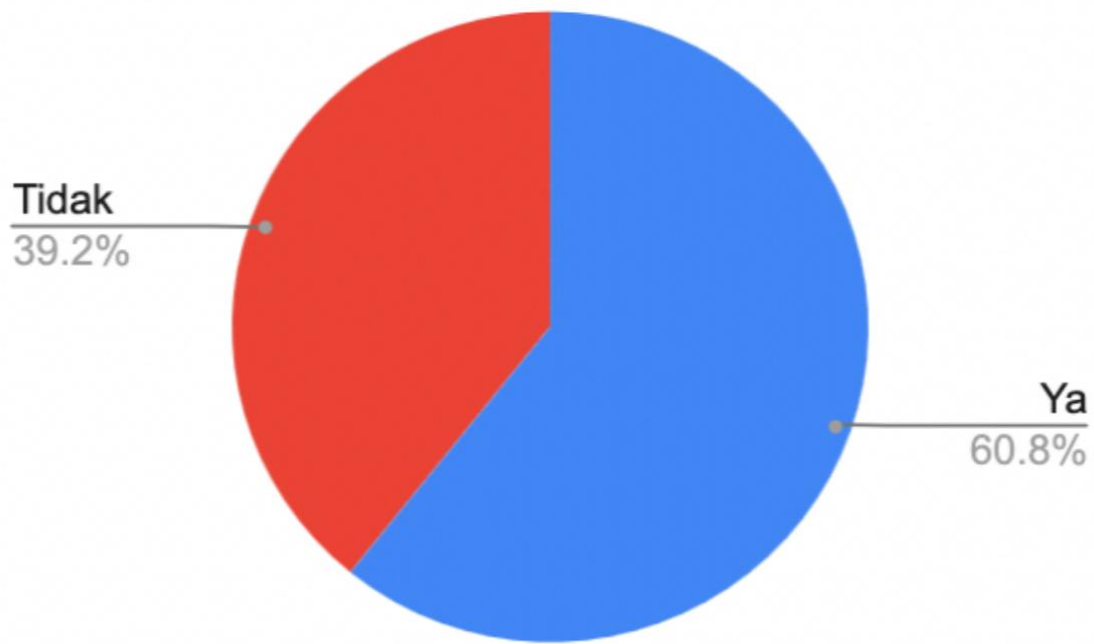

**Gambar 3.8  Pie chart Pengaruh SIAK-NG**

Hasil menunjukkan bahwa jumlah yang besar sebesar 60.8% merasa bahwa kekurangan SIAK-NG berpengaruh terhadap kenyamanan dalam melakukan pekerjaan.

### 3.3.1.3. *User Behavior* Responden

Kategori ketiga dari kuesioner berisi pertanyaan yang menyangkut *user-behavior* dan kegunaan SIAK-NG yang terdiri dari 8 pertanyaan antara lain:

1. Frekuensi penggunaan keseluruhan

Hasil menunjukkan bahwa populasi responden terbesar sebesar 26% mengakses SIAK-NG sebanyak 2-3 kali per minggu. Kategori ini adalah kategori kedua tertinggi dan menunjukkan bahwa penggunaan SIAK-NG cukup sering di kalangan *user role* mahasiswa.

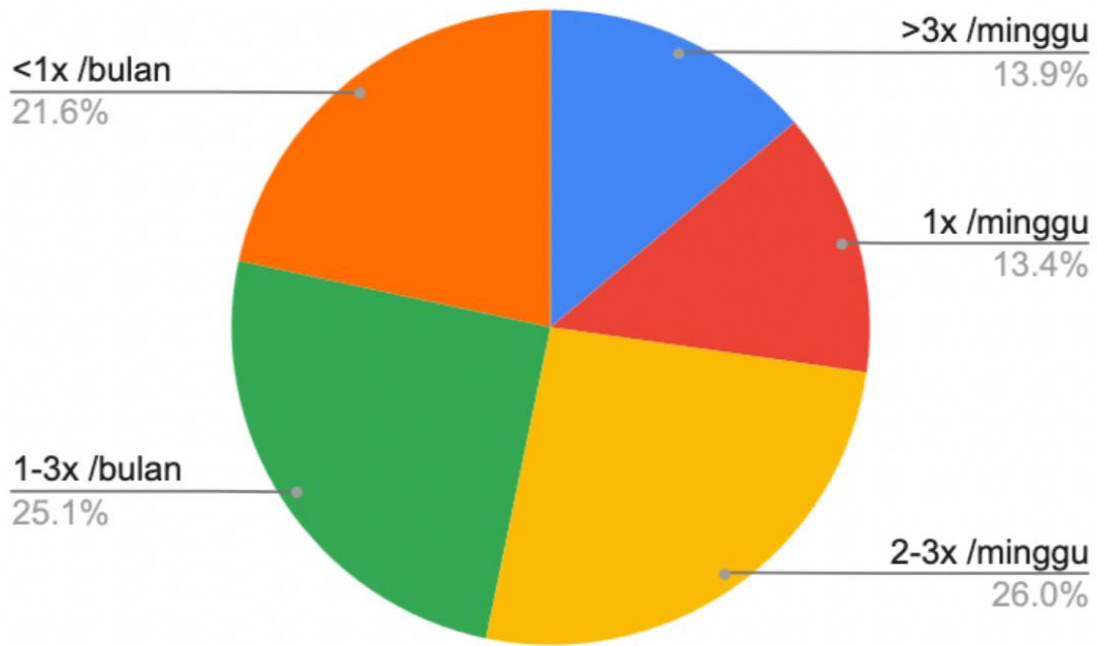

**Gambar 3.9  Pie Chart Frekuensi Penggunaan Keseluruhan Responden**

2.  Frekuensi penggunaan harian

Hasil menunjukkan bahwa populasi responden terbesar sebesar 73.4% mengakses SIAK-NG sebanyak 1-2 kali per hari, sementara sisanya sebesar 26.6% mengakses lebih dari 2 kali per hari.

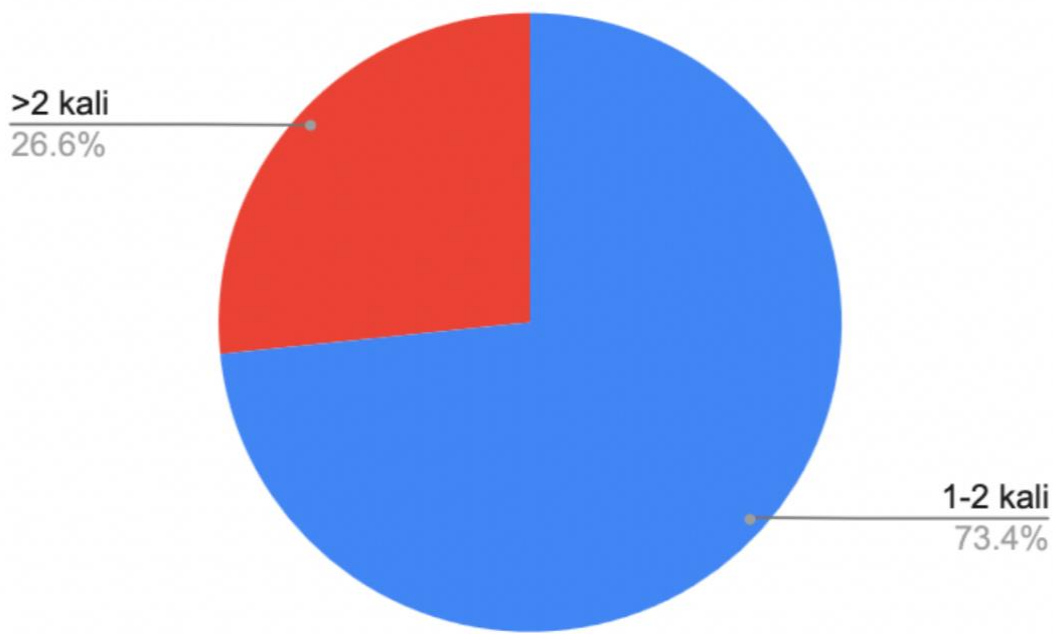

**Gambar 3.10  Pie Chart Frekuensi Penggunaan Harian Responden**

3. Frekuensi penggunaan di desktop site vs. mobile site

Pertanyaan ini dirancang untuk mencari tahu user-behavior tentang preferensi gawai yang dipakai saat mengakses SIAK-NG. Pertanyaan menggunakan medium visual untuk mengkategorikan frekuensi dengan pilihan sebagai berikut:

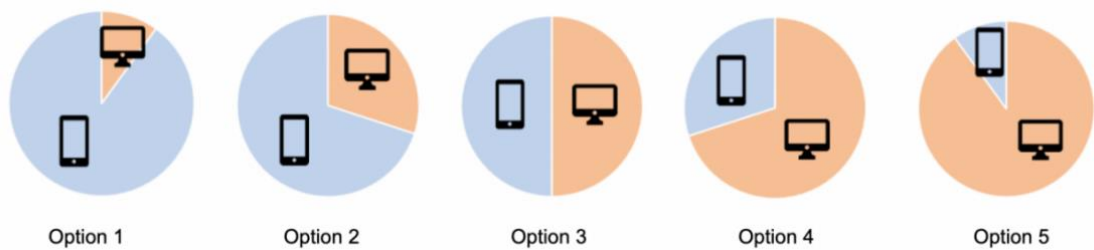

**Gambar 3.11  Gambar yang digunakan sebagai pilihan jawaban pertanyaan frekuensi penggunaan** *mobile vs desktop site*

Hasil menunjukkan bahwa terdapat variasi yang cukup rata antara responden yang sering mengakses melalui *desktop site* dengan *mobile site*. Walau begitu, populasi terbesar sebesar 27.6% mengakses SIAK-NG dengan kisaran 70% pengaksesan melalui mobile site, diikuti dengan populasi terbesar kedua yang beda tipis sebesar 27.1% yang mengakses *desktop* dan *mobile site* secara rata (50-50). Secara keseluruhan terlihat bahwa penggunaan *mobile site* untuk mengakses SIAK-NG mendominasi penggunaan *desktop site.*

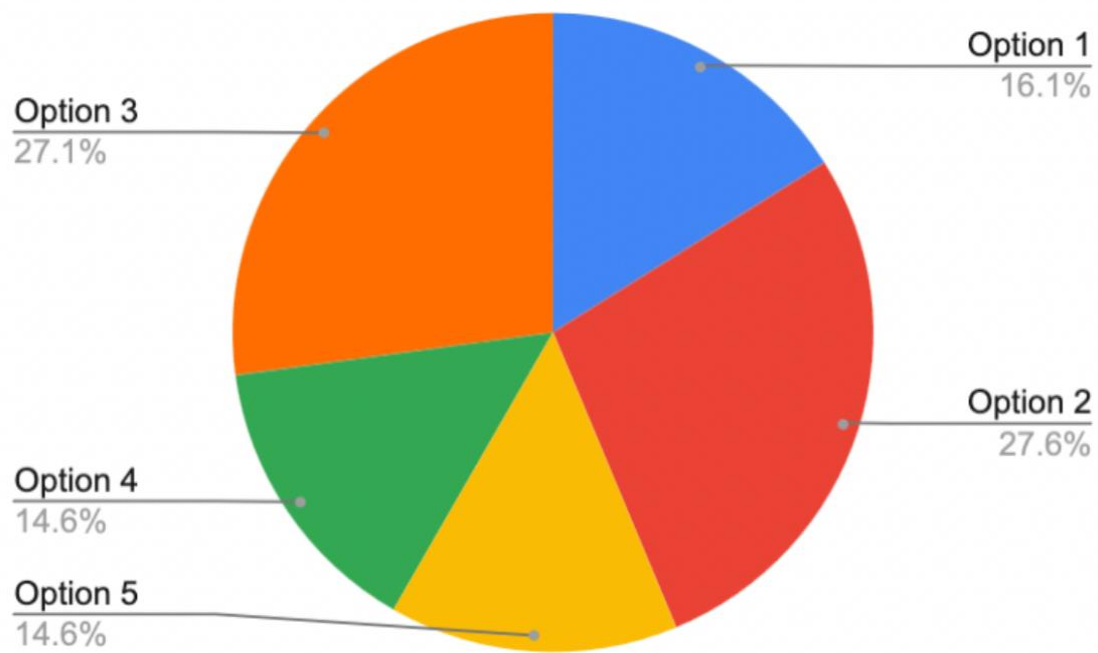

**Gambar 3.12  Pie Chart Frekuensi Penggunaan** *Mobile vs Desktop Site* **Responden**

Pertanyaan selanjutnya berupa kolom pengisian yang digunakan untuk mencari tahu keragaman penggunaan SIAK-NG.

4. Task yang paling sering dilakukan

5. Fitur yang paling sering dipakai

6. Fitur yang paling krusial

7. Fitur yang jarang dipakai

8. Fitur yang kurang efektif

### 3.3.1.4. *User Experience*

Kategori pertanyaan selanjutnya menyangkut perasaan, opini, dan pengalaman pengguna terhadap SIAK-NG. Kategori ini terdiri dari 9 pertanyaan yang mencari tahu goals, pain point, gain point, frustration, first impression, dan feelings:

1. Saran, perbaikan, fitur tambahan yang diharapkan

2. Aspek yang tidak ingin diubah

3. Elemen yang paling pertama menarik perhatian

4. Kesan singkat keseluruhan website

5. Perasaan saat memakai website

6. Kekurangan dan sumber frustasi

7. Pengalaman tidak menyenangkan

8. Penyebab pengalaman tidak menyenangkan

9. Manfaat yang diharapkan

### 3.3.2. In-depth Interview

Dalam penelitian ini, dilakukan *in-depth interview* untuk mendapatkan *insight* dan pemahaman yang lebih mendalam mengenai sudut pandang pengguna website. Terdapat 15 pengguna yang terlibat dalam sesi *usability testing*, yang mewakili tiga persona yang telah ditentukan sebelumnya. Wawancara dilakukan dengan 5 responden untuk setiap user persona karena menurut Nielsen, pengujian dengan 5 orang dapat mengidentifikasi masalah kegunaan yang hampir sebanyak yang ditemukan dengan menggunakan lebih banyak peserta tes (Nielsen, 2012).Wawancara dilakukan secara online melalui platform Google Meets.

Wawancara ini terdiri dari dua bagian, yaitu pertanyaan umum dan pertanyaan khusus. Pertanyaan umum mencakup profil dan informasi umum responden, seperti nama, pekerjaan, dan frekuensi penggunaan website. Sedangkan pertanyaan khusus berkaitan dengan pengalaman pengguna. Untuk pertanyaan khusus bersifat *open-ended* agar memberikan kesempatan bagi responden untuk memberikan jawaban yang personal. Beberapa pertanyaan yang diajukan meliputi kesan pertama saat melihat halaman, elemen-elemen yang menarik perhatian, hambatan utama yang dirasakan, serta persepsi terhadap fungsi SIAK-NG. Selain itu, responden juga diminta untuk memberikan pendapat mengenai alur tugas, kesulitan yang dirasakan, perasaan setelah menyelesaikan tugas, dan saran untuk perbaikan. Mereka juga memberikan pendapat mengenai kelebihan dan kekurangan yang mereka temukan.

Selanjutnya, responden juga berpartisipasi dalam diskusi *storyboarding*, yang bertujuan untuk menghasilkan ide dan gagasan perbaikan lebih lanjut. Setelah sesi wawancara, responden diminta untuk memberikan umpan balik terhadap SIAK-NG. Wawancara dilakukan sesuai dengan persona yang diwakili oleh responden, dan dilakukan setelah sesi *usability testing*

### 3.4. Pengolahan Data

### 3.4.1. *Empathize*

Pada tahap awal, data dikumpulkan melalui survei untuk menentukan jumlah dan kualitas kelompok persona. Melalui analisis *hierarchical cluster* dan K-modes, hasil survei digunakan untuk mengidentifikasi persona.

#### 3.4.1.1. *Clustering*

Dalam penelitian ini, pendekatan *cluster analysis* digunakan untuk mengolah data yang diperoleh dari survey. Data tersebut direpresentasikan sebagai angka untuk mengubah data kualitatif menjadi kuantitatif. *Cluster analysis* dimulai dengan pengolahan data menggunakan perangkat lunak IBM SPSS Statistics 26. Dengan menggunakan *hierarchical clustering* dengan Metode Ward yang melibatkan penghubungan pasangan *cluster* secara berkala hingga semua objek data termasuk dalam hierarki. Linkage Ward digunakan untuk menentukan dan menghitung jarak antara dua cluster sebagai peningkatan jumlah kesalahan kuadrat setelah menggabungkan dua cluster menjadi satu. Hasil dari prosedur *hierarchical cluster* disajikan dalam bentuk *agglomeration schedule*, seperti yang terlihat dalam Gambar 3.13.

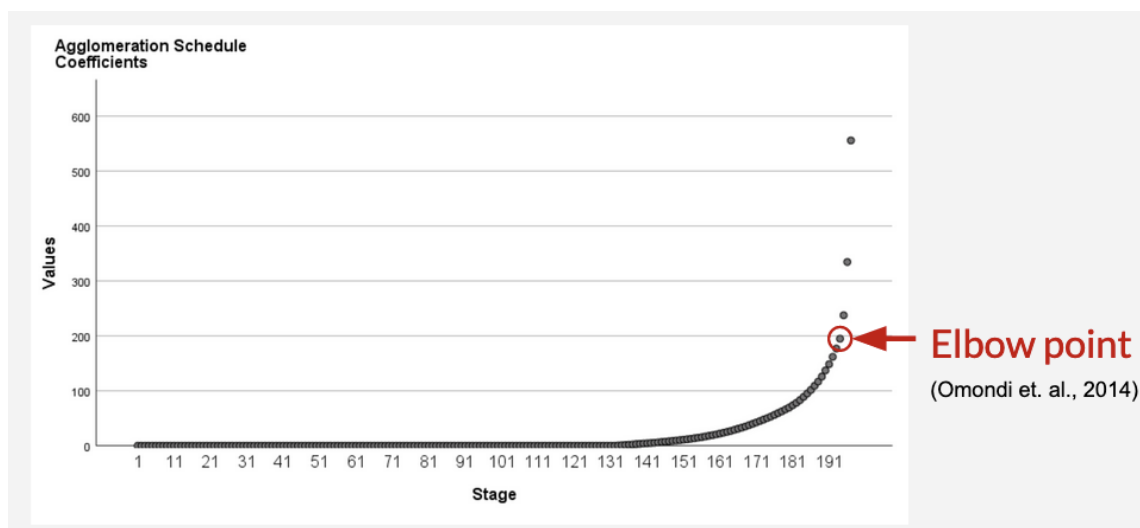

**Gambar 3.13  Grafik Agglomeration Schedule (Ward's Method)**

Penulis menggunakan pendekatan *elbow point* dari *agglomeration schedule* untuk mengidentifikasi jumlah *cluster* persona dalam *dataset*, yang dapat dilihat pada grafik sebagai titik yang ditandai dengan panah. Peningkatan signifikan, dalam hal ini pada tahap ke-97, menandai *elbow point* dari koefisien keseluruhan pada tahap ke-101. Jumlah total *cluster* sekarang dapat dihitung menggunakan rumus berikut:

$$\text{Jumlah stage- Elbow point = cluster persona} \qquad (3.1)$$

$$\text{101 stage- 97 stage = 3 Persona}$$

Selanjutnya, karakteristik per cluster persona harus ditentukan. Definisi karakteristik persona dalam penelitian ini dilakukan dengan menggunakan Microsoft Excel dan metode *cluster* k-modes. Variabel-variabel yang digunakan untuk menentukan persona pengguna sama dengan variabel pada *hierarchical cluster*, seperti data diri dan *user-behavior*. Jawaban dari masing-masing karakteristik pada kuesioner kemudian direpresentasikan menjadi angka-angka seperti yang terlihat pada Tabel di bawah

**Tabel 3.1 Pendefinisian Karakteristik**

| Q1 Dukungan terhadap perubahan | |
|---|---|
| Tidak boleh, sudah sempurna | 1 |
| Sedikit keberatan | 2 |
| Tidak berpengaruh bagi saya | 3 |
| Sedikit mendukung | 4 |
| Mendukung | 5 |
| **Q2 Urgensi perbaikan** | |
| Secepatnya | 1 |
| Nanti saja, periode selanjutnya (>4 tahun) | 2 |
| **Q3 Frekuensi penggunaan keseluruhan** | |
| >3 kali per minggu | 1 |
| 2-3 kali per minggu | 2 |
| Sekali seminggu | 3 |
| 1-3 kali sebulan | 4 |

Tabel 3.1 Pendefinisian Karakteristik (Lanjutan)

| Q4 Frekuensi penggunaan harian | |
|---|---|
| >2 kali | 1 |
| 1-2 kali | 2 |
| **Q5 Frekuensi penggunaan di desktop site vs. mobile site** | |
| Option 1 | 1 |
| Option 2 | 2 |
| Option 3 | 3 |
| Option 4 | 4 |
| Option 5 | 5 |
| **Q6 Pengaruh SIAK-NG terhadap kenyaman perkuliahan** | |
| Ya | 1 |
| Tidak | 2 |
| **Q7 Jenis Kelamin** | |
| Perempuan | 1 |
| Laki-Laki | 2 |
| **Q8 Fakultas** | |
| FT | 1 |
| FMIPA | 2 |
| FASILKOM | 3 |
| FK | 4 |
| FKG | 5 |
| Farmasi | 6 |
| FKM | 7 |
| Ilmu Keperawatan | 8 |
| FEB | 9 |
| FH | 10 |
| FISIP | 11 |
| FIB | 12 |
| FPSI | 13 |

**Tabel 3.1 Pendefinisian Karakteristik (Lanjutan)**

| Q9 Departemen | |
|---|---|
| Departemen Akuntansi | 1 |
| Departemen Arsitektur | 2 |
| Departemen Ilmu Ekonomi | 3 |
| Departemen Teknik Kimia | 4 |
| Departemen Teknik Mesin | 5 |
| Departemen Teknik Metalurgi dan Material | 6 |
| Farmasi | 7 |
| Hubungan Internasional | 8 |
| Ilmu komputer | 9 |
| Manajemen | 10 |
| Psikologi | 11 |
| Sistem Informasi | 12 |

Menggunakan Microsoft Excel dilakukan iterasi sebanyak 4 kali hingga diperoleh hasil yang konsisten:

**Tabel 3.2 Tabel Iterasi K-Means**

| | Q1 | Q2 | Q3 | Q4 | Q5 | Q6 | Q7 | Q8 | Q9 |
|---|---|---|---|---|---|---|---|---|---|
| **Cluster 1** | 3 | 1 | 2 | 2 | 3 | 2 | 1 | 1 | 7 |
| **Cluster 2** | 5 | 1 | 2 | 2 | 2 | 1 | 2 | 1 | 4 |
| **Cluster 3** | 5 | 1 | 4 | 2 | 2 | 1 | 1 | 3 | 12 |
| **I=2** | | | | | | | | | |
| **Cluster 1** | 3 | 1 | 2 | 2 | 2 | 2 | 1 | 6 | 7 |
| **Cluster 2** | 5 | 1 | 2 | 2 | 2 | 1 | 2 | 1 | 4 |
| **Cluster 3** | 5 | 1 | 4 | 2 | 2 | 1 | 1 | 3 | 12 |
| **I=3** | | | | | | | | | |
| **Cluster 1** | 3 | 1 | 2 | 2 | 2 | 2 | 1 | 6 | 7 |
| **Cluster 2** | 5 | 1 | 2 | 2 | 2 | 1 | 2 | 1 | 4 |
| **Cluster 3** | 5 | 1 | 4 | 2 | 3 | 1 | 1 | 3 | 12 |
| **I=4** | | | | | | | | | |
| **Cluster 1** | 3 | 1 | 2 | 2 | 2 | 2 | 1 | 6 | 7 |
| **Cluster 2** | 5 | 1 | 2 | 2 | 2 | 1 | 2 | 1 | 4 |
| **Cluster 3** | 5 | 1 | 4 | 2 | 3 | 1 | 1 | 3 | 12 |

Hasil karakteristik dari I=4 dapat dilihat pada tabel di bawah:

**Tabel 3.3 Hasil Karakteristik Per Persona Cluster**

| Variabel | Cluster 1 | Cluster 2 | Cluster 3 |
|---|---|---|---|
| Dukungan perbaikan | Netral | Mendukung | Mendukung |
| Urgensi perbaikan | Secepatnya | Secepatnya | Secepatnya |
| Frekuensi akses keseluruhan | 2-3 kali per minggu | 2-3 kali per minggu | 1-3 kali sebulan |
| Frekuensi akses/hari | 1-2 kali | 1-2 kali | 1-2 kali |
| Mempengaruhi pekerjaan | Tidak | Ya | Ya |
| Mobile vs Desktop | 70% Mobile | 70% Mobile | 50% Mobile |
| Jenis Kelamin | Perempuan | Laki-Laki | Perempuan |
| Umur | 18 | 19 | 19 |
| Angkatan | 2022 | 2022 | 2022 |
| Fakultas | Farmasi | FT | FASILKOM |
| Jurusan | Farmasi | Teknik Kimia | Sistem Informasi |

### 3.4.1.2. *User Persona*

Setiap ciri-ciri persona yang telah dijelaskan sebelumnya kemudian diubah menjadi diagram yang mengadung deskripsi lebih lengkap tentang persona tersebut, termasuk profil, deskripsi, penggunaan, masalah/frustrasi, dan tujuan/kebutuhan persona tersebut. Diagram user persona untuk masing-masing user dapat dilihat pada Gambar 3.14, Gambar 3.15, dan Gambar 3.16.

## Persona Cluster 1

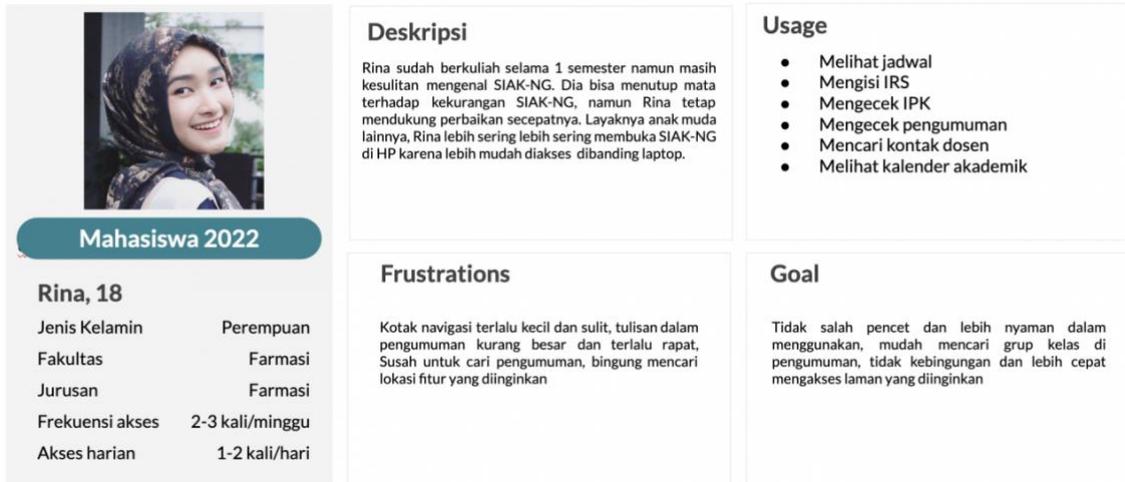

**Gambar 3.14 Diagram User Persona 1**

## Persona Cluster 2

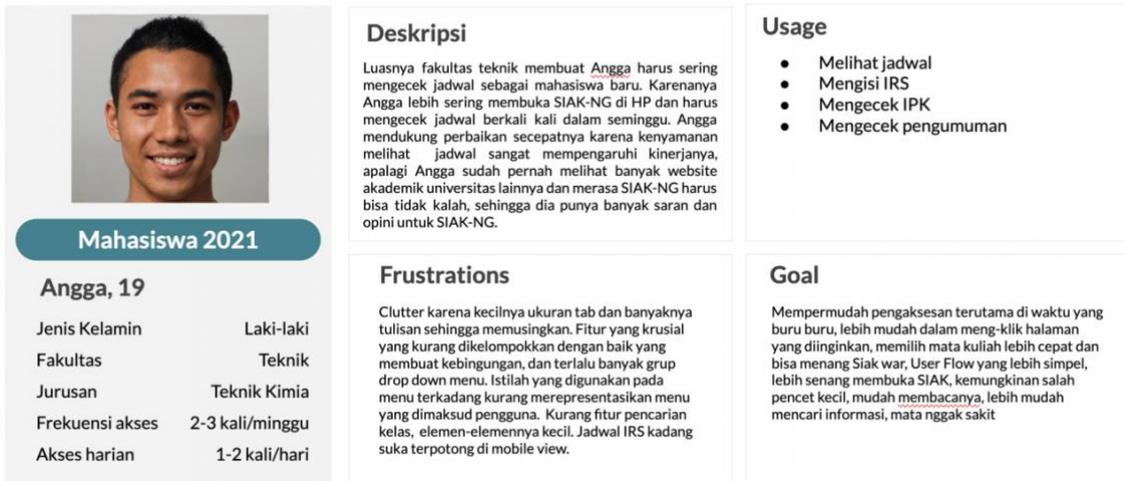

**Gambar 3.15 Diagram User Persona 2**

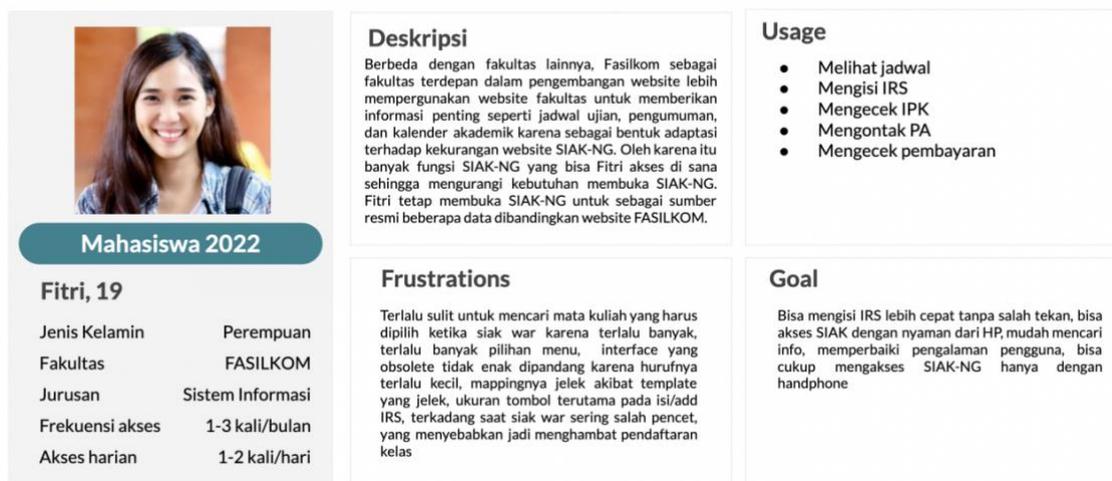

**Gambar 3.16 Diagram User Persona 3**

### 3.4.2. *Define*

#### 3.4.2.1. *Empathy Mapping*

*Empathy Mapping* tahap *define* melibatkan penggunaan *empathy map* untuk melakukan visualisasi kolaboratif guna menggambarkan secara jelas apa yang kita ketahui tentang jenis pengguna tertentu (Nielsen, 2018). Pada tahap ini, setelah mendapatkan jumlah dan mendefinisikan *user persona*, penulis melakukan *empathy map* untuk membantu mengklarifikasi masalah atau kesulitan yang mungkin dialami oleh masing-masing persona. Kategori empathy map mengacu pada apa yang dikatakan (*say*), dilakukan (*do*), dirasakan (*feel*) dan dipikirkan (*think*) oleh persona. Semua kategori *empathy map* kemudian dianalisis untuk memahami kesulitan dan keuntungan persona terkait dengan website.

Berdasarkan jumlah *user persona* yang diperoleh pada tahap sebelumnya, penulis dapat mengembangkan tiga *empathy map* berdasarkan semua pendapat setiap persona. *Empathy map* pertama didasarkan pada karakter Rina (persona 1), karakter Angga (persona 2), dan karakter Fitri (persona 3). Visualisasi ketiga *empathy map* ini dapat dilihat pada Gambar 3.17, Gambar 3.18, dan Gambar 3.19.

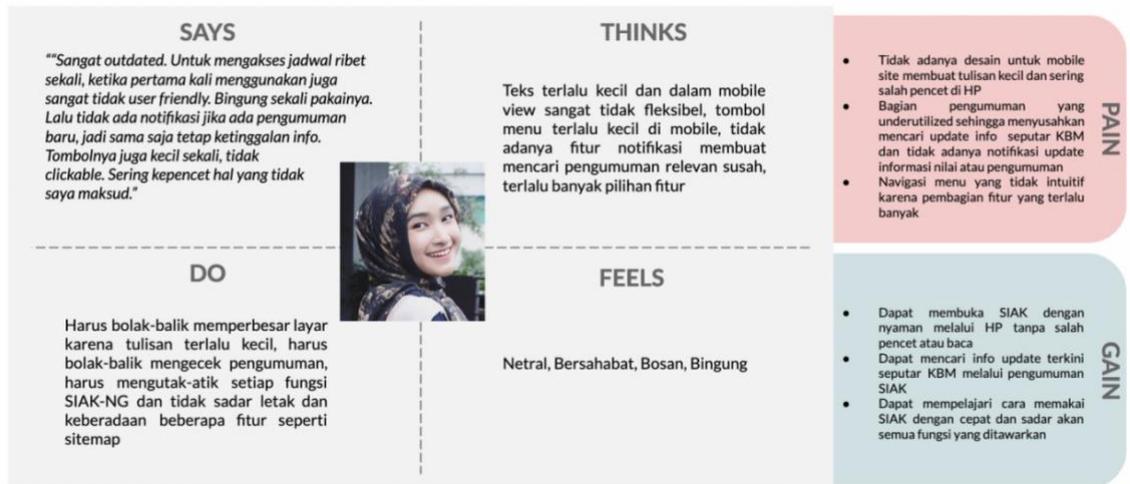

**Gambar 3.17 Diagram Empathy Map Persona 1**

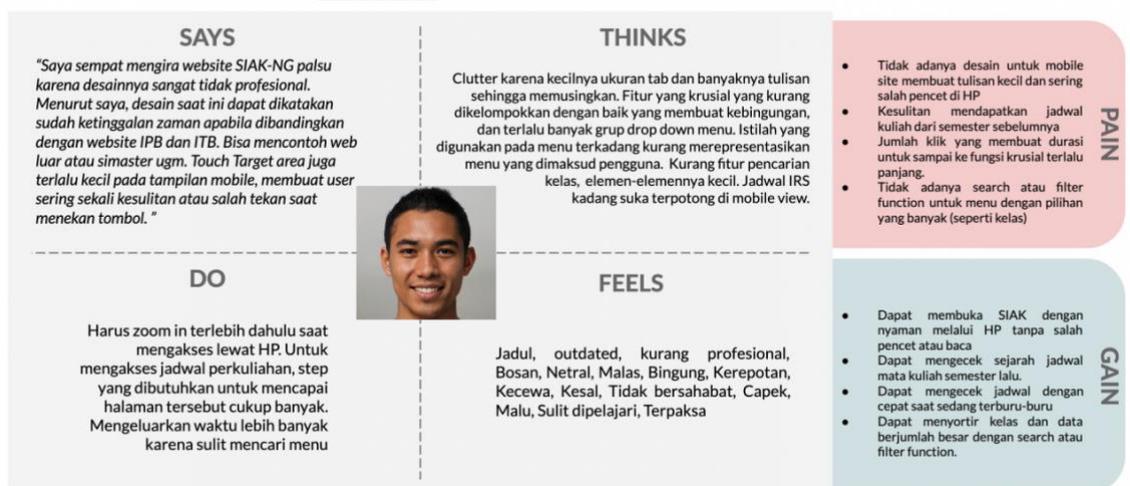

**Gambar 3.18 Diagram Empathy Map Persona 2**

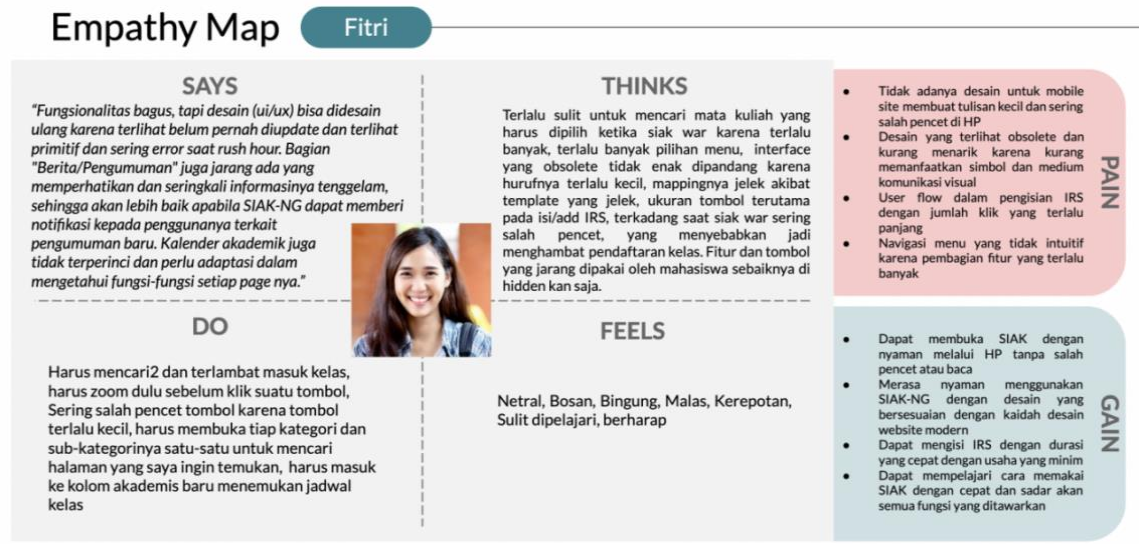

**Gambar 3.19 Diagram Empathy Map Persona 3**

### 3.4.3. *Ideate*

Kemudian, melalui *in-depth interview* diperoleh pandangan dari responden tentang website. *Empathy map* digunakan untuk mengembangkan *storyboard* untuk skenario. Hasil kuesioner bagian *user experience* dianalisis sebagai pain point dan dijadikan bahan diskusi dalam *in-depth interview* kepada masing-masing persona untuk diulik lebih dalam. Kemudian, dicari solusi paling ideal dari sudut pandang semua *stakeholder*.

### 3.4.4. *Prototype*

Tahap selanjutnya adalah mengubah ide-ide menjadi *prototype*. Pembuatan *prototype* akan dilakukan berdasarkan *insight* yang diperoleh pada tahap sebelumnya. Penyesuaian elemen, warna, dan tampilan juga diadaptasi dari *21 Laws of UX* dan 5 *Principles of Visual Design*. *Prototype* akan terdiri dari beberapa UI dan *path* baru untuk beberapa menu dari website SIAK-NG. Tahap *prototyping* terdiri dari 2 bagian yaitu *low-fidelity wireframe* dan *high-fidelity rapid prototype.*

*Low-fidelity wireframe* digunakan untuk menerjemahkan ide layout awal dan membantu membangun konsep desain. *Wireframe* tersebut dibuat menggunakan program Figma. *Low-fidelity wireframe* kemudian dikembangkan menjadi *prototype* yang dapat diklik pada tahap kedua pengembangan *prototype*.

Penulis membangun prototype website dengan menggabungkan fitur-fitur baik dari halaman-halaman website yang sudah ada, kemudian meningkatkan dan menambahkan fitur-fitur dan desain website baru berdasarkan hasil *brainstorm storyboarding*. Meskipun *prototype* yang dapat diklik dibuat dengan fungsi atau fitur baru yang ditingkatkan, fungsi-fungsi tersebut masih identik dengan website lama dengan harapan mempermudah transisi dan adaptasi pengguna. Sekalipun tujuan utama penulis adalah untuk mengembangkan *desktop site*, hasil survei menunjukkan kebutuhan atas *mobile site* yang baik, oleh karena itu penelitian ini mengusahakan pengembangan keduanya.

### 3.4.5. *Testing*

Masuk ke bagian terakhir dari *design thinking* yaitu tahap *testing*. *Testing* terbagi 2 menjadi *usability testing* dan PSSUQ. Kedua tes ini digunakan bersamaan untuk mengukur tidak hanya performa desain tapi juga perasaan pengguna. Setelah selesai merancang *user interface* menggunakan Figma dan mengembangkan *prototype* yang dapat diklik, *prototype* siap untuk diuji. Mengacu pada Jakob Nielsen (2012), pengujian cukup dilakukan dengan 5 individu karena ini memungkinkan untuk mengidentifikasi hampir semua masalah kegunaan seperti halnya pengujian dengan jumlah peserta yang lebih banyak. Penulis melibatkan 5 peserta untuk setiap persona, baik untuk *usability testing* UI lama maupun baru, dengan target total 15 peserta. Aturan testing dijelaskan secara online menggunakan layanan Google Meets dan dilakukan tanpa dibantu.

Tugas yang diberikan kepada responden bertujuan untuk mengumpulkan data metrik performa untuk *user interface* yang lama dan yang baru. Evaluasi hasil *testing* akan membandingkan metrik yang diperoleh dari versi website lama dan versi prototype baru untuk menentukan apakah prototype baru merupakan perbaikan dari SIAK-NG yang lama.

### 3.4.5.1. *Usability Testing*

Uji coba kegunaan (*usability testing*) bertujuan untuk menilai performa desain dalam penggunaan dan bersifat kuantitatif. Ini melibatkan pengamatan terhadap pengguna saat mereka mencoba menyelesaikan tugas dan dapat dilakukan untuk berbagai jenis desain.

Dilakukan *usability testing* pada total 15 responden, dengan 5 responden dari setiap persona yang ditentukan. Jumlah responden ditentukan berdasarkan penelitian yang menunjukkan bahwa pengujian pada 5 responden sudah cukup untuk mengidentifikasi masalah kegunaan, dibandingkan dengan mengevaluasi jumlah yang lebih banyak (Nielsen, 2012). Pengujian ini dilakukan untuk menentukan performa setiap halaman yang diuji, yang meliputi halaman login, halaman utama, halaman profil/akademik, halaman IRS, halaman administrasi, dan halaman jadwal.

Untuk skenario yang digunakan sebagai dasar untuk pengujian diambil dari hasil penggunaan website SIAK-NG dari kuesioner, yang kemudian diverifikasi dalam wawancara dengan perwakilan dari setiap *cluster*. Hasilnya menunjukkan bahwa meskipun terdapat perbedaan dalam perilaku pengguna, secara keseluruhan tugas yang dilakukan di antara setiap kelompok adalah sama, namun dengan prioritas yang beragam. Daftar skenario yang diuji antara lain:

- Task 1: Mengecek pengumuman
- Task 2: Mencari nilai per mata kuliah
- Task 3: Mengecek status pembayaran BOP
- Task 4: Mengecek jadwal waktu dan lokasi kelas
- Task 5: Melakukan registrasi IRS
- Task 6: Mencari informasi mata kuliah yang diselenggarakan fakultas saya

Hasil interview memberikan *insight* mengenai urgensi, prioritas task, dan kekurangan layout. Oleh karena itu, *prototype* SIAK-NG baru memberikan solusi dengan mengatur ulang kategorisasi dan layout untuk beberapa fungsi. Perbandingan flowchart pengerjaan task dengan UI lama dan UI baru dijadikan dasar untuk *expected path* dalam menilai metrik *direct success*, dan dapat dilihat di bawah:

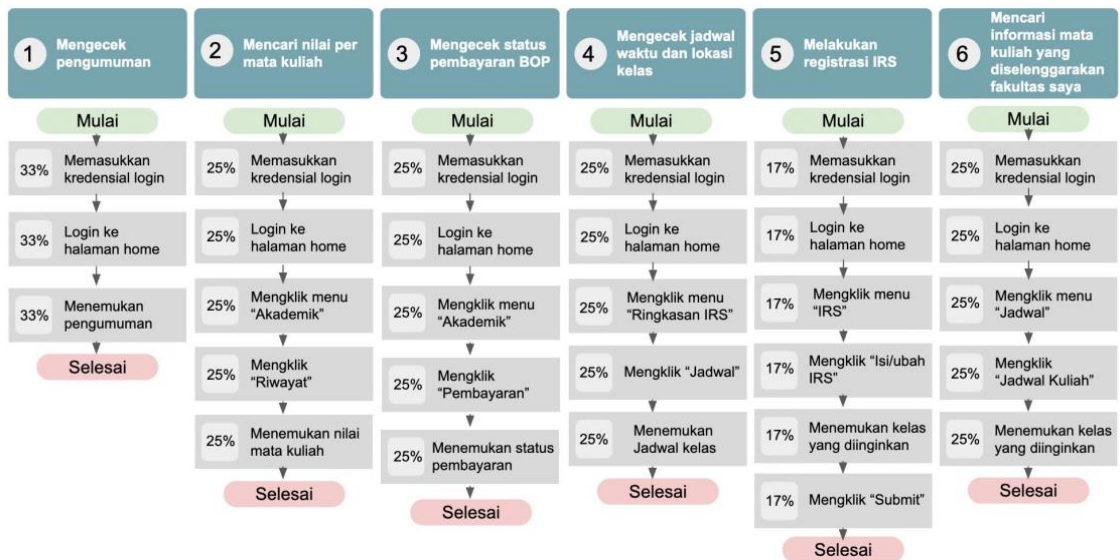

**Gambar 3.20 Flowchart Task untuk SIAK-NG Lama (Mobile dan Desktop)**

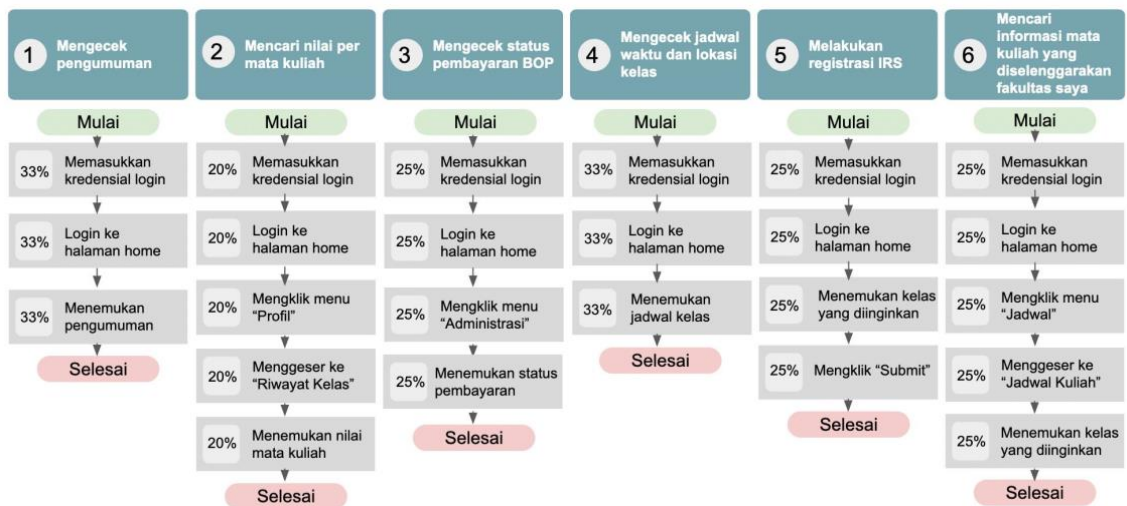

**Gambar 3.21 Flowchart Task untuk SIAK-NG Baru (Mobile)**

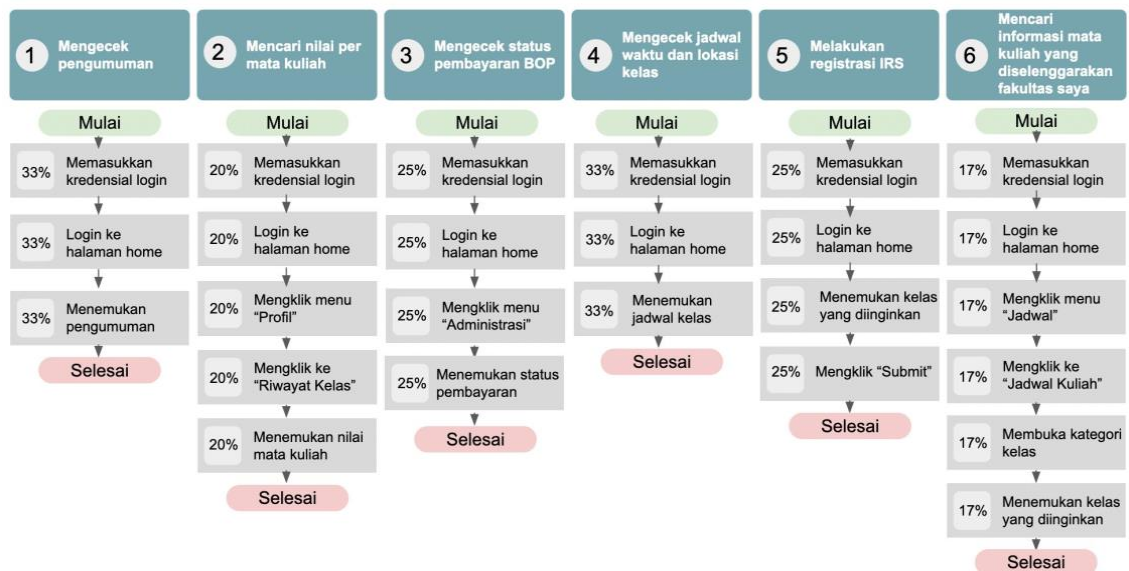

**Gambar 3.22 Flowchart Task untuk SIAK-NG Lama (Desktop)**

Sesi *usability testing* dilakukan dengan menggunakan Maze, sebuah aplikasi product testing profesional untuk mengukur, mencatat, dan mengkalkulasi variabel performa masing masing *prototype*. Penulis memilih menggunakan Maze untuk sesi *supervised testing* karena Maze dapat merekam dan memproses data *usability* kualitatif menjadi kuantitatif secara otomatis dengan sistem perhitungan yang telah diuji dan diterima di lingkungan website development; menggabungkan metrik sehingga mengurangi kemungkinan perhitungan yang inkonsisten yang dapat terjadi dengan pemrosesan data kualitatif secara manual. Selain itu Maze juga menyediakan opsi untuk membuat *heatmap* yang dapat memvisualisasi efektivitas UI sehingga mempermudah komunikasi antar tim dalam proses pengembangan website nantinya. Variabel kinerja yang diukur adalah sebagai berikut:

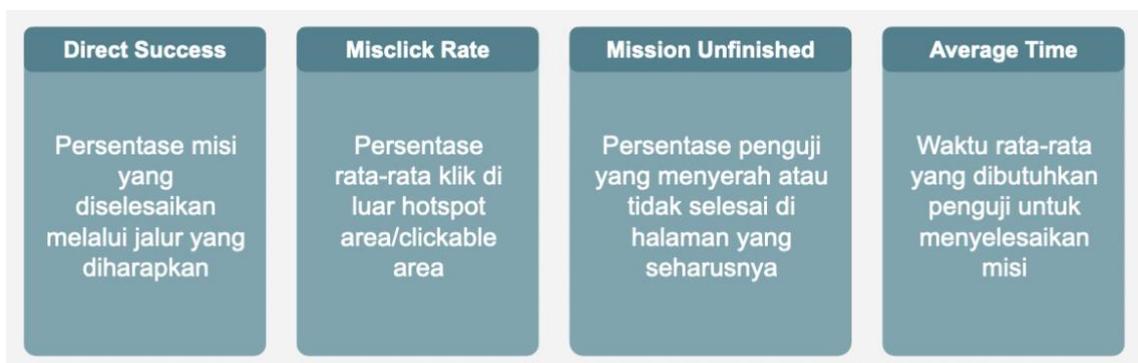

**Gambar 3.23 Variabel Dasar untuk Perhitungan Usability Testing (Usability Score)**

Metrik di atas selaras dengan variabel *usability testing* yang direkomendasikan dari hasil studi literatur. Efektivitas mencakup metrik *direct success, misclick rate,* dan *mission unfinished*. Efisiensi diukur dari *average time. Direct success* menunjukkan kemudahan pengguna untuk menemukan halaman yang dicari dan mengevaluasi seberapa intuitif layout dalam berkomunikasi secara visual. *Mission unfinished* menunjukkan kemampuan responden untuk menyelesaikan tugas yang diberikan dan seberapa merepotkan melakukan suatu task. *Misclick rate* mencatat kesalahan yang dilakukan oleh responden selama menyelesaikan tugas. Efisiensi dievaluasi melalui pengukuran waktu yang diperlukan oleh responden untuk menyelesaikan tugas yang diberikan. Waktu yang lebih singkat menunjukkan efisiensi yang lebih tinggi.

Maze menggabungkan seluruh variabel di atas menjadi *usability score* dengan 3 sistem pemrosesan data (Cunha, 2023). *Usability score* sendiri adalah angka antara 0-100 yang memberikan penilaian terhadap setiap desain UI, tugas, dan path. Tiga jenis perhitungan untuk menemukan usability score adalah

1. *Screen Usability Score* (SCUS)
2. *Mission Usability Score* (MIUS)
3. *Maze Usability Score* (MAUS)

### 3.4.5.1.1. *Screen Usability Score (SCUS)*

*Screen Usability Score* (SCUS) mengukur seberapa mudah pengguna dapat melakukan tugas pada setiap layar yang ada dalam jalur yang diharapkan. Skor ini mencerminkan kemudahan penggunaan dan intuitivitas desain prototype. Prinsip perhitungan yang akan mengurangi skor adalah:

1. Mengklik *hotspot* yang berbeda dengan yang diharapkan. Hal ini menunjukkan bahwa pengguna keluar dari jalur yang diharapkan, dan dapat menyebabkan frustrasi atau kehilangan ketertarikan. Selain itu, jika pengguna menghentikan tugas sebelum mencapai layar akhir, ini menjadi indikasi bahwa ada sesuatu yang tidak beres dan perlu diperiksa.
2. Menghentikan tugas tanpa mencapai layar akhir. Hal ini merupakan indikasi pengguna merasa kesulitan dan lelah.

3. Mengklik secara tidak sengaja. kesalahan klik juga dapat mengurangi *usability score*. Kesalahan klik terjadi ketika pengguna mengklik area yang bukan hotspot atau tidak dapat diklik. Dalam produk yang sudah diluncurkan, kesalahan klik dapat membawa pengguna ke halaman yang salah, yang berhubungan dengan poin sebelumnya. Dalam konteks *prototype*, kesalahan klik tidak mengarahkan tester ke mana pun karena ini merupakan klik di luar area yang dapat diklik.

4. Menghabiskan terlalu banyak waktu pada layar tertentu. Terlalu banyak waktu yang dihabiskan oleh pengguna di satu layar juga dapat mengindikasikan adanya masalah yang perlu diperbaiki. Di dalam *prototype*, waktu yang terlalu lama menandakan adanya kebingungan dari perspektif tester.

Prinsip-prinsip ini kemudian diterjemahkan ke dalam data kuantitatif. *Exit rate* dan *misclick rate* digunakan untuk menggambarkan prinsip (1) dan (2). Setiap persen pengguna yang menyerah atau berhenti sebelum mencapai layar sukses akan mengurangi 1 poin *usability score*. Sementara itu, prinsip (3) diwakili oleh tingkat *misclick rate*. Namun, tidak setiap kesalahan klik merupakan indikasi tindakan yang salah, oleh karena itu setiap persen kesalahan klik akan mengurangi 0,5 poin *usability*. Prinsip (4) tentang durasi rata-rata juga dihitung dengan aturan tertentu, yaitu tidak ada pengurangan poin untuk durasi 0 hingga 5 detik, kemudian setiap 2 detik akan mengurangi 1 poin kegunaan dari durasi 5 hingga 25 detik, dan setiap durasi di atas 25 detik akan mengurangi 10 poin kegunaan. Rumus untuk menghitung SCUS (*Screen Usability Score*) adalah sebagai berikut:

$$SCUS = MAX(0, 100 - (DOR * dW) - (MCR * mW) - (MIN(10, MAX(0, (AVGD - 5) / 2)))) \qquad (3.2)$$

dengan variabel:

- SCUS = Screen usability score
- DOR = tingkat keluar (exit rate)
- dW = bobot DOR; dW bernilai 1 poin untuk setiap keluar
- MCR = tingkat kesalahan klik (misclick rate)
- mW = bobot MCR; mW bernilai 0,5 poin untuk setiap kesalahan klik
- AVGD = Durasi Rata-rata dalam detik

dan dengan fungsi:

- MAX: MAX(VALUE, {EXPRESSION}) => mengembalikan maksimum antara VALUE dan EXPRESSION
- MAX: MIN(VALUE, {EXPRESSION}) => mengembalikan nilai minimum antara VALUE dan EXPRESSION

### 3.4.5.1.2. *Mission Usability Score (MIUS)*

*Mission Usability Score* (MIUS) yang mengukur seberapa mudah pengguna dapat menyelesaikan misi. Prinsip-prinsip yang digunakan untuk menghitung MIUS adalah sebagai berikut:

- Keberhasilan langsung harus berkorelasi kuat dengan *usability score*.
- Keberhasilan tidak langsung tidak harus dianggap sebagai kegagalan.
- Metrik kegunaan rata-rata harus mempengaruhi MIUS.

Untuk menghitung MIUS, setiap persentase tingkat keberhasilan langsung akan memberikan 1 poin, sementara setiap persentase tingkat keberhasilan tidak langsung akan memberikan 0,5 poin. Selain itu, pengurangan poin yang terkait dengan kesalahan klik dan durasi rata-rata layar akan diterapkan pada MIUS. Sebagai contoh, jika misi memiliki tingkat keberhasilan 100%, tetapi memiliki rata-rata tingkat kesalahan klik sebesar 50% pada setiap layar, maka akan dikurangi 25 poin. Sehingga, MIUS adalah hasil dari penjumlahan tingkat keberhasilan langsung dan tidak langsung, dikurangi dengan pengurangan yang berkaitan dengan kesalahan klik dan durasi rata-rata.

Rumus untuk menghitung *Mission Usability Score* adalah sebagai berikut:

$$\text{MIUS} = \text{DSR} + (\text{IDSR} / 2) - \text{avg(MC\_P)} - \text{avg(DU\_P)} \qquad (3.3)$$

- MIUS = Mission Usability Score
- DSR = tingkat keberhasilan langsung (Direct Success Rate)
- IDSR = tingkat keberhasilan tidak langsung (Indirect Success Rate) rata-rata untuk rata-rata
- MC_P = pengurangan poin kesalahan klik (misclick penalty) = MCR * 0.5
- DU_P = pengurangan poin durasi (duration penalty) = MIN(10, MAX(0, (AVGD - 5) / 2))

### 3.4.5.1.3. *Maze Usability Score (MAUS)*

*Maze Usability Score* (MAUS) merupakan *usability score* untuk setiap *path* yang diuji. MAUS tidak berdampak pada skor misi lainnya. Oleh karena itu, SCUS adalah rata-rata dari MIUS. Rumus untuk menghitung MAUS adalah sebagai berikut:

$$MAUS = avg(MIUS) \qquad\qquad (3.4)$$

- Maze Usability Score (MAUS)
- MIUS = Mission Usability Score

Dengan menggunakan sistem pengukuran ini, dapat diperoleh informasi yang lebih terperinci tentang kegunaan dari setiap layar, misi, dan path dalam pengembangan prototype. Hal ini dapat membantu meningkatkan desain agar lebih mudah digunakan, intuitif, dan efisien bagi pengguna. Hasil akhir *usability score* beserta efisiensi waktu dapat dilihat di tabel berikut:

**Tabel 3.4 Nilai Keseluruhan SIAK-NG Desktop Site (Lama)**

| Task | Direct success | Mission unfinished | Misclick rate | Avg duration | Usability Score | Sub-Task | Avg. Time (s) | Misclick Rate | Usability Score |
|---|---|---|---|---|---|---|---|---|---|
| 1. Mengecek pengumuman | 34.6% | 30.8% | 61.7% | 42.0s | 30 | Memasukkan kredensial login | 57s | 35% | 69 |
| | | | | | | Login ke halaman home | 2s | 12% | 94 |
| | | | | | | Menemukan pengumuman | 21s | 48% | 5 |
| 2. Mencari nilai per mata kuliah | 60.9% | 8.7% | 61.5% | 52.1s | 63 | Memasukkan kredensial login | 3s | 9% | 92 |
| | | | | | | Login ke halaman home | 1s | 9% | 96 |
| | | | | | | Mengklik menu "Akademik" | 8s | 36% | 63 |
| | | | | | | Mengklik "Riwayat" | 7s | 28% | 74 |
| | | | | | | Menemukan nilai mata kuliah | 19s | 31% | 65 |
| 3. Mengecek status pembayaran BOP | 45.5% | 9.1% | 59.3% | 44.2s | 57 | Memasukkan kredensial login | 2s | 14% | 88 |
| | | | | | | Login ke halaman home | 1s | 10% | 95 |
| | | | | | | Mengklik menu "Akademik" | 6s | 24% | 83 |
| | | | | | | Mengklik "Pembayaran" | 5s | 10% | 90 |
| | | | | | | Menemukan status pembayaran | 17s | 42% | 26 |
| 4. Mengecek jadwal waktu dan lokasi kelas | 59.1% | 18.2% | 68.4% | 31.4s | 59 | Memasukkan kredensial login | 2s | 9% | 91 |
| | | | | | | Login ke halaman home | 1s | 0% | 100 |
| | | | | | | Mengklik menu "Ringkasan IRS" | 6s | 33% | 55 |
| | | | | | | Mengklik "Jadwal" | 8s | 60% | 56 |
| | | | | | | Menemukan jadwal kelas | 4s | 8% | 96 |
| 5. Melakukan registrasi IRS | 5.0% | 35.0% | 35.5% | 63.0s | 24 | Memasukkan kredensial login | 2s | 15% | 88 |
| | | | | | | Login ke halaman home | 1s | 11% | 95 |
| | | | | | | Mengklik menu "IRS" | 5s | 47% | 72 |
| | | | | | | Mengklik "Isi/ubah IRS" | 4s | 33% | 73 |
| | | | | | | Menemukan kelas yang diinginkan | 12s | 19% | 0 |
| | | | | | | Mengklik "Submit" | 3s | 0% | 100 |
| 6. Mencari informasi mata kuliah yang diselenggarakan fakultas saya | 68.4% | 10.5% | 35.6% | 26.6s | 72 | Memasukkan kredensial login | 2s | 0% | 95 |
| | | | | | | Login ke halaman home | 1s | 6% | 97 |
| | | | | | | Mengklik menu "Jadwal" | 5s | 28% | 58 |
| | | | | | | Mengklik "Jadwal Kuliah" | 1s | 0% | 100 |
| | | | | | | Menemukan kelas yang diinginkan | 14s | 31% | 81 |

**Tabel 3.5 Nilai Keseluruhan SIAK-NG Desktop Site (Baru)**

| Task | Direct success | Mission unfinished | Misclick rate | Avg duration | Usability Score | Sub-Task | Avg. Time (s) | Misclick Rate | Usability Score |
|---|---|---|---|---|---|---|---|---|---|
| 1. Mengecek pengumuman | 89.5% | 5.3% | 51.1% | 36.3s | 77 | Memasukkan kredensial login | 6s | 5% | 98 |
| | | | | | | Login ke halaman home | 6s | 32% | 84 |
| | | | | | | Menemukan pengumuman | 22s | 37% | 63 |
| 2. Mencari nilai per mata kuliah | 73.7% | 5.3% | 41.6% | 45.8s | 78 | Memasukkan kredensial login | 3s | 11% | 95 |
| | | | | | | Login ke halaman home | 1s | 0% | 95 |
| | | | | | | Mengklik menu "Profil" | 4s | 11% | 78 |
| | | | | | | Mengklik ke "Riwayat Kelas" | 8s | 7% | 96 |
| | | | | | | Menemukan nilai mata kuliah | 16s | 20% | 78 |
| 3. Mengecek status pembayaran BOP | 84.2% | 5.3% | 17.9% | 22.6s | 78 | Memasukkan kredensial login | 2s | 11% | 95 |
| | | | | | | Login ke halaman home | 1s | 5% | 93 |
| | | | | | | Mengklik menu "Administrasi" | 6s | 28% | 75 |
| | | | | | | Menemukan status pembayaran | 11s | 38% | 78 |
| 4. Mengecek jadwal waktu dan lokasi kelas | 84.2% | 5.3% | 33.3% | 27.3s | 76 | Memasukkan kredensial login | 2s | 11% | 95 |
| | | | | | | Login ke halaman home | 2s | 11% | 90 |
| | | | | | | Menemukan jadwal kelas | 18s | 44% | 61 |
| 5. Melakukan registrasi IRS | 89.5% | 5.3% | 55.9% | 35.3s | 76 | Memasukkan kredensial login | 2s | 11% | 95 |
| | | | | | | Login ke halaman home | 2s | 0% | 95 |
| | | | | | | Menemukan kelas yang diinginkan | 23s | 67% | 53 |
| | | | | | | Mengklik "Submit" | 8s | 35% | 82 |
| 6. Mencari informasi mata kuliah yang diselenggarakan fakultas saya | 84.2% | 5.3% | 23.1% | 34.4s | 81 | Memasukkan kredensial login | 4s | 11% | 95 |
| | | | | | | Login ke halaman home | 2s | 5% | 93 |
| | | | | | | Mengklik menu "Jadwal" | 6s | 17% | 86 |
| | | | | | | Mengklik "Jadwal Kuliah" | 8s | 29% | 79 |
| | | | | | | Membuka kategori kelas | 6s | 19% | 91 |
| | | | | | | Menemukan kelas yang diinginkan | 6s | 25% | 88 |

**Tabel 3.6 Nilai Keseluruhan SIAK-NG Mobile Site (Lama)**

| Task | Direct success | Mission unfinished | Misclick rate | Avg duration | Usability Score | Sub-Task | Avg. Time (s) | Misclick Rate | Usability Score |
|------|----------------|--------------------|---------------|--------------|-----------------|----------|---------------|---------------|-----------------|
| 1. Mengecek pengumuman | 73.7% | 5.3% | 13.0% | 19.1s | 75 | Memasukkan kredensial login | 4s | 37% | 82 |
| | | | | | | Login ke halaman home | 3s | 5% | 93 |
| | | | | | | Menemukan pengumuman | 10s | 6% | 73 |
| 2. Mencari nilai per mata kuliah | 33.3% | 50.0% | 79.2% | 52.8s | 22 | Memasukkan kredensial login | 2s | 22% | 89 |
| | | | | | | Login ke halaman home | 1s | 6% | 86 |
| | | | | | | Mengklik menu "Akademik" | 6s | 25% | 82 |
| | | | | | | Mengklik "Riwayat" | 5s | 47% | 77 |
| | | | | | | Menemukan nilai mata kuliah | 31s | 80% | 0 |
| 3. Mengecek status pembayaran BOP | 12.5% | 37.5% | 44.7% | 23.8s | 24 | Memasukkan kredensial login | 2s | 31% | 85 |
| | | | | | | Login ke halaman home | 2s | 19% | 85 |
| | | | | | | Mengklik menu "Akademik" | 5s | 33% | 84 |
| | | | | | | Mengklik "Pembayaran" | 3s | 20% | 90 |
| | | | | | | Menemukan status pembayaran | 10s | 40% | 0 |
| 4. Mengecek jadwal waktu dan lokasi kelas | 43.8% | 37.5% | 62.4% | 23.8s | 39 | Memasukkan kredensial login | 4s | 38% | 75 |
| | | | | | | Login ke halaman home | 1s | 7% | 90 |
| | | | | | | Mengklik menu "Ringkasan IRS" | 5s | 29% | 57 |
| | | | | | | Mengklik "Jadwal" | 6s | 50% | 45 |
| | | | | | | Menemukan jadwal kelas | 5s | 14% | 93 |
| 5. Melakukan registrasi IRS | 6.3% | 62.5% | 50.8% | 40.1s | 13 | Memasukkan kredensial login | 2s | 13% | 88 |
| | | | | | | Login ke halaman home | 1s | 7% | 90 |
| | | | | | | Mengklik menu "IRS" | 4s | 29% | 57 |
| | | | | | | Mengklik "Isi/ubah IRS" | 6s | 50% | 35 |
| | | | | | | Menemukan kelas yang diinginkan | 10s | 17% | 7 |
| | | | | | | Mengklik "Submit" | 1s | 0% | 100 |
| 6. Mencari informasi mata kuliah yang diselenggarakan fakultas saya | 37.5% | 43.8% | 30.2% | 27.1s | 34 | Memasukkan kredensial login | 3s | 25% | 88 |
| | | | | | | Login ke halaman home | 2s | 19% | 72 |
| | | | | | | Mengklik "Jadwal Kuliah" | 6s | 31% | 70 |
| | | | | | | Menemukan kelas yang diinginkan | 12s | 27% | 39 |

**Tabel 3.7 Nilai Keseluruhan SIAK-NG Mobile Site (Baru)**

| Task | Direct success | Mission unfinished | Misclick rate | Avg duration | Usability Score | Sub-Task | Avg. Time (s) | Misclick Rate | Usability Score |
|------|------|------|------|------|------|------|------|------|------|
| 1. Mengecek pengumuman | 81.8% | 18.2% | 27.7% | 11.7s | 72 | Memasukkan kredensial login | 4s | 37% | 86 |
| | | | | | | Login ke halaman home | 3s | 5% | 96 |
| | | | | | | Menemukan pengumuman | 10s | 6% | 69 |
| 2. Mencari nilai per mata kuliah | 60.0% | 10.0% | 50.3% | 46.7s | 54 | Memasukkan kredensial login | 1s | 0% | 100 |
| | | | | | | Login ke halaman home | 1s | 10% | 85 |
| | | | | | | Mengklik menu "Profil" | 6s | 89% | 45 |
| | | | | | | Menggeser ke "Riwayat Kelas" | 20s | 75% | 56 |
| | | | | | | Menemukan nilai mata kuliah | 7s | 25% | 63 |
| 3. Mengecek status pembayaran BOP | 80.0% | 10.0% | 33.3% | 22.2s | 74 | Memasukkan kredensial login | 2s | 20% | 90 |
| | | | | | | Login ke halaman home | 1s | 10% | 85 |
| | | | | | | Mengklik menu "Administrasi" | 8s | 22% | 77 |
| | | | | | | Menemukan status pembayaran | 7s | 25% | 87 |
| 4. Mengecek jadwal waktu dan lokasi kelas | 60.0% | 10.0% | 34.9% | 23.3s | 70 | Memasukkan kredensial login | 5s | 10% | 95 |
| | | | | | | Login ke halaman home | 1s | 10% | 85 |
| | | | | | | Menemukan jadwal kelas | 4s | 11% | 62 |
| 5. Melakukan registrasi IRS | 80.0% | 10.0% | 39.4% | 27.9s | 73 | Memasukkan kredensial login | 2s | 10% | 95 |
| | | | | | | Login ke halaman home | 1s | 10% | 85 |
| | | | | | | Menemukan kelas yang diinginkan | 18s | 33% | 67 |
| | | | | | | Mengklik "Submit" | 2s | 25% | 88 |
| 6. Mencari informasi mata kuliah yang diselenggarakan fakultas saya | 60.0% | 10.0% | 46.4% | 62.2s | 62 | Memasukkan kredensial login | 2s | 10% | 95 |
| | | | | | | Login ke halaman home | 1s | 10% | 85 |
| | | | | | | Mengklik menu "Jadwal" | 4s | 22% | 89 |
| | | | | | | Menggeser ke "Jadwal Kuliah" | 10s | 44% | 65 |
| | | | | | | Menemukan kelas yang diinginkan | 8s | 38% | 55 |

### 3.4.5.2. PSSUQ

Dalam penelitian ini, kuesioner *Post-Study System Usability Questionnaire* (PSSUQ) diberikan sebanyak empat kali untuk masing-masing versi *user interface* (UI), yaitu lama versi desktop, baru versi desktop, lama versi mobile, dan baru versi mobile. Tujuan dari pemberian kuesioner PSSUQ lama adalah untuk mengevaluasi sejauh mana tingkat kegunaan UI yang sedang digunakan, sementara evaluasi UI baru bertujuan untuk melihat perbedaan kegunaan setelah dilakukan revisi.

Hasil dari kuesioner PSSUQ dihitung dengan cara mengambil nilai rata-rata dari tanggapan responden untuk setiap pertanyaan. Hal ini bertujuan untuk mendapatkan gambaran keseluruhan tentang persepsi, perasaan, dan tingkat kepuasan responden terhadap UI yang dievaluasi, berbeda dengan usability testing yang lebih fokus pada performa responden. Pertanyaan yang terdapat pada PSSUQ antara lain adalah:

**Tabel 3.8 Daftar Pertanyaan PSSUQ**

| Kategori | No. | Pertanyaan |
|---|---|---|
| SYSUSE | 1 | Secara keseluruhan, saya puas dengan betapa mudahnya menggunakan sistem ini |
| | 2 | Sangat mudah untuk menggunakan sistem ini |
| | 3 | Saya dapat menyelesaikan tugas dan skenario dengan cepat menggunakan sistem ini |
| | 4 | Saya merasa nyaman menggunakan sistem ini |
| | 5 | Mudah untuk belajar menggunakan sistem ini |
| | 6 | Saya yakin saya dapat menjadi produktif dengan cepat menggunakan sistem ini |
| INFOQUAL | 7 | Sistem memberi pesan kesalahan yang dengan jelas memberi tahu saya cara memperbaiki masalah |
| | 8 | Setiap kali saya melakukan kesalahan menggunakan sistem, saya dapat mencari solusi dengan mudah dan cepat |
| | 9 | Informasi (seperti bantuan online, pesan di layar, dan dokumentasi lainnya) yang disediakan dengan sistem ini jelas |
| | 10 | Mudah menemukan informasi yang saya butuhkan |
| | 11 | Informasi tersebut efektif dalam membantu saya menyelesaikan tugas dan skenario |
| | 12 | Informasi pada layar sistem terorganisir dengan jelas |
| INTERQUAL | 13 | User-interface sistem ini menyenangkan |
| | 14 | Saya suka menggunakan user-interface sistem ini |
| | 15 | Semua fungsi dan kemampuan yang saya harapkan sudah ada desainnya dalam sistem ini |
| | 16 | Secara keseluruhan, saya puas dengan sistem ini |

### 3.4.5.2.1. *Desktop Site*

Berikut hasil PSSUQ untuk desktop site baru dan lama.

Tabel 3.9 Hasil PSSUQ desktop site lama untuk cluster 1

| Cluster 1 | | | | | | | | | | | | | | | | |
|-----------|-----|-----|-----|-----|-----|-----|-----|-----|-----|---------|---------|---------|---------|---------|---------|------|
| Responden | Q1 | Q2 | Q3 | Q4 | Q5 | Q6 | Q7 | Q8 | Q9 | Q10 | Q11 | Q12 | Q13 | Q14 | Q15 | Q16 | Mean |
| X1 | 5 | 5 | 6 | 3 | 3 | 4 | 5 | 4 | 3 | 4 | 4 | 5 | 6 | 4 | 4 | 3 | 4.3 |
| X2 | 4 | 4 | 4 | 4 | 4 | 4 | 4 | 4 | 4 | 5 | 5 | 4 | 3 | 3 | 4 | 4 | 4.0 |
| X3 | 6 | 6 | 4 | 7 | 7 | 7 | 6 | 7 | 3 | 4 | 6 | 4 | 6 | 7 | 7 | 7 | 5.9 |
| X4 | 6 | 6 | 6 | 6 | 5 | 3 | 5 | 6 | 6 | 6 | 6 | 6 | 6 | 6 | 6 | 5 | 5.6 |
| X5 | 3 | 4 | 3 | 5 | 3 | 3 | 4 | 3 | 4 | 4 | 3 | 4 | 5 | 4 | 4 | 4 | 3.8 |
| Mean | System Use | | | | | | Information Quality | | | | | | Interface Quality | | | TOTAL |
| | 4.7 | | | | | | 4.6 | | | | | | 4.9 | | | 4.7 |

Tabel 3.10 Hasil PSSUQ desktop site lama untuk cluster 2

| Cluster 2 | | | | | | | | | | | | | | | | |
|-----------|-----|-----|-----|-----|-----|-----|-----|-----|-----|---------|---------|---------|---------|---------|---------|------|
| Responden | Q1 | Q2 | Q3 | Q4 | Q5 | Q6 | Q7 | Q8 | Q9 | Q10 | Q11 | Q12 | Q13 | Q14 | Q15 | Q16 | Mean |
| X1 | 7 | 7 | 7 | 7 | 7 | 7 | 7 | 7 | 7 | 7 | 7 | 7 | 7 | 7 | 7 | 7 | 7.0 |
| X2 | 7 | 7 | 7 | 7 | 7 | 6 | 7 | 7 | 7 | 6 | 5 | 7 | 7 | 6 | 5 | 7 | 6.6 |
| X3 | 5 | 6 | 6 | 7 | 5 | 6 | 7 | 5 | 4 | 5 | 3 | 6 | 7 | 7 | 6 | 6 | 5.7 |
| X4 | 4 | 5 | 4 | 4 | 5 | 5 | 5 | 5 | 4 | 6 | 5 | 5 | 6 | 6 | 5 | 4 | 4.9 |
| X5 | 3 | 4 | 4 | 4 | 4 | 4 | 4 | 4 | 5 | 5 | 4 | 3 | 6 | 6 | 5 | 5 | 4.4 |
| Mean | System Use | | | | | | Information Quality | | | | | | Interface Quality | | | TOTAL |
| | 5.6 | | | | | | 5.5 | | | | | | 6.1 | | | 5.7 |

Tabel 3.11 Hasil PSSUQ desktop site lama untuk cluster 3

| Cluster 3 | | | | | | | | | | | | | | | | |
|-----------|-----|-----|-----|-----|-----|-----|-----|-----|-----|---------|---------|---------|---------|---------|---------|------|
| Responden | Q1 | Q2 | Q3 | Q4 | Q5 | Q6 | Q7 | Q8 | Q9 | Q10 | Q11 | Q12 | Q13 | Q14 | Q15 | Q16 | Mean |
| X1 | 6 | 6 | 7 | 7 | 6 | 7 | 7 | 7 | 6 | 6 | 6 | 7 | 7 | 7 | 7 | 7 | 6.6 |
| X2 | 4 | 5 | 4 | 5 | 5 | 5 | 5 | 5 | 5 | 4 | 5 | 3 | 6 | 6 | 4 | 5 | 4.8 |
| X3 | 4 | 5 | 5 | 4 | 3 | 5 | 4 | 3 | 4 | 5 | 5 | 5 | 6 | 6 | 4 | 5 | 4.6 |
| X4 | 4 | 4 | 3 | 5 | 3 | 5 | 6 | 3 | 6 | 5 | 5 | 6 | 5 | 5 | 3 | 5 | 4.6 |
| X5 | 4 | 4 | 3 | 5 | 5 | 5 | 5 | 4 | 5 | 4 | 4 | 3 | 5 | 5 | 5 | 4 | 4.4 |
| Mean | System Use | | | | | | Information Quality | | | | | | Interface Quality | | | TOTAL |
| | 4.8 | | | | | | 4.9 | | | | | | 5.4 | | | 5.0 |

Tabel 3.12 Hasil PSSUQ desktop site baru untuk cluster 1

| | Cluster 1 | | | | | | | | | | | | | | | | |
|---|---|---|---|---|---|---|---|---|---|---|---|---|---|---|---|---|---|
| Responden | Q1 | Q2 | Q3 | Q4 | Q5 | Q6 | Q7 | Q8 | Q9 | Q10 | Q11 | Q12 | Q13 | Q14 | Q15 | Q16 | Mean |
| X1 | 3 | 3 | 3 | 3 | 3 | 4 | 4 | 3 | 4 | 2 | 2 | 2 | 3 | 3 | 3 | 3 | 3.0 |
| X2 | 1 | 1 | 1 | 1 | 1 | 1 | 1 | 1 | 1 | 1 | 1 | 1 | 1 | 1 | 1 | 1 | 1.0 |
| X3 | 2 | 2 | 2 | 2 | 3 | 3 | 3 | 4 | 3 | 3 | 3 | 2 | 1 | 3 | 2 | 2 | 2.5 |
| X4 | 1 | 4 | 4 | 3 | 4 | 3 | 7 | 7 | 2 | 1 | 2 | 1 | 4 | 3 | 2 | 3 | 3.2 |
| X5 | 3 | 3 | 3 | 2 | 2 | 2 | 4 | 3 | 3 | 3 | 3 | 2 | 2 | 2 | 3 | 3 | 2.7 |
| Mean | System Use | | | | | | Information Quality | | | | | | Interface Quality | | | | TOTAL |
| | 2.4 | | | | | | 2.6 | | | | | | 2.3 | | | | 2.5 |

Tabel 3.13 Hasil PSSUQ desktop site baru untuk cluster 2

| | Cluster 2 | | | | | | | | | | | | | | | | |
|---|---|---|---|---|---|---|---|---|---|---|---|---|---|---|---|---|---|
| Responden | Q1 | Q2 | Q3 | Q4 | Q5 | Q6 | Q7 | Q8 | Q9 | Q10 | Q11 | Q12 | Q13 | Q14 | Q15 | Q16 | Mean |
| X1 | 3 | 3 | 3 | 3 | 3 | 4 | 5 | 4 | 3 | 4 | 4 | 2 | 2 | 2 | 4 | 2 | 3.2 |
| X2 | 3 | 2 | 2 | 3 | 2 | 3 | 4 | 3 | 2 | 2 | 3 | 2 | 2 | 2 | 3 | 2 | 2.5 |
| X3 | 1 | 2 | 3 | 3 | 3 | 2 | 5 | 2 | 2 | 2 | 1 | 3 | 2 | 1 | 5 | 2 | 2.4 |
| X4 | 3 | 3 | 1 | 1 | 1 | 2 | 3 | 2 | 2 | 2 | 2 | 1 | 1 | 1 | 2 | 2 | 1.8 |
| X5 | 2 | 1 | 2 | 1 | 2 | 2 | 4 | 2 | 1 | 2 | 2 | 1 | 1 | 1 | 2 | 2 | 1.8 |
| Mean | System Use | | | | | | Information Quality | | | | | | Interface Quality | | | | TOTAL |
| | 2.3 | | | | | | 2.6 | | | | | | 2.1 | | | | 2.3 |

Tabel 3.14 Hasil PSSUQ desktop site baru untuk cluster 3

| | Cluster 3 | | | | | | | | | | | | | | | | |
|---|---|---|---|---|---|---|---|---|---|---|---|---|---|---|---|---|---|
| Responden | Q1 | Q2 | Q3 | Q4 | Q5 | Q6 | Q7 | Q8 | Q9 | Q10 | Q11 | Q12 | Q13 | Q14 | Q15 | Q16 | Mean |
| X1 | 1 | 1 | 1 | 2 | 2 | 2 | 5 | 2 | 2 | 2 | 2 | 1 | 1 | 1 | 1 | 2 | 1.8 |
| X2 | 2 | 1 | 1 | 1 | 1 | 1 | 3 | 2 | 3 | 1 | 1 | 1 | 2 | 2 | 2 | 2 | 1.6 |
| X3 | 2 | 1 | 1 | 1 | 1 | 1 | 4 | 1 | 3 | 2 | 1 | 2 | 2 | 1 | 1 | 1 | 1.6 |
| X4 | 1 | 1 | 1 | 1 | 1 | 1 | 3 | 1 | 1 | 1 | 1 | 1 | 1 | 1 | 2 | 1 | 1.2 |
| X5 | 1 | 2 | 1 | 1 | 1 | 1 | 1 | 1 | 1 | 1 | 1 | 1 | 1 | 1 | 1 | 1 | 1.1 |
| Mean | System Use | | | | | | Information Quality | | | | | | Interface Quality | | | | TOTAL |
| | 1.2 | | | | | | 1.7 | | | | | | 1.4 | | | | 1.4 |

**Tabel 3.15 Hasil PSSUQ mobile site lama untuk cluster 1**

| Responden | Q1 | Q2 | Q3 | Q4 | Q5 | Q6 | Q7 | Q8 | Q9 | Q10 | Q11 | Q12 | Q13 | Q14 | Q15 | Q16 | Mean |
|---|---|---|---|---|---|---|---|---|---|---|---|---|---|---|---|---|---|
| | | | | | | | **Cluster 1** | | | | | | | | | | |
| X1 | 7 | 7 | 7 | 7 | 7 | 7 | 7 | 7 | 7 | 7 | 7 | 7 | 7 | 7 | 7 | 7 | 7.0 |
| X2 | 4 | 5 | 6 | 6 | 6 | 6 | 7 | 6 | 5 | 4 | 4 | 4 | 7 | 6 | 6 | 6 | 5.5 |
| X3 | 5 | 4 | 5 | 4 | 3 | 4 | 5 | 4 | 5 | 5 | 4 | 4 | 5 | 5 | 4 | 4 | 4.4 |
| X4 | 1 | 1 | 1 | 1 | 1 | 1 | 1 | 1 | 1 | 1 | 1 | 1 | 1 | 1 | 1 | 1 | 1.0 |
| Mean | System Use | | | | | | Information Quality | | | | | | Interface Quality | | | | TOTAL |
| | 4.4 | | | | | | 4.4 | | | | | | 4.7 | | | | 4.5 |

**Tabel 3.16 Hasil PSSUQ mobile site lama untuk cluster 2**

| Responden | Q1 | Q2 | Q3 | Q4 | Q5 | Q6 | Q7 | Q8 | Q9 | Q10 | Q11 | Q12 | Q13 | Q14 | Q15 | Q16 | Mean |
|---|---|---|---|---|---|---|---|---|---|---|---|---|---|---|---|---|---|
| | | | | | | | **Cluster 2** | | | | | | | | | | |
| X1 | 7 | 7 | 7 | 7 | 7 | 7 | 7 | 7 | 7 | 7 | 7 | 7 | 7 | 7 | 7 | 7 | 7.0 |
| X2 | 7 | 7 | 7 | 7 | 7 | 7 | 7 | 7 | 7 | 7 | 7 | 7 | 7 | 7 | 7 | 7 | 7.0 |
| X3 | 6 | 7 | 7 | 6 | 5 | 6 | 7 | 5 | 4 | 6 | 4 | 3 | 7 | 7 | 6 | 6 | 5.8 |
| X4 | 5 | 5 | 6 | 5 | 5 | 5 | 6 | 6 | 4 | 5 | 5 | 4 | 6 | 6 | 6 | 5 | 5.3 |
| X5 | 7 | 7 | 6 | 6 | 7 | 5 | 4 | 3 | 3 | 5 | 5 | 3 | 7 | 7 | 2 | 6 | 5.2 |
| Mean | System Use | | | | | | Information Quality | | | | | | Interface Quality | | | | TOTAL |
| | 6.3 | | | | | | 5.5 | | | | | | 6.4 | | | | 6.0 |

**Tabel 3.17 Hasil PSSUQ mobile site lama untuk cluster 3**

| Responden | Q1 | Q2 | Q3 | Q4 | Q5 | Q6 | Q7 | Q8 | Q9 | Q10 | Q11 | Q12 | Q13 | Q14 | Q15 | Q16 | Mean |
|---|---|---|---|---|---|---|---|---|---|---|---|---|---|---|---|---|---|
| | | | | | | | **Cluster 3** | | | | | | | | | | |
| X1 | 7 | 7 | 7 | 7 | 7 | 7 | 7 | 7 | 7 | 6 | 7 | 7 | 7 | 7 | 7 | 7 | 6.9 |
| X2 | 7 | 7 | 7 | 7 | 2 | 7 | 6 | 6 | 7 | 7 | 7 | 7 | 7 | 7 | 5 | 7 | 6.4 |
| X3 | 7 | 7 | 5 | 7 | 4 | 7 | 4 | 4 | 4 | 6 | 4 | 7 | 7 | 7 | 7 | 6 | 5.8 |
| X4 | 6 | 4 | 6 | 6 | 5 | 5 | 5 | 4 | 4 | 5 | 5 | 5 | 6 | 6 | 5 | 5 | 5.1 |
| X5 | 2 | 3 | 2 | 3 | 5 | 3 | 1 | 3 | 4 | 5 | 3 | 3 | 2 | 3 | 4 | 2 | 3.0 |
| Mean | System Use | | | | | | Information Quality | | | | | | Interface Quality | | | | TOTAL |
| | 5.5 | | | | | | 5.2 | | | | | | 5.7 | | | | 5.5 |

**Universitas Indonesia**

**Tabel 3.18 Hasil PSSUQ mobile site lama untuk cluster 3**

| Responden | Q1 | Q2 | Q3 | Q4 | Q5 | Q6 | Q7 | Q8 | Q9 | Q10 | Q11 | Q12 | Q13 | Q14 | Q15 | Q16 | Mean |
|---|---|---|---|---|---|---|---|---|---|---|---|---|---|---|---|---|---|
| | | | | | | | | **Cluster 1** | | | | | | | | | |
| X1 | 1 | 2 | 2 | 1 | 1 | 1 | 2 | 2 | 2 | 2 | 2 | 1 | 1 | 1 | 2 | 1 | 1.5 |
| Mean | System Use | | | | | | Information Quality | | | | | | Interface Quality | | | | TOTAL |
| | 1.3 | | | | | | 1.8 | | | | | | 1.3 | | | | 1.5 |

**Tabel 3.19 Hasil PSSUQ mobile site baru untuk cluster 2**

| Responden | Q1 | Q2 | Q3 | Q4 | Q5 | Q6 | Q7 | Q8 | Q9 | Q10 | Q11 | Q12 | Q13 | Q14 | Q15 | Q16 | Mean |
|---|---|---|---|---|---|---|---|---|---|---|---|---|---|---|---|---|---|
| | | | | | | | | **Cluster 2** | | | | | | | | | |
| X1 | 2 | 3 | 3 | 2 | 4 | 2 | 4 | 4 | 3 | 2 | 3 | 2 | 3 | 3 | 3 | 3 | 2.9 |
| X2 | 2 | 2 | 3 | 2 | 1 | 2 | 4 | 3 | 1 | 2 | 2 | 1 | 1 | 1 | 1 | 1 | 1.8 |
| X3 | 1 | 1 | 1 | 1 | 1 | 1 | 1 | 1 | 1 | 1 | 1 | 2 | 1 | 1 | 1 | 1 | 1.1 |
| X4 | 1 | 1 | 1 | 1 | 1 | 1 | 1 | 1 | 1 | 1 | 1 | 1 | 1 | 1 | 1 | 1 | 1.0 |
| X5 | 1 | 1 | 1 | 1 | 1 | 1 | 1 | 1 | 1 | 1 | 1 | 1 | 1 | 1 | 1 | 1 | 1.0 |
| Mean | System Use | | | | | | Information Quality | | | | | | Interface Quality | | | | TOTAL |
| | 1.5 | | | | | | 1.7 | | | | | | 1.4 | | | | 1.6 |

**Tabel 3.20 Hasil PSSUQ mobile site baru untuk cluster 3**

| Responden | Q1 | Q2 | Q3 | Q4 | Q5 | Q6 | Q7 | Q8 | Q9 | Q10 | Q11 | Q12 | Q13 | Q14 | Q15 | Q16 | Mean |
|---|---|---|---|---|---|---|---|---|---|---|---|---|---|---|---|---|---|
| | | | | | | | | **Cluster 3** | | | | | | | | | |
| X1 | 6 | 6 | 7 | 6 | 6 | 6 | 1 | 3 | 4 | 7 | 4 | 7 | 7 | 7 | 6 | 6 | 5.6 |
| X2 | 3 | 4 | 2 | 2 | 1 | 1 | 4 | 2 | 2 | 2 | 1 | 2 | 2 | 1 | 1 | 1 | 1.9 |
| X3 | 1 | 1 | 2 | 2 | 1 | 2 | 5 | 2 | 2 | 2 | 2 | 2 | 2 | 2 | 1 | 2 | 1.9 |
| X4 | 3 | 3 | 3 | 2 | 3 | 1 | 1 | 3 | 1 | 1 | 1 | 1 | 2 | 1 | 3 | 1 | 1.9 |
| X5 | 1 | 1 | 1 | 1 | 1 | 1 | 3 | 4 | 1 | 3 | 1 | 1 | 2 | 2 | 2 | 1 | 1.6 |
| Mean | System Use | | | | | | Information Quality | | | | | | Interface Quality | | | | TOTAL |
| | 2.7 | | | | | | 2.5 | | | | | | 2.6 | | | | 2.6 |

**BAB 4**

**ANALISIS DATA**

Bab ini akan membahas secara mendalam mengenai analisis terhadap pengumpulan dan pengolahan data yang telah dijelaskan pada Bab 3. Mulai dari identifikasi permasalahan hingga penemuan solusi. Selain itu, bagian ini juga akan membahas hasil akhir dari penelitian ini. Penjelasan mengenai desain dan analisis keseluruhan *user interface* akan disampaikan untuk menggambarkan kesimpulan akhir dari bab ini dan berfungsi sebagai penghubung menuju kesimpulan.

## 4.1. Analisis *Empathize*

### 4.1.1. *User Persona*

Dengan menggabungkan hasil karakteristik yang didapatkan untuk masing-masing persona melalui K-Modes *Clustering* dan hasil *in-depth interview*, penulis dapat membuat koneksi dari karakteristik pengguna dengan perilakunya, yang kemudian menghasilkan pengertian mendalam tentang latar belakang penggunaan SIAK-NG bagi setiap persona. Hal ini membantu dalam memahami sudut pandang yang berbeda akan kebutuhan, preferensi, dan harapan pengguna terhadap website SIAK-NG. Pengembangan persona merupakan salah satu metode yang sangat berguna dalam menggambarkan pengguna potensial dari sebuah website. Kesimpulan yang dapat diambil dari setiap persona adalah sebagai berikut:

1. Rina - *Persona Cluster 1*

Rina merepresentasikan mahasiswa pengguna SIAK-NG yang sering membuka SIAK-NG (2-3 kali/minggu), dan dapat menerima kekurangan SIAK-NG (walau begitu tetap mendukung pembaharuan SIAK-NG secepatnya). Dapat disimpulkan bahwa Rina datang dari fakultas yang cukup membutuhkan banyak fungsionalitas SIAK-NG, namun karena tidak memiliki tekanan dalam mengakses fitur tersebut, kekurangan SIAK-NG tidak menimbulkan ketidaknyamanan yang berarti bagi Rina.

2. Angga - *Persona Cluster 2*

Angga juga merepresentasikan mahasiswa yang sering membuka SIAK-NG dengan intensitas yang sama dengan Rina (2-3 kali/minggu). Perbedaannya dengan Rina adalah dia memiliki opini kuat tentang kekurangan SIAK-NG dengan cukup banyak umpan balik yang disampaikan. Hal ini dapat dikaitkan dengan perbedaan utama Angga dengan Rina yaitu perilaku sering membuka SIAK-NG di bawah tekanan (waktu yang terbatas, membuka sambil multitasking) sehingga kekurangan-kekurangan SIAK-NG terasa jauh lebih signifikan dan mengganggu dibanding untuk Rina.

3. Fitri - *Persona Cluster 3*

Fitri berbeda dengan persona sebelumnya, yaitu dia sangat jarang membuka SIAK-NG. Sekilas hal tersebut cukup membingungkan karena SIAK-NG memiliki banyak fitur yang krusial baik di fakultas manapun, namun hal ini dapat dijelaskan dengan karakteristik utama Fitri yaitu datang dari Fakultas Ilmu Komputer (FASILKOM). Sebagai fakultas dengan banyak sumber daya yang cukup familiar dengan desain website, FASILKOM berhasil memindahkan banyak fungsionalitas SIAK-NG ke website fakultasnya sendiri yang memberikan keuntungan yaitu lebih mudah dikelola dan diubah oleh fakultas tanpa perlu melalui DSTI. Sedikit contoh fitur yang telah dipindahkan ke website FASILKOM adalah kalender akademik, jadwal ujian, kontak dosen, dan pengumuman. Sehingga, dapat disimpulkan bahwa Fitri merepresentasikan mahasiswa pengguna SIAK-NG yang sebenarnya sama besar kebutuhannya untuk mengakses SIAK-NG, namun jarang dikarenakan bisa disubstitusi dengan website lokal fakultas yang memiliki sistem yang lebih baik

## 4.2.    Analisis *Define*

Pada Analisis Define, didefinisikan *pain point* general dari setiap *user persona* dengan mempertimbangkan latar belakang dan perilaku persona dari hasil *empathize* dengan analisis *empathy map*.

Ada tiga persona, masing-masing dengan *empathy map* mereka sendiri. *Empathy map* tersebut merangkum *pain points* dan *gain points* untuk setiap persona. *Pain points* menggambarkan hambatan potensial yang persona dapat temui saat menggunakan situs web. Sementara itu, *gain points* menggambarkan harapan yang persona ingin capai di masa depan saat mengunjungi situs web.

*User persona* membantu penelitian dalam memfokuskan target penelitian pada persona utama dari target pengguna utama SIAK-NG, dan membantu menemukan masalah-masalah utama berdasarkan pandangan persona terhadap website tersebut. Secara keseluruhan, kebutuhan yang dapat ditarik dari seluruh user persona adalah sebagai berikut:

1. Kemudahan mengakses SIAK-NG melalui mobile site
2. Letak fitur yang mudah diakses, terutama untuk fitur yang sering dipakai.
3. Layout yang lebih besar dan mudah dibaca
4. Layout yang lebih ergonomis dan mencegah salah klik
5. Tampilan yang segar dan modern
6. Kategorisasi fitur yang lebih intuitif
7. Layout dan tulisan yang ringkas dan padat
8. Pengingat penggantian password yang lebih terlihat
9. Layout yang mendukung penggunaan yang tidak terfokus
10. Desain yang lebih memberdayakan komunikasi visual.

Seluruh *user persona* menyampaikan keluhan dari aspek yang sama, perbedaanya adalah intensitas gangguan yang disebabkan. Sebagai contoh, terdapat keinginan kuat dari Angga untuk perbaikan SIAK-NG dari sisi keringkasan (*concision*) dan kemudahan dipelajari (*intuitivity*) jika dibandingkan dengan Rina dikarenakan ketidaknyamanan yang jauh lebih terasa dengan kebutuhan Angga mengakses SIAK-NG secara terburu-buru. Contoh lain adalah keinginan yang cukup kuat dari Fitri untuk memperbaiki komposisi dan estetika SIAK-NG dikarenakan sudah terbiasa dengan website FASILKOM yang memiliki desain lebih baik karena dikembangkan oleh lingkungan yang berpengetahuan dalam bidang komputer, namun bagi Angga, sekalipun dia juga menyuarakan kebutuhan estetika, hal tersebut lebih ditujukan untuk mendukung keringkasan dan kecepatan bekerja.

Seluruh pain poin dari kebutuhan di atas yang tidak terpenuhi dapat disimpulkan menjadi 3 aspek perubahan sebagai berikut:

1. Keringkasan (*concision*)

Mencakup *pain point* tentang efisiensi tampilan dan penggunaan fitur SIAK-NG. Poin ini dapat dicapai terutama dengan membuang elemen distraktif (elemen yang tidak memberi nilai tambahan ke kemudahan fungsionalitas) sesuai dengan prinsip *data-ink ratio*, dan membuat desain UI mengacu kembali kepada kegunaan dasarnya dan kebutuhan user yang didukung oleh *framework* yang dipakai dalam penelitian ini yaitu *design thinking*.

2. Kemudahan dipelajari (*intuitivity*)

Mencakup *pain point* tentang kesulitan mengerti penggunaan SIAK-NG, ketidaktahuan tentang keberadaan suatu fitur, perasaan terintimidasi untuk mengeksplorasi fitur-fitur, dan waktu untuk membiasakan diri dengan tampilan SIAK-NG. Poin ini dapat dicapai terutama dengan menggunakan pendekatan *Atomic Design* agar terdapat pola yang bisa membuat user mudah bernavigasi karena mendukung pengembang untuk menggunakan template/ elemen yang konsisten. Hal tersebut mendukung kemudahan mempelajari *user interface* karena elemen yang konsisten memberikan sebuah pola yang berulang sehingga mempercepat dan mempermudah pengguna untuk mengidentifikasi fungsi sebuah fitur.

3. Kenyamanan (*comfort*)

Mencakup *pain point* mengenai flow yang masih bisa di-*streamline* dan tampilan yang tidak didesain untuk mempermudah pengguna semaksimal mungkin. Poin ini dapat dicapai dengan mengatur ulang posisi agar disesuaikan dengan frekuensi kegunaan masing-masing fitur, dan menggunakan desain yang estetik namun tidak mengganggu, sebaliknya mendukung kemudahan penggunaan melalui sistem komunikasi visual.

### 4.3. Analisis *Ideate*

Setelah mengumpulkan *pain point* dari pengguna, tahap ideate dalam *design thinking* dilakukan setelah tahap ideate untuk mencari solusi bagi setiap pain point. Dalam tahap *ideate* ini, digunakan beberapa *tools* seperti *in-depth interview, empathy map,* dan *storyboard,* dengan juga mempertimbangkan seluruh hukum UX, desain visual, dan kaidah perancangan *user interface* yang telah disebut dalam studi literatur. *Storyboard* tersebut digunakan sebagai suplemen dalam mengkomunikasikan perilaku user dalam *in-depth interview* dengan perwakilan *user-persona* dari seluruh *cluster*. Diskusi dilakukan untuk merangsang ide-ide dari responden dengan menggunakan pertanyaan yang terkait dengan setiap komponen alat yang digunakan.

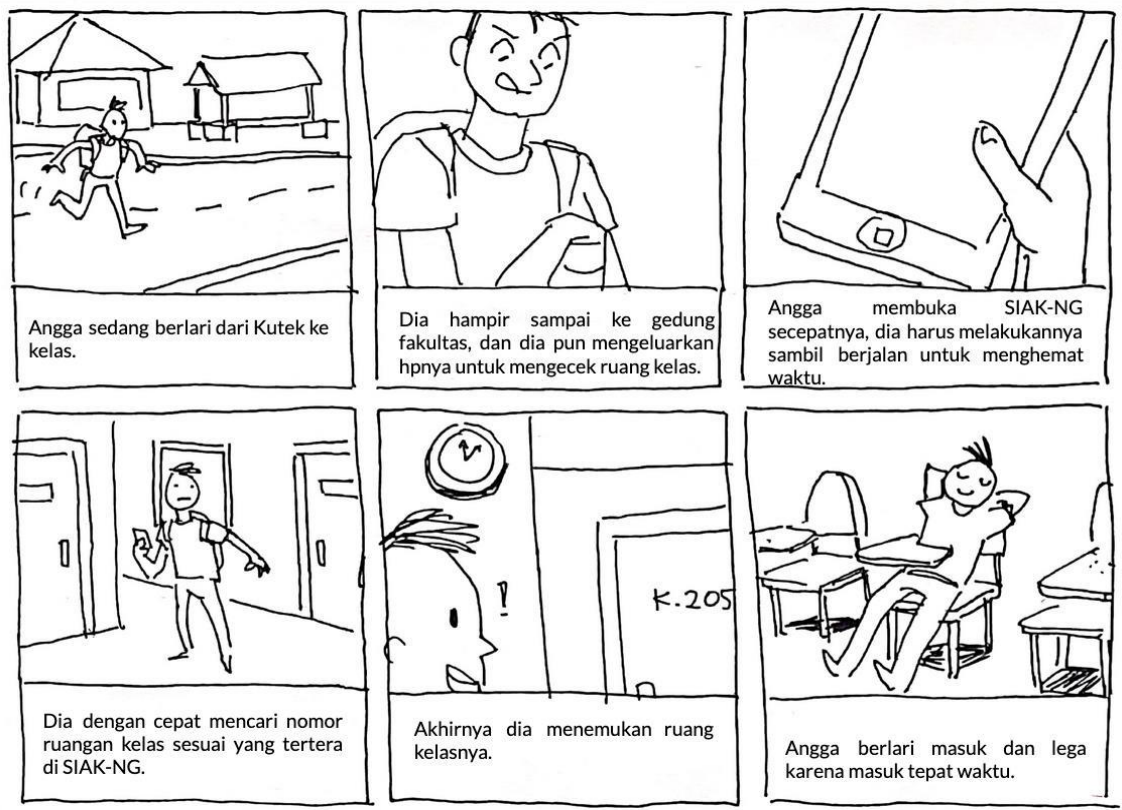

**Gambar 4.1 Ilustrasi skenario 1**

Hasil skenario 1 menggambarkan kebutuhan utama untuk dapat mengakses website dengan cepat. Salah satu skenario yang menggambarkan hal tersebut adalah saat mahasiswa sedang terburu-buru mencari kelas. Dalam kasus seperti ini, penggunaan website SIAK-NG melalui *desktop site* tidak memungkinkan, sehingga menaruh peran signifikan pada *mobile site*. Selain itu dalam skenario ini ada kemungkinan penggunaan SIAK-NG saat sedang beraktivitas seperti berlari. Oleh karena itu penting untuk membuat UI yang memiliki sensitivitas tinggi dan akurat. Menurut banyak laporan dari pengguna, SIAK-NG versi mobile saat ini sudah cukup sulit diakses walaupun tidak sambil mengerjakan hal lain, sehingga membuat penggunaan website lebih sulit lagi jika digunakan sambil melakukan aktivitas lain.

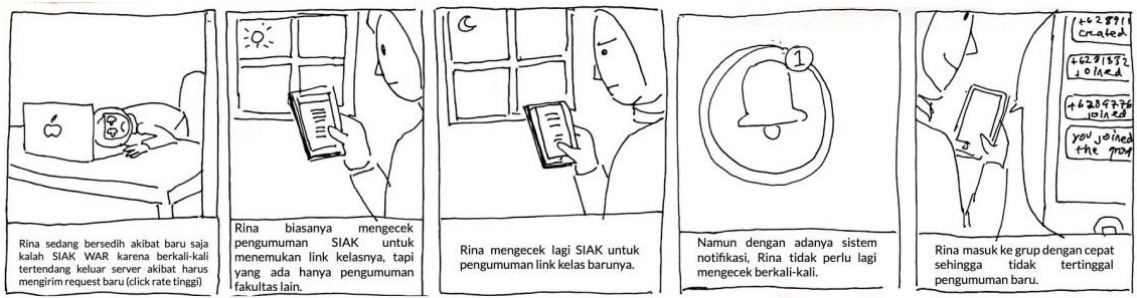

**Gambar 4.2 Ilustrasi skenario 2**

Hasil skenario 2 menggambarkan kebutuhan utama untuk sebuah website dengan sistem komunikasi yang lebih baik. Salah satu skenario yang menggambarkan hal tersebut adalah saat masa pengisian IRS dan menggunakan salah satu fitur SIAK-NG yaitu mencari pengumuman. Dalam kasus seperti ini, dibutuhkan sebuah website yang responsif dan komunikatif terhadap perubahan. Banyak fungsi dari SIAK- NG sendiri yang secara intrinsik merupakan fungsi *dashboard*, yaitu fungsi yang tujuan utamanya untuk mengkomunikasikan kondisi sistem. Sistem SIAK-NG saat ini membutuhkan user untuk memantau segalanya secara manual, dan hal ini memberikan ruang besar untuk *human-error* yang dapat berakibat signifikan seperti tidak mendapat kursi untuk mendaftar kelas. Sementara itu, akibat ringannya adalah hal seperti ketidaknyamanan, seperti harus melakukan pengecekan secara manual seperti mencari pengumuman yang dibutuhkan di antara pengumuman lainnya.

Jawaban dari responden kemudian digunakan sebagai dasar ide untuk pembuatan prototype baru SIAK-NG. Meskipun ide-ide dari responden mungkin berbeda-beda, tidak ada ide yang saling bertentangan. Penulis kemudian menyusun ide-ide tersebut menjadi ide final dengan mempertimbangkan prinsip desain user interface yang dapat dilihat pada Tabel 4.1. Penulis juga berhati-hati untuk memastikan bahwa perubahan dilakukan hanya sebatas kekurangan, dan tidak mengubah aspek yang tidak secara langsung mengganggu. Hal ini karena menurut Jakob's Law, pengguna menyukai desain yang mereka sudah familiar, dan penulis menjaga seluruh desain SIAK-NG lama yang memenuhi fungsinya dengan baik seperti *flow* dan *layout.*

Tabel 4.1 Mapping Solusi Desain Baru untuk setiap *Pain Point*

| No. | Pain Point | Solusi | Dasar |
|---|---|---|---|
| 1 | Terlihat tidak profesional, yang menyebabkan keraguan dan kesan bahwa website tersebut mungkin tidak asli. | Memberikan visual dalam bentuk foto, menggunakan layout login dengan pembagian menggunakan golden ratio dan benchmarking login page Simaster, KU Loket, dan layout login page populer lainnya | Aesthetic-Usability Effect, Golden Ratio, Jakob's law |
| 2 | Halaman login yang terlalu berantakan dengan semua informasi kontak dan bantuan | Membuat informasi kontak ke dalam 1 link agar elemen login page hanya foto, form login, bahasa, dan kontak bantuan | Hick's law, occam's razor |
| 3 | Kesulitan mengklik form kredensial login karena kecil | Membuat kolom kredensial login lebih besar dengan ukuran optimal di sisi sebelah kiri setelah welcome image | Fitts's law, golden ratio, Peak-End Rule, sinistrodextrality |
| 4 | Terlalu lama dan menyusahkan untuk mengakses informasi mengenai jadwal kelas (task dengan frekuensi tinggi) | Mengurangi click rate dengan menaruh mini widget di home page, dan diposisikan di first upper third dari div utama | Hick's Law, Golden ratio, goal gradient effect, serial position effect, |
| 5 | Posisi Isi IRS yang membutuhkan terlalu banyak klik menyusahkan saat pengisian IRS karena membuang waktu dan kemungkinan terkena isu teknis | Menaruh mini widget Isi IRS di home page | Fitts's law |

**Tabel 4.1 Mapping Solusi Desain Baru untuk setiap *Pain Point* (Lanjutan)**

| | | | |
|---|---|---|---|
| 6 | Menemukan kelas di daftar kelas di halaman Isi IRS membuat pusing karena tulisan kecil dan data banyak serta kemungkinan salah mengklik row | Mempertahankan layout namun membuat kursor menghighlight row di tabel daftar kelas saat dihover, menghilangkan kolom dengan informasi yang redundant agar setiap row terbaca seolah menjadi 1 kalimat, memberikan fitur sort by di setiap judul kolom | Jakob's law, Doherty Threshold, Fitts's Law, Law of Common Region, Tesler's Law, Von Restorff Effect, sinistrodextrality |
| 7 | Border kurang efektif memisahkan antar fitur dalam 1 page | Menggunakan icon sebelum judul fitur | Aesthetic-Usability Effect, Law of Prägnanz, Occam's Razor, Serial Position Effect |
| 8 | Pengingat password yang tidak terlihat | Menaruh pengingat di paling kanan sebelah fitur krusial (profil) | Serial Position Effect |
| 9 | Kebingungan membaca grafik dan statistik nilai | Menggunakan grafik yang lebih pantas untuk tujuannya (IP untuk kuantitas dan IPK untuk trend), mengganti judul indikator dari singkatan menjadi kata yang singkat namun representatif, menyembunyikan statistik mendalam agar tidak membuat bingung rata-rata mahasiswa. Memberikan kolom tersendiri untuk indikator nilai paling penting seperti IPK dan SKS. | Jakob's law, Hick's law, Miller's law, Occam's razor, Pareto principle, Parkinson's law, Von Restorff Effect |

**Tabel 4.1 Mapping Solusi Desain Baru untuk setiap *Pain Point* (Lanjutan)**

| 10 | Riwayat kelas yang terlalu berantakan dan rapat | Memberikan icon (membuat nomor menjadi icon), memberikan spacing lebih | Serial Position Effect, Law of Prägnanz |
|---|---|---|---|
| 11 | Terlalu banyak white space sehingga tulisan kecil | Mengoptimasikan penggunaan white space, membesarkan tulisan | Golden Ratio, color value weight |
| 12 | Melihat jadwal kelas kurang jelas bordernya | Menggunakan sistem data ink conservation dan color scheme Call To Action | Jakob's law, Law of Common Region, Von Restorff Effect |
| 13 | Jadwal kuliah tidak mengundang dan membuat malas mencari kelas diluar informasi kelas wajib | Menggunakan sistem collapsible category per jenis mata kuliah dengan membenchmark ke KU Loket, memberikan lebih banyak jarak antar baris, meletakkan mata kuliah fakultas mahasiswa yang bersangkutan sebagai tab default, lalu baru mata kuliah eksternal | JaKob's law, Doherty Threshold, Hick's law, Miller's law, Goal Gradient Effect |

**Tabel 4.1 Mapping Solusi Desain Baru untuk setiap *Pain Point* (Lanjutan)**

| 14 | Menu Pembayaran dengan indikator yang tidak tahu gunanya sehingga membingungkan | Menyembunyikan indikator sekunder ke bagian details, menggunakan warna hijau dan merah untuk indikator status dan memberikan tab tersendiri untuk informasi tagihan semester saat ini agar mempercepat pencarian informasi | Aesthetic-Usability Effect, Fitts's law, Hick's law, Serial Position Effect, Von Restorff Effect |
|---|---|---|---|
| 15 | Mata kuliah spesial terasa tidak berguna untuk mengingatkan tugas dan setoran laporan | Memberikan indikator tugas yang butuh dikerjakan, menaruh tombol Call to Action | Doherty Threshold, Von Restorff Effect |

Matriks mapping di atas berfokus kepada pengubahan spesifik per halaman SIAK-NG. Selain pengubahan per halaman, penulis juga melakukan pengubahan template tampilan website secara keseluruhan berdasarkan prinsip desain yang telah disebutkan dalam studi literatur.

### 4.3.1. Warna

Terdapat banyak peningkatan yang harus dilakukan terhadap pemilihan warna SIAK-NG lama berdasarkan kaidah *data-ink ratio*. Penulis melakukan analisis *gamut map* menggunakan KGamut dari desain SIAK-NG yang lama dan mendapati *gamut* dengan 6 sudut yang merepresentasikan 6 *hue* berbeda. Hal ini tidak selaras dengan kaidah desain *user interface* modern seperti *data-ink ratio* dan *call to action color*. Efek dari penggunaan *hue* yang hanya bersifat dekoratif menurut *data-ink ratio* akan menjadi distraksi, terutama untuk website dashboard yang fungsi utamanya adalah sebagai pusat pemberian informasi, dan akan jauh terasa pada mobile website dikarenakan areanya yang sempit sehingga meningkatkan densitas warna.

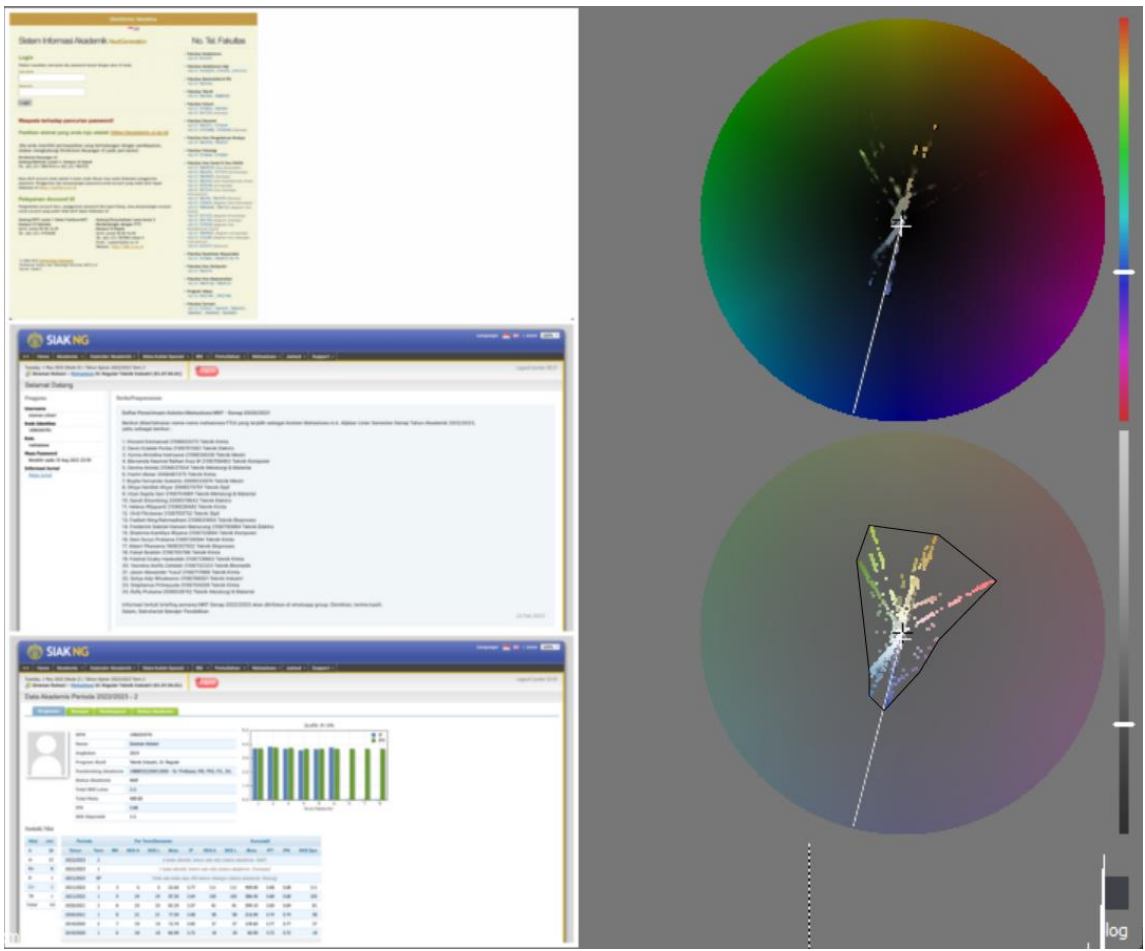

**Gambar 4.3 Gamut Map untuk SIAK-NG Lama**

Sebagai solusi, penulis menetapkan untuk membatasi *hue* website baru menjadi 2. Jumlah ini sesuai dengan kaidah *data-ink ratio* dan *call to action*, dimana salah satu warna akan menjadi warna mayoritas, dan warna kedua menjadi warna aksen yang berfungsi untuk menarik mata pengguna ke titik tertentu.

Sistematika pemilihan warna dimulai dengan permintaan dari pengembangan dan user persona untuk mempertahankan warna kuning sebagai ciri khas dari Universitas Indonesia.

Sebelumnya, penulis harus menetapkan warna kuning spesifik yang akan digunakan.

Warna kuning adalah warna dengan properti yang cukup ketat. Hal ini dikarenakan warna kuning adalah warna dengan *value* atau kecerahan yang tinggi di skala *grayscale*. Warna kuning dengan *value* pertengahan harus dihindari karena menurut riset warna tersebut merupakan warna terjelek bagi mata manusia (Ferrier, 2016), sehingga dipakai sebagai warna barang-barang yang diharapkan dijauhi seperti bungkus rokok.

Menggunakan warna kuning pertengahan bertentangan dengan *aesthetic-usability effect* yang menyatakan bahwa agar sesuatu terlihat terpercaya, hal tersebut harus terlihat cantik dan estetik. Sementara itu, tidak dimungkinkan menggunakan warna kuning dengan *value* gelap karena spektrum warna dengan hue kuning tidak cukup kuat untuk melawan *value* gelap, sehingga warna kuning dengan value gelap cenderung diasosiasikan dengan hitam saja. Hal ini berbeda dengan warna seperti biru yang memiliki spektrum *hue* cukup kuat walaupun dengan value gelap seperti warna biru dongker ataupun hijau phthalo. Oleh karena itu, disimpulkan warna kuning yang dapat digunakan terbatas pada kuning dalam kotak merah pada skala munsel di bawah.

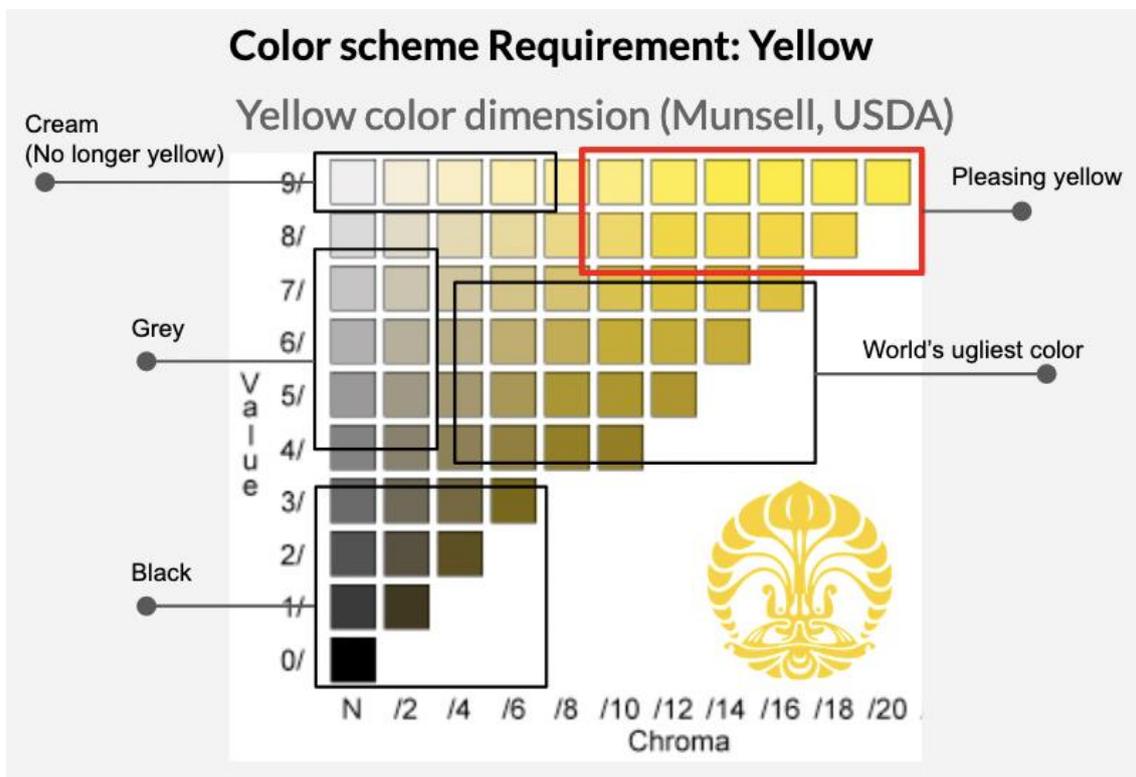

**Gambar 4.4 Warna Kuning Layak Pakai**

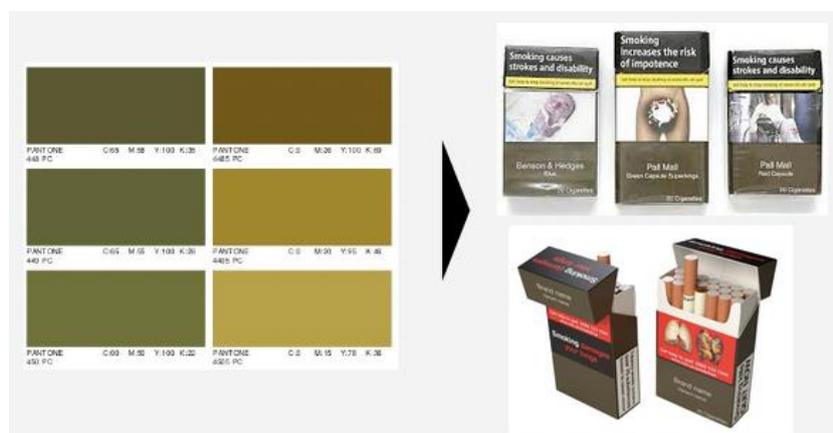

**Gambar 4.5 Warna Terjelek bagi Mata Manusia (Ferrier, 2016)**

Langkah selanjutnya adalah untuk memilih warna utama atau mayoritas yang cocok dengan warna kuning. Menurut kaidah *call to action*, kedua warna yang terpilih harus memiliki kontras tinggi. Semakin tinggi kontras antara dua warna dalam sebuah desain, semakin efektif warna aksen dalam menarik perhatian pengguna untuk mengklik atau menginspeksi elemen dengan warna tersebut. Dikarenakan warna kuning yang terpilih berada di spektrum ekstrim dari *grayscale value*, maka warna kedua diharuskan berada pada ekstrim gelap di *grayscale value*. Kemudian, untuk pemilihan *hue*, digunakan warna komplementer dari warna kuning, yakni warna biru. Warna komplementer didefinisikan sebagai pasangan warna yang saling berseberangan pada *color wheel*. Cirinya adalah warna komplementer selalu memiliki kontras tertinggi dibanding gabungan hue manapun.

Hasil akhir dari pertimbangan dengan kaidah-kaidah tersebut adalah warna biru dongker adalah warna terbaik sebagai pasangan warna kuning yang merupakan ciri khas Universitas Indonesia, dengan kelebihan antara lain:

- Kontras tinggi (mendukung aspek keringkasan atau *concision* dari user interface)
- Harmoni tinggi (warna yang saling mendukung dan tidak saling berkompetisi, mendukung aspek kenyamanan atau *comfort*)

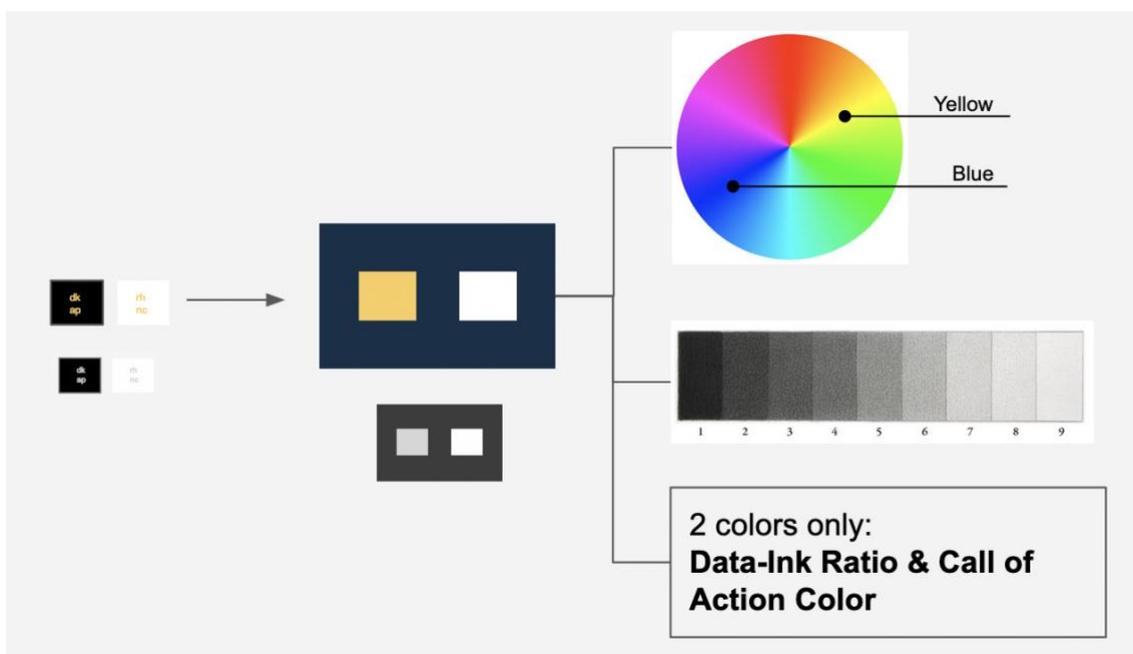

**Gambar 4.6 Pertimbangan Pemilihan Warna**

Hasil akhirnya adalah *color palette* yang terdiri dari kuning dan biru dongker, yang berhasil menghasilkan gamut map dengan 2 sudut ynag merepresentasikan 2 *hue* sesuai kaidah *data-ink ratio*.

### 4.3.1.1. *Dark Mode vs. Light Mode*

Pada akhirnya, penulis memilih mengembangkan *prototype* dengan *dark mode*. Hal ini berdasarkan pertimbangan teknis dari properti bawaan spektrum warna biru dan kuning. Warna kuning memiliki ciri bawaan yaitu saturasi yang tinggi, sehingga sangat cocok digunakan sebagai *Call to Action color* karena warna dengan saturasi tinggi selalu menarik mata manusia. Warna kuning tersebut tidak dimungkinkan diturunkan saturasinya (disebut juga chroma dalam diagram Munsel) karena properti bawaan dari warna kuning adalah kekuatan spektrum *hue* yang lemah, sehingga tanpa saturasi yang tinggi akan kehilangan asosiasinya dengan warna kuning dan diasosiasikan dengan warna krem atau putih. Oleh karena itu, warna yang dapat diturunkan saturasinya dan dibuat netral adalah warna biru. Warna biru mampu diberikan saturasi yang lebih rendah, sehingga menyebabkan warna biru menjadi warna netral dan pantas digunakan sebagai latar belakang.

Selain itu, telah terdapat banyak riset yang konsisten mendukung penggunaan *dark mode,* yang menunjukkan bahwa user interface yang menggunakan *dark mode* memberikan *visual acuity* yang lebih tinggi dengan *fatigue* yang lebih rendah. Penelitian oleh Kim (2019) dan Dillon (1998) menunjukkan bahwa teks putih pada latar belakang gelap secara signifikan mengurangi kelelahan visual yang akan membantu kesehatan mata pengguna. Selain itu, penggunaan *dark mode* telah menjadi best practice pada rata-rata website dashboard seperti Coinbase dan e-Trade. Palet warna desain SIAK-NG yang baru terbatas pada navy dengan aksen putih dan kuning sebagai warna pemanggilan aksi, berbeda jika dibandingkan dengan website lama yang menggunakan skema warna biru, hijau, putih, abu-abu, dan kuning dengan luas area penggunaan yang saling bersaing satu sama lain.

Hasil akhir analisis gamut untuk pemilihan warna website baru menunjukkan pengurangan hue yang signifikan, dan secara efektif berhasil memberi solusi bagi 2 dari 3 aspek kebutuhan pengguna yaitu keringkasan (*concision*) dan kenyamanan (*comfort*).

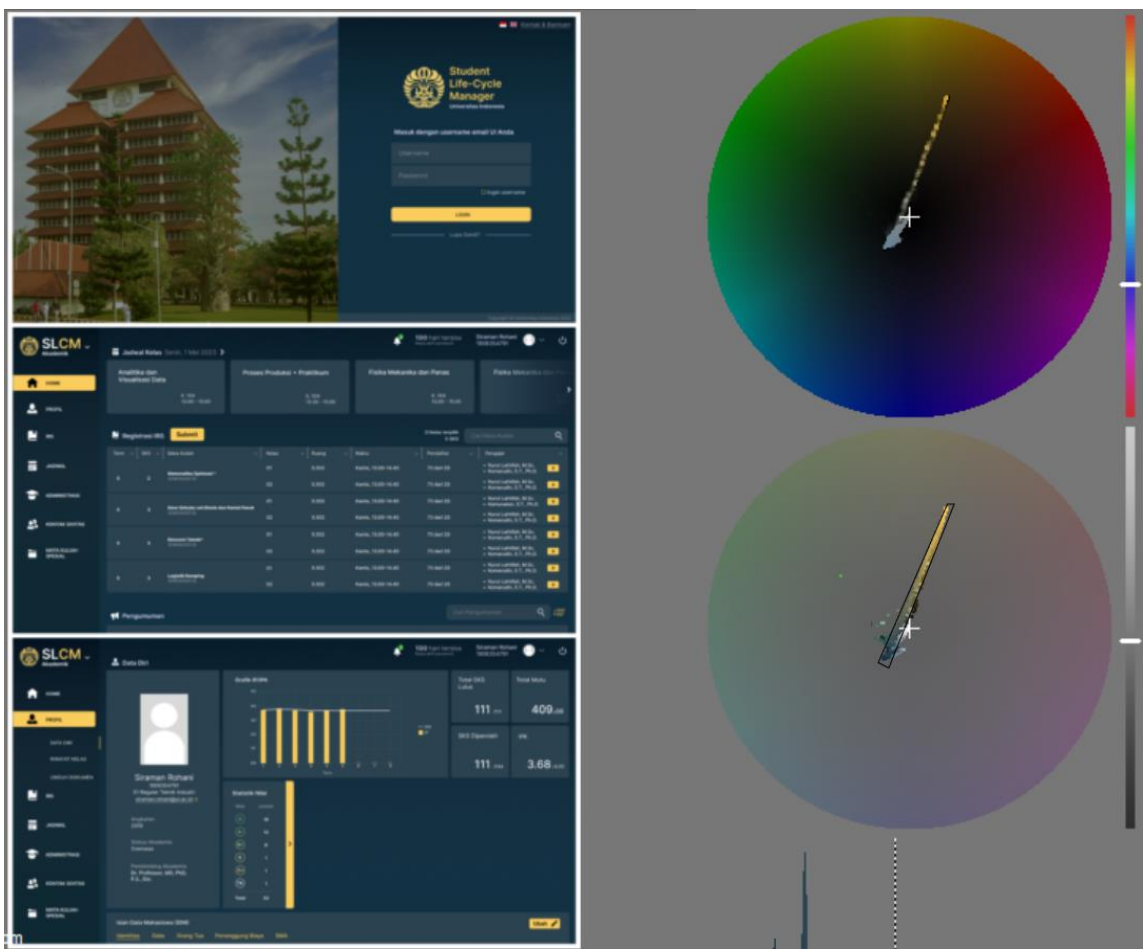

**Gambar 4.7 Gamut Map untuk SIAK-NG Baru**

### 4.3.2. Layout utama

Menurut Fu et. al. (2023), hasil eyetracking paling mendukung pencarian data dengan cepat adalah *F-Shaped Click Model* atau FSCM. Prinsip FSCM merupakan prinsip ergonomi kognitif yang disebabkan oleh kebiasaan manusia membaca dari kiri ke kanan untuk mencari detail (*sinistrodextrality*), dan atas ke bawah dalam membuat sebuah daftar. Pada website lama, dapat dilihat bahwa *navigation bar* terbalik 180 derajat dengan kaidah FSCM. Hal tersebut membuat pencarian dan pemetaan kategori kurang efektif, dan dapat menjelaskan mengapa *layout navigation bar* yang terpakai di SIAK-NG lama jarang ditemukan di *website-website* besar.  Hal ini diubah agar mendukung aspek perbaikan yang diinginkan oleh *user persona* yakni kenyamanan dan kemudahan dipelajari.

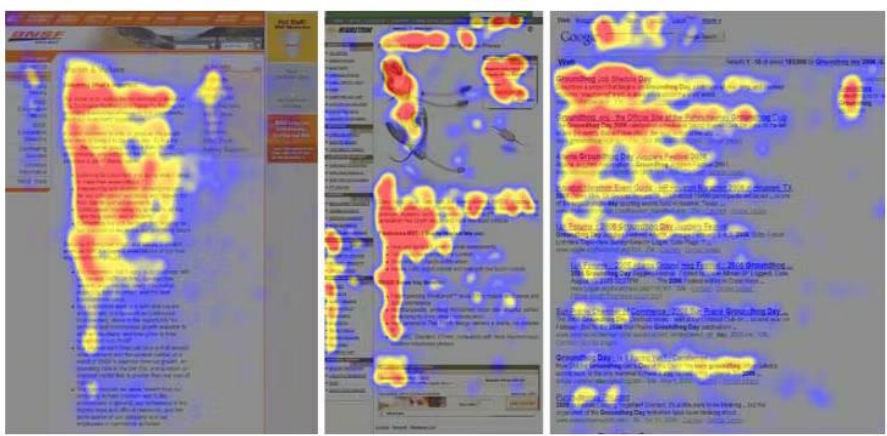

**Gambar 4.8 FSCM**

Sumber: Fu et. al. (2023)

### 4.3.2.1. Hamburger Menu

Sebuah opsi lain yang dipertimbangkan oleh penulis adalah *hamburger menu* yang populer pada tahun 2012, namun segera hilang pada akhir dekade tersebut (Suleiman, 2023). Desain *hamburger menu* pada hakikatnya menyembunyikan menu navigasi. Tendensi desainer adalah untuk menyembunyikan menu navigasi supaya dapat memberi ruang untuk konten, padahal menu navigasi tidak boleh disembunyikan karena dengan menyembunyikan menu navigasi, maka konteks pun juga tersembunyi. Akibatnya pengguna tidak tahu posisi mereka di dalam website secara bawah sadar, dan cenderung takut mengeksplorasi halaman lain.

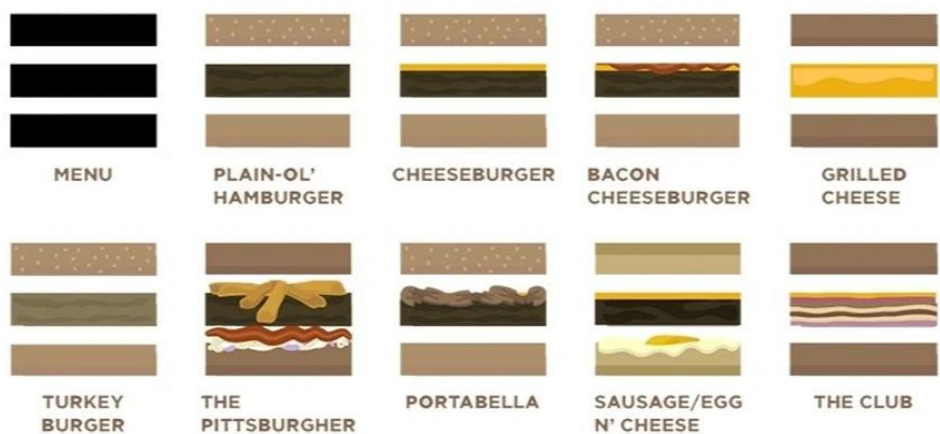

**Gambar 4.9 Hamburger Menu**

Sumber: Suleiman (2023)

Berdasarkan data pada bab 4.3, ditemukan bahwa pain poin utama SIAK-NG adalah ketidaktahuan mahasiswa dengan fitur-fitur berguna namun tersembunyi dalam SIAK-NG, yang menyebabkan mahasiswa berpikir bahwa website tersebut kurang berguna. Sekalipun *layout navigation bar* SIAK-NG saat ini dapat dibilang kurang efektif, menggunakan *hamburger menu* untuk menggantikannya hanya akan memperparah *user experience* karena menukar fungsionalitas dengan estetika dan kebersihan desain. Terdapat batas maksimum untuk simplifikasi, dan memaksa simplifikasi lebih lanjut akan menukar fungsionalitas dengan estetika menurut Tesler's Law.

Hasil penurunan performa diukur oleh Pernice, K., & Budiu, R. (2016), dan menunjukkan bahwa dengan menggunakan *menu hamburger,* efektivitas pencarian dalam website mengalami penurunan hampir 50% karena navigasi utama situs yang disembunyikan. Selain itu, waktu penyelesaian tugas lebih lama dan tingkat kesulitan tugas yang dirasakan meningkat.

Hal ini karena navigasi yang tersembunyi lebih sulit ditemukan daripada navigasi yang terlihat atau sebagian terlihat. Ketika navigasi tersembunyi, pengguna cenderung tidak menggunakan menu navigasi.

Kuncinya adalah untuk tidak pernah menyembunyikan navigasi. Navigasi yang tersembunyi memberikan user experience yang lebih buruk dibandingkan dengan navigasi yang terlihat atau sebagian terlihat, baik pada *user interface mobile site* ataupun *desktop site*. Tingkat kesulitan tugas, waktu yang dihabiskan untuk menyelesaikan tugas, dan keberhasilan tugas.

Berbeda dengan menu FSCM yang memiliki layout yang didesain sedemikian rupa sehingga pengguna selalu memiliki indikator di halaman mana mereka berada yang di-*highlight*, dan halaman apa saja yang bisa mereka akses (tidak di-*highlight*), memberikan kenyamanan psikologis dan rasa aman untuk berinteraksi lebih banyak dengan website tanpa tersesat dan meningkatkan engagement.

## 4.4. Analisis *Prototype*

Tahap *prototype* dalam proses perancangan merupakan langkah penting dalam menghadirkan ide-ide dari tahap ideate ke dalam bentuk desain konkret. Dalam hal ini, penulis menggunakan sistem *desain atomic*, yang memberikan kerangka kerja yang terstruktur dan efisien. Penulis memulai proses desain dengan pendekatan dari yang besar ke yang kecil. Pertama-tama, penulis menganalisis dan menentukan layout terbaik untuk setiap div dalam desain, memperhatikan ukuran dan posisinya. Dalam melakukan ini, penulis banyak mengandalkan *golden ratio* yang telah terbukti menghasilkan *layout* yang estetis dan seimbang.

Selain itu, penulis juga melakukan *benchmarking* terhadap *layout* yang telah berhasil sebelumnya, sebagai panduan dalam memilih *layout* yang optimal. Setelah menyelesaikan analisis *layout*, penulis kemudian fokus pada elemen- elemen yang lebih kecil dalam desain. Dalam proses ini, penulis mempertimbangkan kecerahan masing-masing elemen, dengan memperhatikan prinsip *color value weight*. Hal ini penting untuk menciptakan keseimbangan visual dan memastikan elemen-elemen tersebut dapat dengan jelas dilihat dan mempengaruhi pengalaman pengguna. Hasil akhir dari proses perancangan ini kemudian ditampilkan kepada tester. Hal ini dilakukan untuk mendapatkan umpan balik terakhir sebelum desain tersebut diuji secara keseluruhan.

Dengan melibatkan tester, penulis dapat memperoleh wawasan yang berharga mengenai keefektifan desain dan melakukan perbaikan yang diperlukan sebelum desain tersebut diimplementasikan secara penuh. Secara keseluruhan, website baru memiliki beberapa perubahan visual. Tujuan utama dari desain website baru ini dapat dirangkum menjadi 3 hal: meningkatkan keringkasan (*concision*), kemudahan dipelajari (*intuitiveness*), dan kenyamanan (comfort), karena semua masalah yang dilaporkan dapat diselesaikan dengan tiga isu tersebut berdasarkan tahap *define*. Hal ini dicapai melalui metode berikut:

Keseluruhan desain website dibuat menggunakan sistem *atomic design* yang mendorong penggunaan *space* maksimum namun tetap menjaga pentingnya penggunaan *white space*. Desain *atomic* membantu menciptakan ukuran div besar dengan banyak *white space* yang secara keseluruhan membantu memberikan ruang lebih untuk *font* yang lebih besar namun tetap mencegah *layout* untuk tidak terlihat berantakan. Sistem *atomic design* juga mendorong pengembang untuk memiliki elemen yang konsisten untuk seluruh website yang memperkuat penggunaan *Jakob's Law* dengan mengulangi pola desain elemen dan membantu pengguna mengidentifikasi desain elemen yang mirip sehingga menciptakan website yang lebih intuitif.

*Prototype* baru ini juga menciptakan sistem klasifikasi baru. Kategori yang membingungkan adalah salah satu permasalahan utama antara pengguna, terutama karena website lama memiliki 9 kategori. Berdasarkan Hukum Miller, rata-rata pengguna hanya dapat mengingat 5-7 item. Dengan adanya 9 kategori, batas ini jauh terlampaui. Oleh karena itu, dengan menciptakan kategori yang lebih sedikit, pengguna dapat memetakan seluruh fitur yang ditawarkan oleh *website* dengan lebih baik dan membantu mereka mengingat dengan lebih mudah selama kategori-kategori tersebut mampu menggambarkan fitur website secara akurat.

Kemudahan penggunaan juga dicapai melalui *law of similarity* and *law of uniform connectedness* dengan menciptakan layout dengan susunan fungsi yang kontinyu. Hal tersebut akan menciptakan *fluidity* dalam bernavigasi. *Fluidity* tersebut akan lebih efektif dalam mendorong pengguna untuk mengecek fitur berikutnya jika dibandingkan dengan batasan yang jelas seperti mengklik tab.

Perubahan-perubahan utama tersebut terefleksi di seluruh halaman SIAK-NG yang baru, sementara perubahan yang spesifik ke halaman tertentu didasarkan pada pain poin dan solusi yang terdapat pada tabel. 4.2.

### 4.4.1. Halaman Login

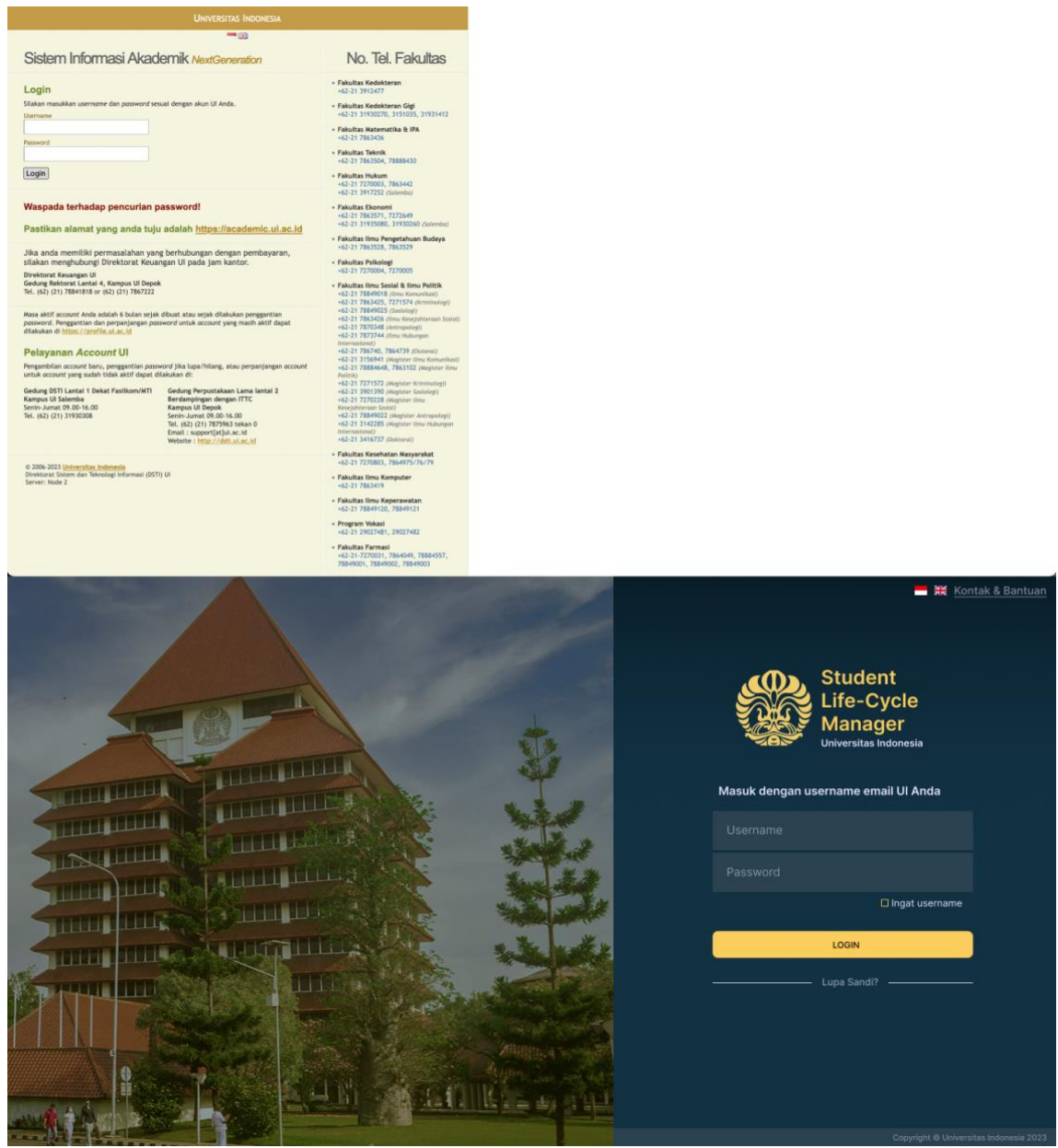

**Gambar 4.10 Login**

Solusi telah diterapkan pada halaman login. Kontras yang kecil dan terang bagi mata diatasi dengan skema warna yang memiliki kontras tinggi dan warna komplementer.

Masalah tulisan yang berantakan diatasi dengan *collapsible* link, layout berhierarki, dan merangkum tulisan dengan prinsip *scannable, concise, dan objective*. Kesalahan dalam memencet username dan password diatasi dengan memperbesar ukuran form, memberikan jarak, dan indikator form yang aktif. Rating reliabilitas yang rendah karena tampilan yang kurang profesional diperbaiki dengan menambahkan gambar sebagai daya tarik, tetapi tetap menjaga *color gamut* minimal. Aplikasi *Jakob's Law* terlihat pada layout login yang mengikuti layout best practice halaman login yang dipakai di website profesional seperti Credit Suisse selain *website academic portal* lainnya seperti KU Loket.

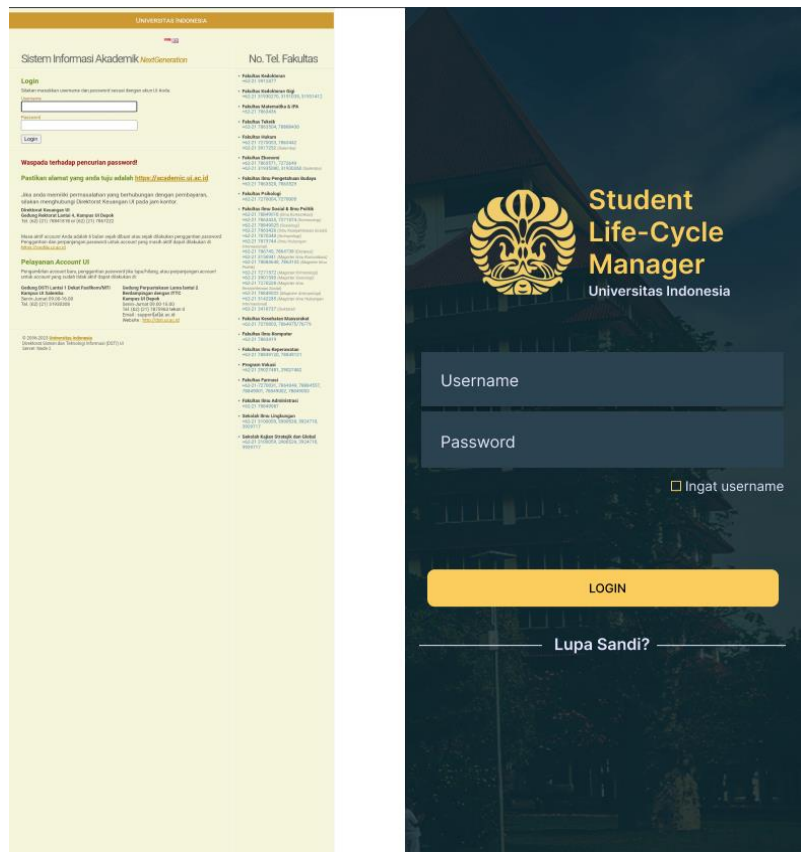

**Gambar 4.11 Login (mobile)**

### 4.4.2. Halaman Home

Gambar 4.12 Home

Masalah yang membutuhkan minimal 3 kali klik untuk melihat jadwal kelas dan mengganggu saat terburu-buru diatasi dengan menempatkan *widget* jadwal kelas di halaman utama. Fitur pengumuman yang kurang diperhatikan karena banyak kata diatasi dengan sistem preview dan merangkum tulisan menggunakan prinsip *scannable, concise, dan objective*. Untuk memudahkan pengguna dalam mengetahui pengumuman baru atau lama, digunakan simbol notifikasi berdasarkan prinsip *Shneiderman's Golden Rules.*

Untuk mengurangi kebingungan pengguna terhadap *widget* di *dashboard*, digunakan simbol dan ikon sebagai komunikasi visual, serta div dengan kontras tinggi. Untuk memudahkan akses ke isi IRS, ditambahkan *shortcut* ke *widget* IRS di dashboard. Terakhir, untuk meningkatkan kesadaran pengguna terhadap pengumuman baru, widget pengumuman ditempatkan di *dashboard* agar mudah terlihat ketika menggunakan fitur-fitur utama SIAK-NG. Pengisian IRS akan ditutup di luar masa pengisian IRS untuk memberikan *dashboard* yang lebih fungsional.

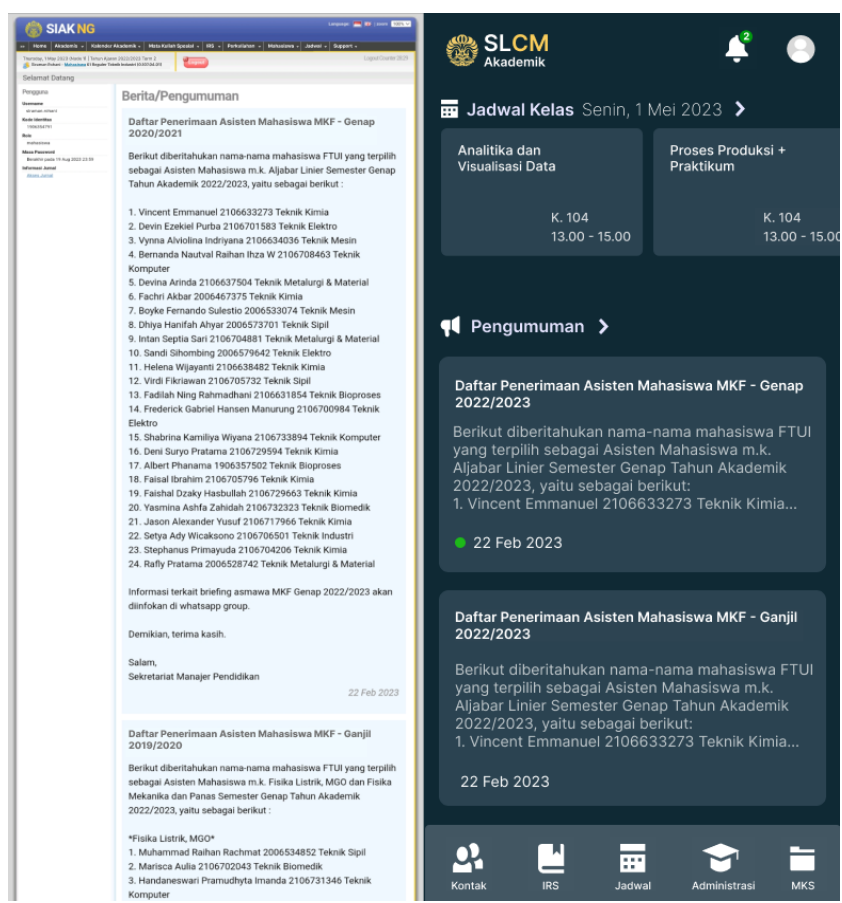

**Gambar 4.13 Home (Mobile)**

### 4.4.3. Kategori Profil

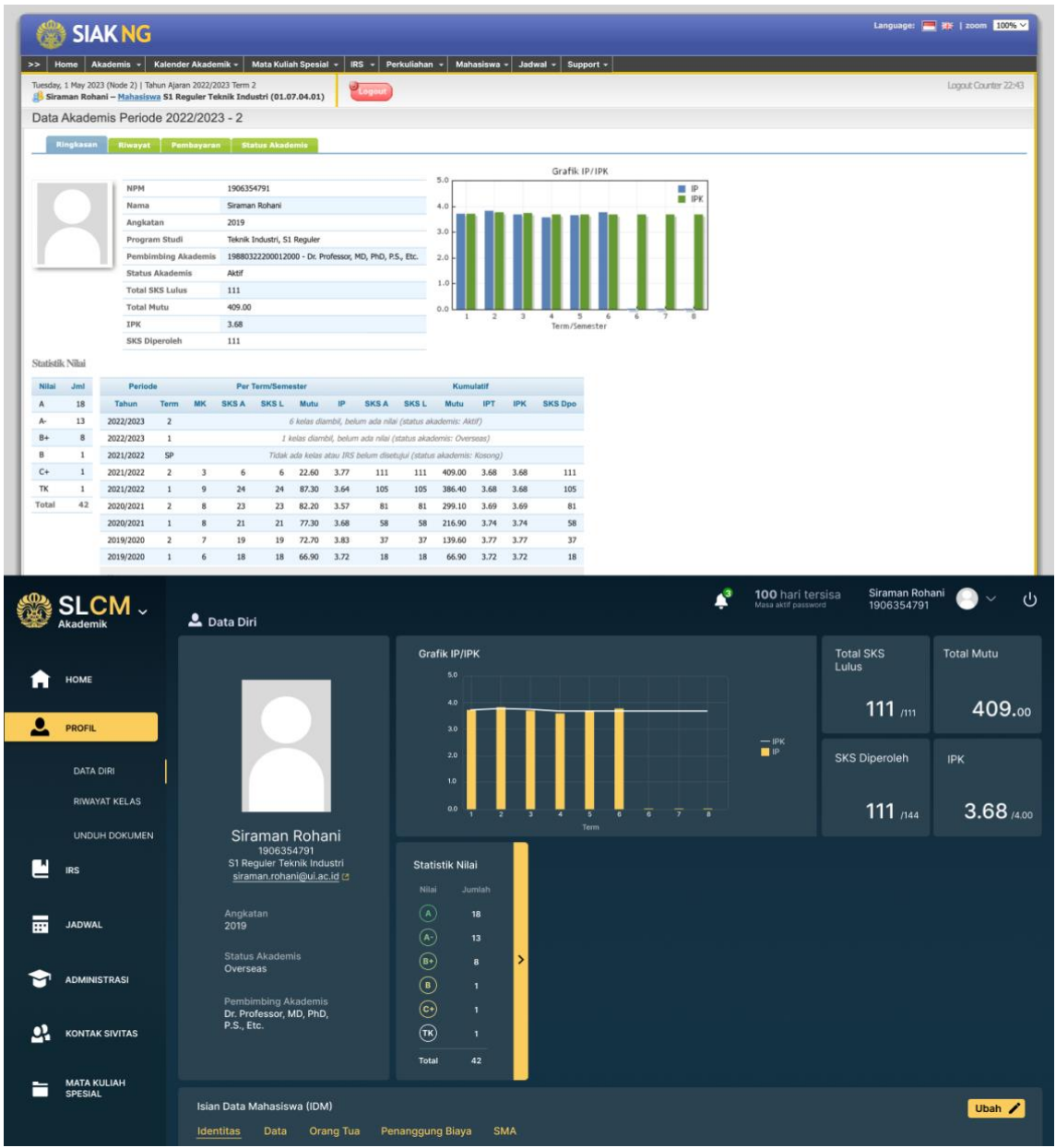

**Gambar 4.14 Profil**

Sebelumnya, kategori ini dikenal sebagai kategori akademis. Namun, nama tersebut dinilai kurang representatif karena semua yang ada di SIAK-NG berhubungan dengan akademis. Oleh karena itu, kategori ini sekarang diberi nama "Profil" yang lebih sesuai, dan hanya menampilkan data yang spesifik untuk masing-masing mahasiswa. Kategori ini terdiri dari beberapa bagian, yaitu "Ringkasan", "Riwayat", dan "Unduh Dokumen". Bagian "Ringkasan" telah digabungkan dengan "IDM", karena fungsinya secara keseluruhan berkaitan dengan data diri pengguna. Bagian "Riwayat" tidak mengalami banyak perubahan dalam tata letaknya, namun dilakukan peningkatan dalam hal tampilan dengan memberikan padding yang lebih nyaman dibaca. Bagian "Unduh Dokumen" telah ditingkatkan dengan memberikan detail file dan opsi untuk melihat dan mengunduh dokumen secara langsung. Halaman ini sekarang menjadi *repository* untuk semua dokumen yang dapat diunduh, sehingga mengatasi masalah tersebarnya dokumen di berbagai halaman sebelumnya.

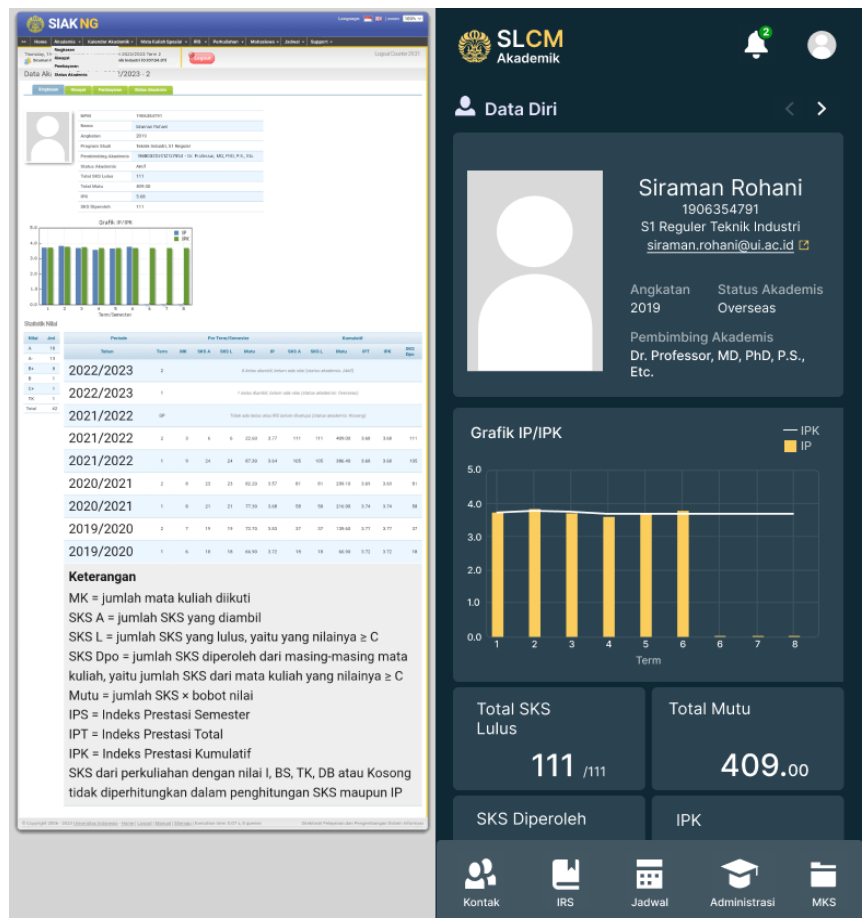

**Gambar 4.15 Profil (mobile)**

### 4.4.4. Kategori IRS

**Gambar 4.16 IRS**

Perubahan signifikan telah diterapkan pada kategori IRS untuk meningkatkan pengalaman pengguna dan meningkatkan efektivitas fitur-fitur yang disediakan. Fitur "Isi IRS" diubah menjadi "Registrasi IRS" untuk memberikan nama yang lebih representatif terhadap fungsi fitur tersebut. Fitur "Add IRS" dan "Drop IRS" digabungkan menjadi satu, mempermudah penggunaan fitur dan menjadikannya lebih ringkas, khusus untuk kasus-kasus khusus. Halaman ringkasan IRS juga mengalami perubahan dengan penambahan indikator yang lebih lengkap untuk memungkinkan pengguna memantau dan melacak IRS dengan lebih mudah dan efisien.

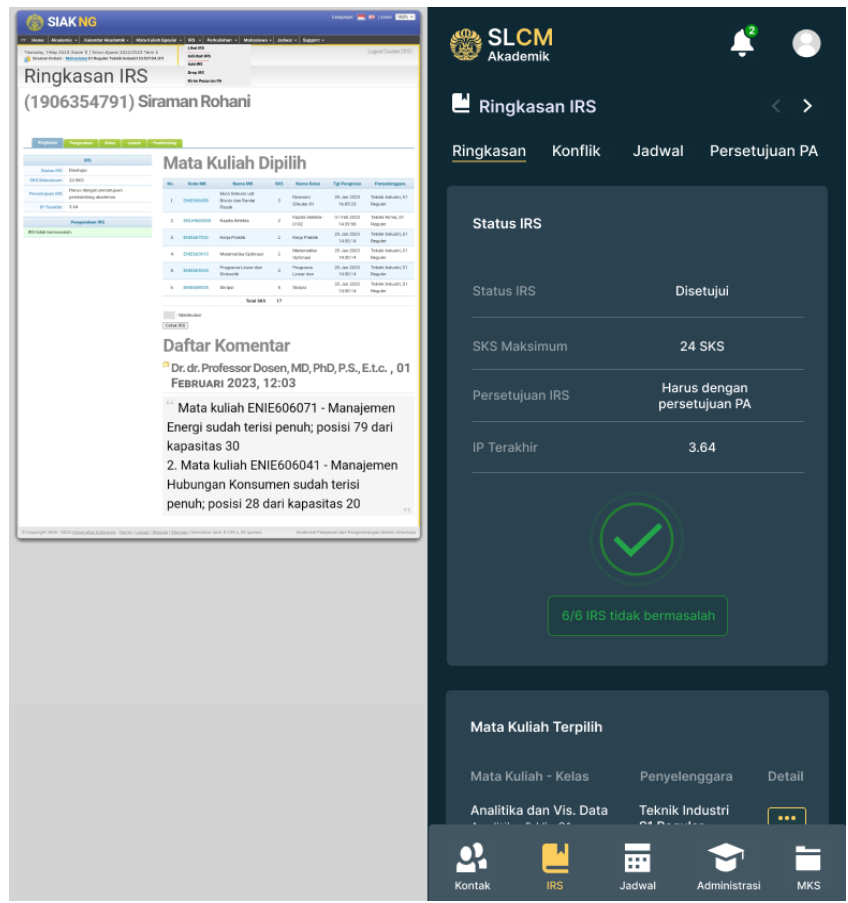

**Gambar 4.17 IRS (mobile)**

### 4.4.5. Kategori Jadwal

**Gambar 4.18 Jadwal**

Perubahan pada kategori Jadwal mencakup: Jadwal kelas yang tidak lagi bersifat statis, tetapi dapat dilihat setiap minggu dan terhubung ke database pemakaian kelas. Ini memudahkan mahasiswa untuk memperoleh informasi terbaru tentang perubahan jadwal, tidak perlu mengandalkan komunikasi informal. Penambahan fitur ekspor kalender berdasarkan rekomendasi dari *persona cluster* 3. Fitur ini memungkinkan pengguna untuk mengunduh jadwal kuliah mereka dalam format kalender yang sesuai dengan aplikasi kalender yang mereka gunakan. Pengembangan halaman Jadwal Kuliah sebagai direktori kelas dengan penggunaan menu *collapsible* untuk klasifikasi kelas. Ini memudahkan pengguna dalam menavigasi dan mencari informasi yang relevan tentang kelas yang mereka ikuti. Penambahan fitur silabus yang terpusat pada halaman Jadwal Kuliah, memungkinkan pengguna untuk mengakses informasi yang lebih rinci tentang materi pembelajaran dan tugas yang terkait dengan setiap kelas. Penggabungan fitur Add IRS dan Drop IRS menjadi satu fitur yang lebih ringkas dan digunakan hanya untuk kasus-kasus khusus, meningkatkan efisiensi penggunaan fitur tersebut. Penambahan indikator lebih banyak pada halaman ringkasan IRS, sehingga pengguna dapat dengan mudah melacak dan memantau IRS dengan lebih efektif sebagai alat pemantauan.

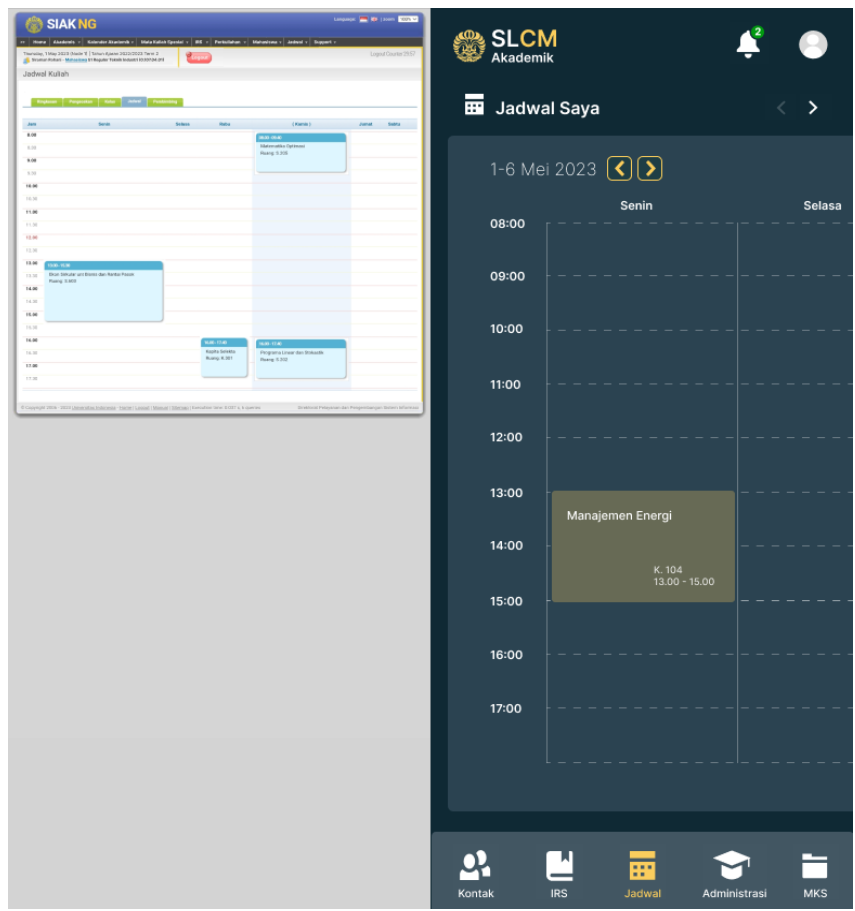

**Gambar 4.19 Jadwal (mobile)**

### 4.4.6.   Kategori Kontak Sivitas

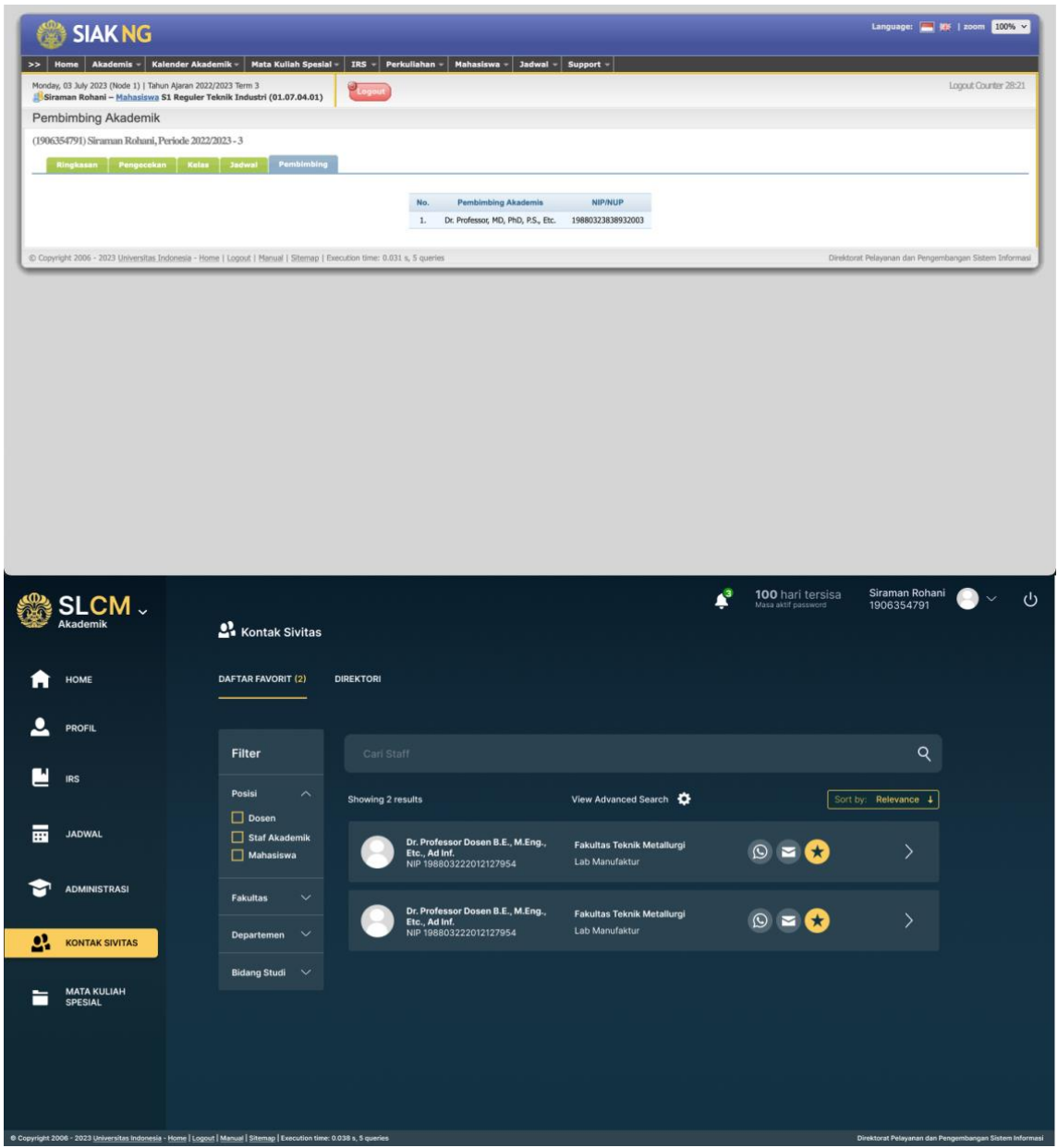

**Gambar 4.20 Kontak**

Halaman ini mengubah sistem lama dimana kontak tercecer dan sulit diakses karena pengguna harus mencari kelas terlebih dahulu sebelum menemukan informasi kontak dosen terkait. Namun, sekarang sistem telah diperbarui sehingga pengguna dapat dengan langsung mencari *personel sivitas* UI melalui halaman yang telah disediakan. Sebelumnya, informasi kontak tersebar di berbagai website fakultas dengan kualitas informasi yang berbeda-beda. Namun, sekarang semua informasi kontak telah disentralisasi di SIAK-NG, sehingga pengguna dapat mengaksesnya dengan mudah dan memperoleh informasi yang konsisten. Selain itu, telah ditambahkan halaman favorit yang menjadi halaman pertama untuk memudahkan pengguna dalam mencari dan menghubungi dosen-dosen yang sering dibutuhkan untuk komunikasi. Hal ini memberikan kemudahan dan efisiensi dalam mencari kontak yang penting bagi pengguna.

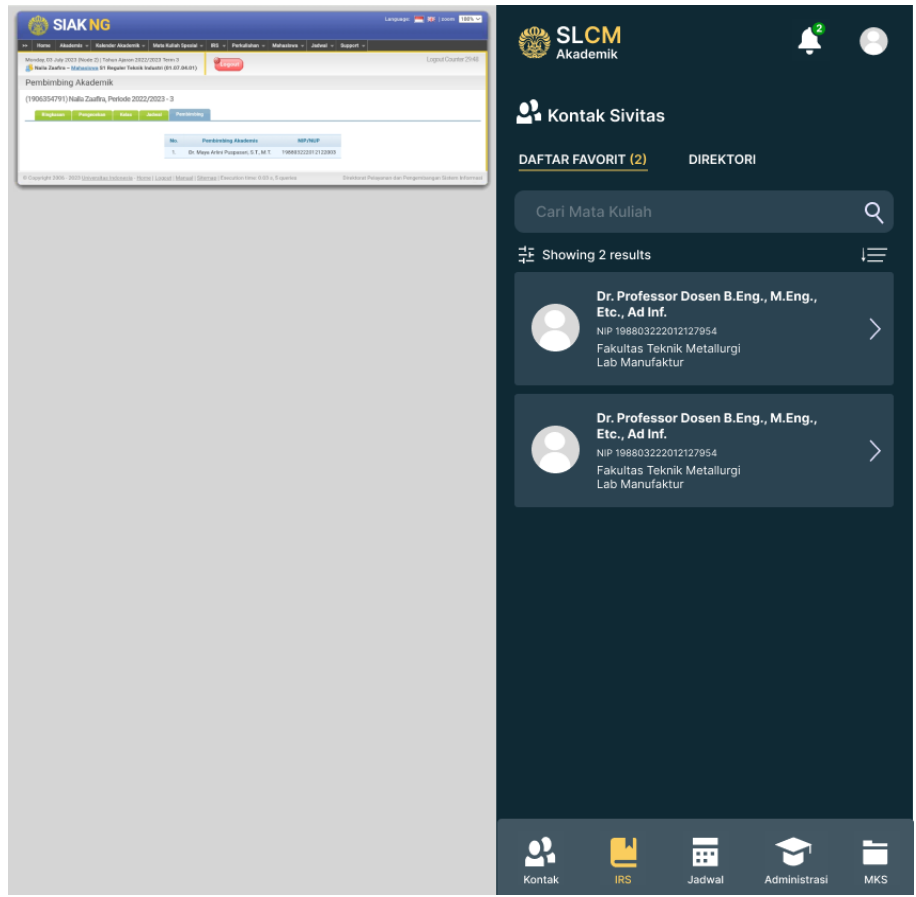

**Gambar 4.21 Kontak (mobile)**

### 4.4.7. Kategori Administrasi

**Gambar 4.22 Pembayaran**

Perubahan yang signifikan telah dilakukan pada kategori "Administrasi" untuk meningkatkan pengalaman pengguna dan menyederhanakan proses administratif. Fitur-fitur administrasi yang sebelumnya digabungkan dengan profil dan data diri pengguna kini dipisahkan untuk memberikan fokus yang lebih jelas. Perubahan utama terjadi pada halaman pembayaran, yang telah diubah menjadi sebuah *dashboard* yang menampilkan status pembayaran per semester. Pengguna dapat dengan mudah melihat status pembayaran terbaru dan melihat gambaran keseluruhan pembayaran untuk semua semester yang telah berlalu melalui indikator yang disediakan. Selain itu, informasi mengenai bank yang bekerja sama untuk pembayaran juga disediakan. Pengguna dapat mengklik bank yang tercantum untuk memperoleh panduan tentang cara pembayaran yang harus dilakukan. Tujuan dari perubahan ini adalah untuk memberikan panduan yang jelas dan memudahkan pengguna dalam melaksanakan proses pembayaran.

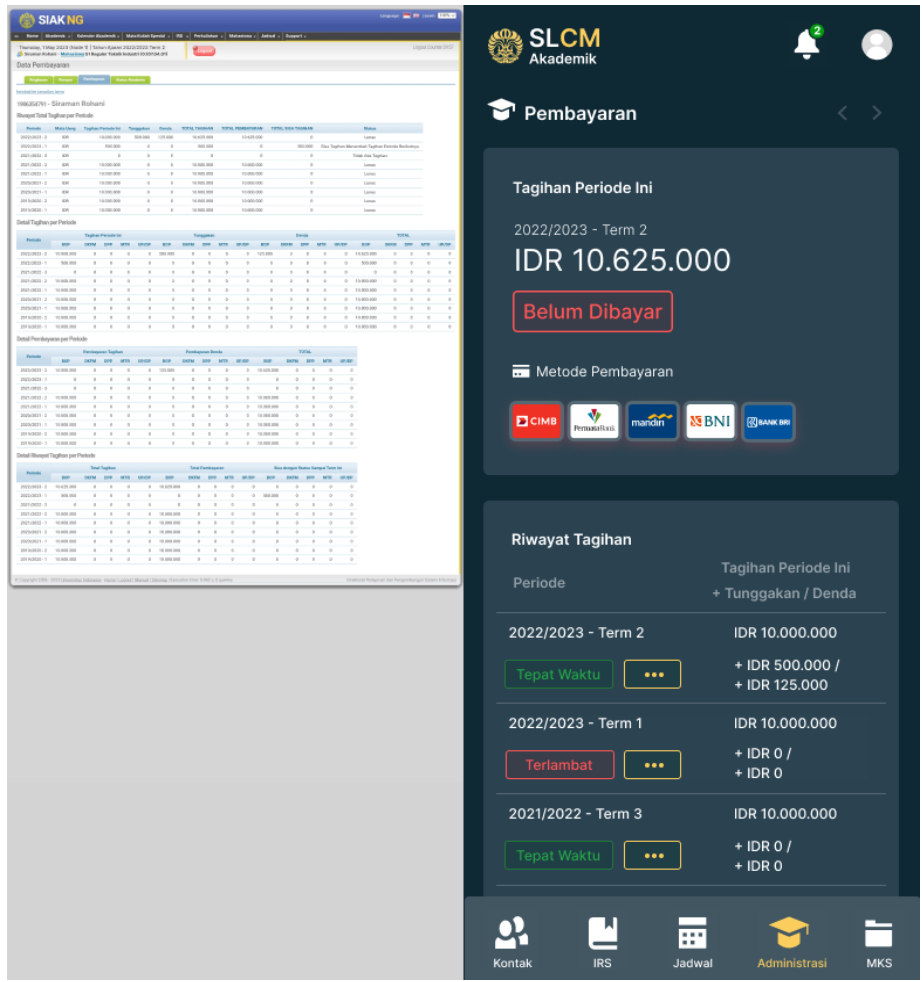

**Gambar 4.23 Pembayaran (mobile)**

### 4.4.8. Kategori Mata Kuliah Spesial

**Gambar 4.24 Mata Kuliah Spesial**

Kategori mata kuliah spesial diubah menjadi pusat pengumpulan laporan untuk memudahkan pengguna dalam mengelola tugas-tugas terkait mata kuliah tersebut. Indikator pengumpulan ditambahkan untuk memantau status dan kemajuan tugas dengan lebih mudah, mengurangi kebingungan dan meningkatkan transparansi dalam proses pengumpulan tugas.Fitur "Edit" diubah menjadi "Pilih Pembimbing" untuk mengkomunikasikan secara spesifik tujuan halaman tersebut, yaitu memilih pembimbing untuk mata kuliah spesial.

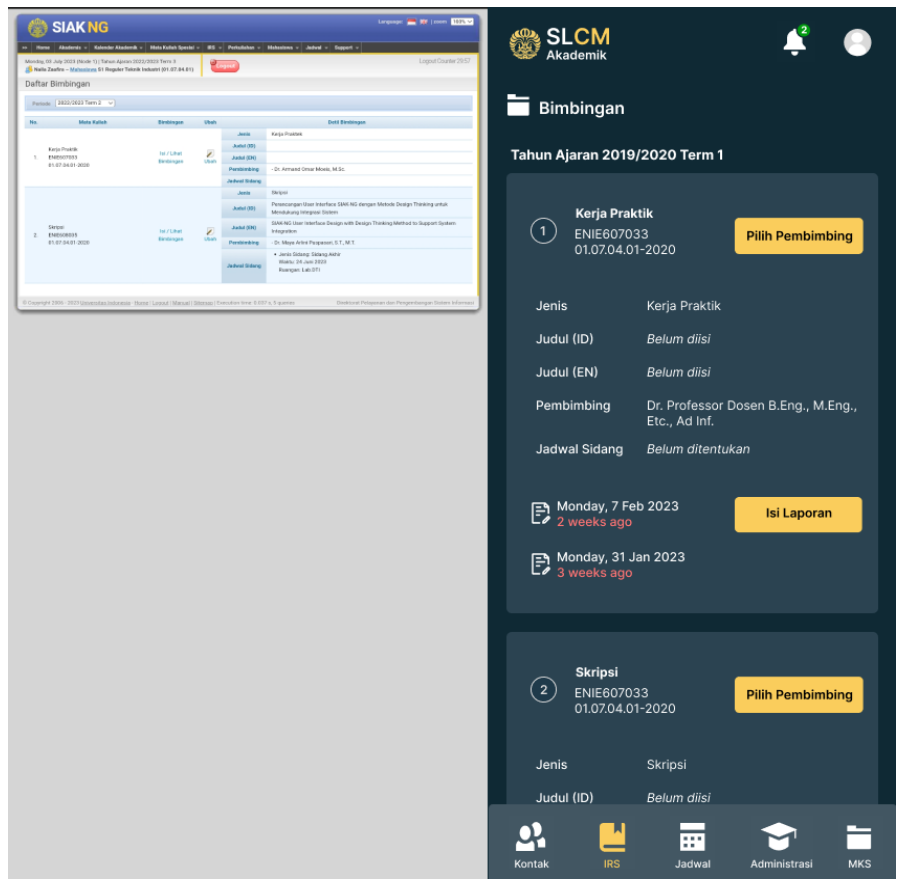

**Gambar 4.25 Mata Kuliah Spesial (mobile)**

## 4.5. Analisis *Testing*

Penelitian ini bertujuan untuk menggunakan *usability testing* guna mengevaluasi efektivitas, efisiensi, dan kepuasan pengguna terhadap desain ulang situs web SIAK-NG. Dalam penelitian ini, dilibatkan 15 partisipan yang terbagi menjadi 5 dari setiap persona, dan setiap partisipan menjalani dua sesi pengujian.

Pada *usability testing* pertama, dilakukan evaluasi terhadap *user interface* (UI) lama dari situs web, sedangkan *usability testing* kedua mengevaluasi UI baru yang telah diredesain. Pengujian dilakukan baik untuk versi *mobile* maupun *desktop*. Hasil dari *usability testing* penggunaan dan hasil dari penggunaan PSSUQ UI lama dibandingkan dengan UI baru untuk melakukan validasi. Dalam analisis penelitian ini, dilakukan evaluasi terhadap *direct success, mission unfinished, misclick rate, duration,* dan *usability score.* Penelitian ini juga menyajikan tabel perhitungan untuk *usability testing* penggunaan serta hasil PSSUQ UI baik untuk versi lama maupun versi baru dari antarmuka pengguna (UI).

### 4.5.1. Direct Success

Hasil perbandingan *direct success* ditampilkan dalam bentuk grafik di bawah. Performa lebih baik didapatkan dengan direct succes yang lebih tinggi. Pada seluruh task, dapat dilihat bahwa versi baru memiliki persentase direct success yang sangat signifikan, terutama untuk task 5. Hal tersebut berarti mayoritas pengguna dapat menemukan menu yang dicari lebih mudah dengan *layout* dan klasifikasi yang baru. Hal ini cukup signifikan karena pertimbangan tester yang belum pernah memakai versi terbaru namun tetap bisa bersaing dengan versi lama.

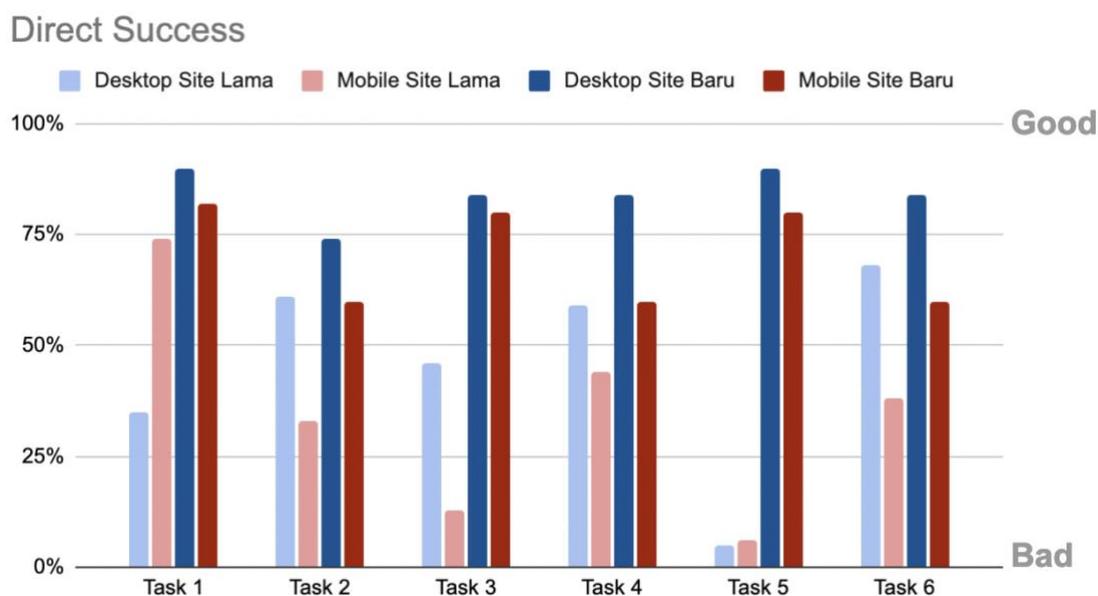

**Gambar 4.26 Perbandingan Metrik: Direct Success**

### 4.5.2. Mission Unfinished

Hasil perbandingan *mission unfinished* ditampilkan dalam bentuk grafik yang dapat dilihat pada Gambar 4.12. Performa lebih baik didapatkan dengan *mission unfinished* yang lebih rendah. Pada seluruh task, dapat dilihat bahwa untuk *desktop site*, versi baru memiliki persentase *mission unfinished* yang jauh lebih sedikit. Hal tersebut berarti mayoritas pengguna tidak frustasi dan menyerah sebelum menyelesaikan task. Hal ini dapat diatribusikan ke kenyamanan estetik dan kemudahan bernavigasi, sebaliknya tingkat *mission unfinished* yang tinggi berarti pengguna merasa kebingungan atau kesulitan. Kesuksesan yang mirip dapat dilihat pada *mobile site*, namun terdapat tingkat *mission unfinished* yang tinggi di task 1 jika dibandingkan dengan versi lamanya. Hal ini mengindikasikan ruang untuk perbaikan.

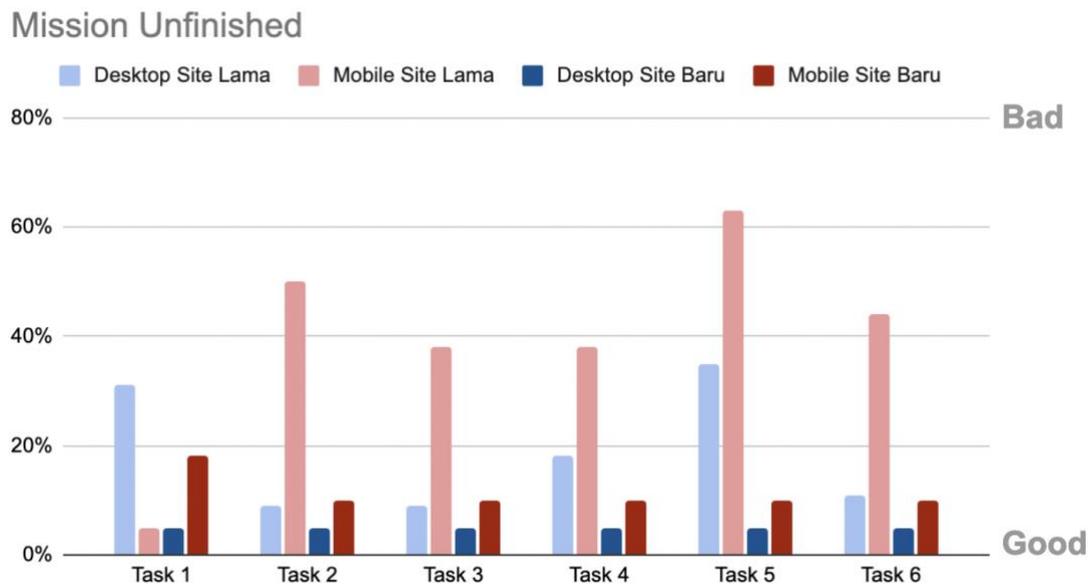

**Gambar 4.27 Perbandingan Metrik: Mission Unfinished**

### 4.5.3. Misclick Rate

Performa lebih baik didapatkan dengan nilai yang lebih rendah. Dapat dilihat bahwa secara keseluruhan, *mislick rate* berkurang pada versi baru. Pencapaian ini terlihat lebih baik pada desktop site dengan hanya 1 task yang memiliki *misclick rate* yang lebih rendah pada versi lama, sementara untuk mobile site terdapat 2 task dengan *misclick rate* lebih rendah pada versi lama.

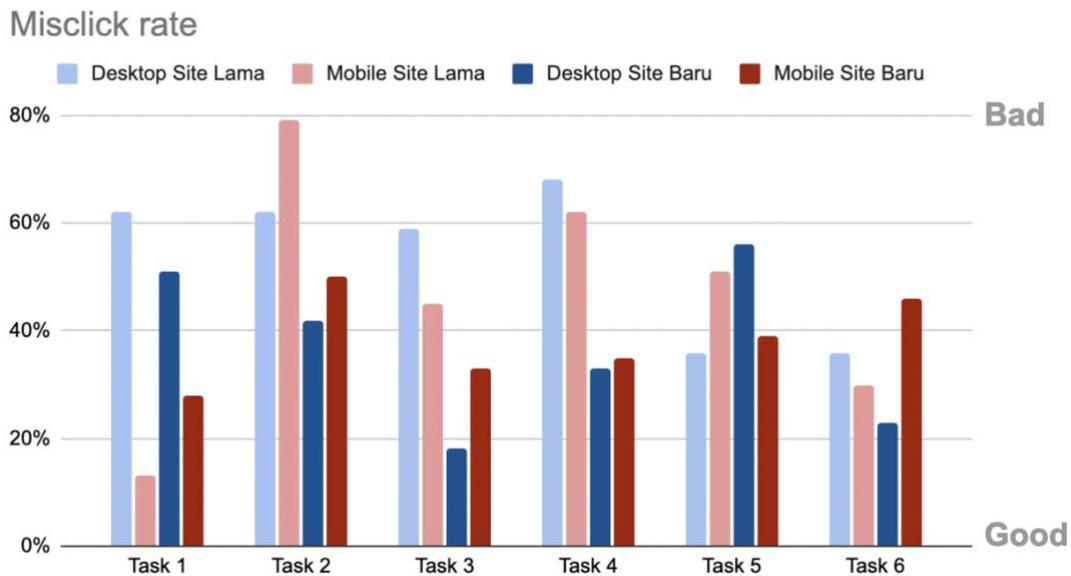

**Gambar 4.28 Perbandingan Metrik: Misclick Rate**

### 4.5.4. Average Duration

Performa lebih baik didapatkan dengan nilai yang lebih rendah. Dapat dilihat bahwa secara keseluruhan, *average duration* berkurang pada versi baru. Terdapat 1 task untuk *mobile* dan *desktop site* yang memiliki *average duration* lebih cepat pada website lama, namun secara keseluruhan menunjukkan kesuksesan mengurangi waktu. Pencapaian ini terlihat lebih baik pada *desktop site* yang memiliki gradien yang lebih rendah untuk task yang kurang suksesnya.

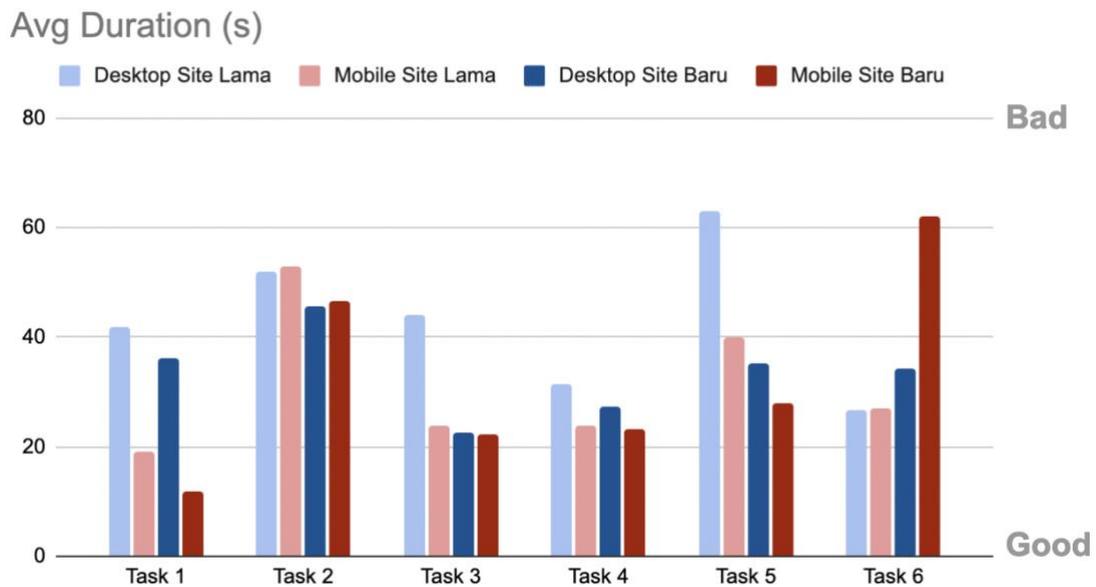

**Gambar 4.29 Perbandingan Metrik: Average Duration**

### 4.5.5. *Usability Score*

Performa lebih baik didapatkan dengan nilai yang lebih tinggi dengan nilai maksimum 100. Dapat dilihat bahwa secara keseluruhan, *usability score* meningkat pada versi baru. Terdapat 1 task untuk *mobile site* yang memiliki *usability score* lebih tinggi pada website lama namun dengan perbedaan tipis.

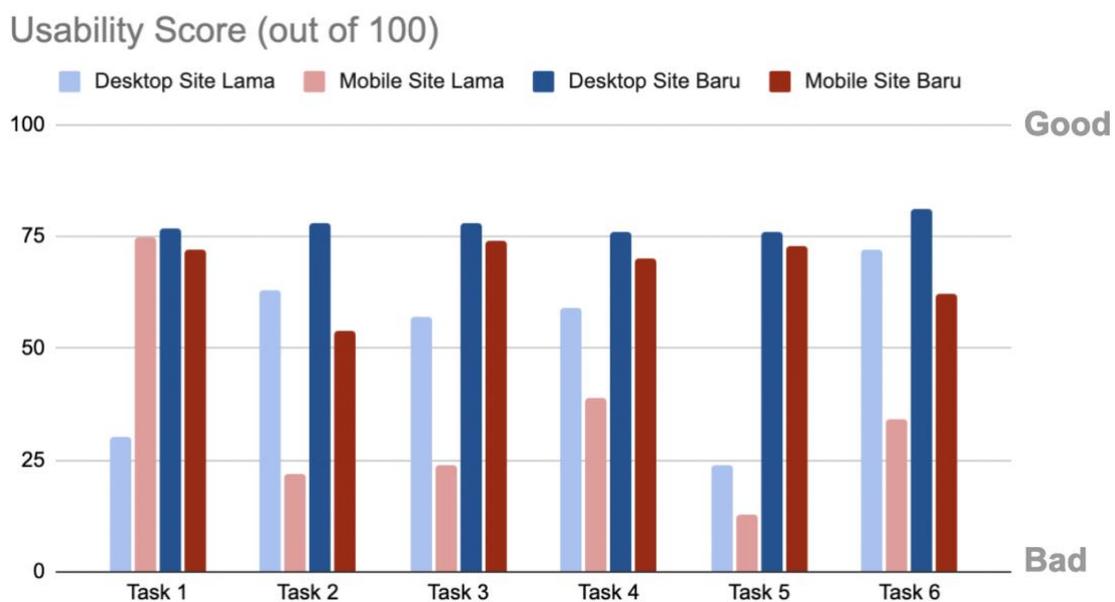

**Gambar 4.30 Perbandingan Metrik: Usability Score**

### 4.5.6. PSSUQ

Performa lebih baik didapatkan dengan nilai yang lebih rendah dengan nilai maksimum 1 dari 7. Dapat dilihat bahwa secara keseluruhan, PSSUQ lebih baik pada versi baru. Terdapat apresiasi dan kepuasan yang sangat signifikan di *cluster* 2 dan 3. Dapat dilihat bahwa *mobile site* menunjukkan peningkatan kepuasan yang sangat signifikan walaupun dengan nilai performa *usability testing* di bawah *desktop*. Secara keseluruhan terdapat gradien yang lebih tinggi pada PSSUQ dibandingkan dengan usability testing. Hal ini dapat diinterpretasikan bahwa secara keseluruhan pengguna menyukai website baru, namun masih belum terbiasa menggunakannya dibandingkan dengan website lama, sehingga menghasilkan skor efisiensi performa yang meskipun masih jauh lebih baik dari website lama, masih lebih rendah dari apa yang sebenarnya mampu mereka capai. Pengguna telah memiliki waktu minimal 1 semester (6 bulan) untuk mempelajari sistem SIAK-NG lama, dan 1 hari untuk mempelajari sistem SIAK-NG baru. Konsep ini didukung oleh teori Nielsen tentang definisi efektivitas, yaitu seberapa cepat pengguna dapat menyelesaikan aktivitas setelah belajar menggunakan suatu sistem.

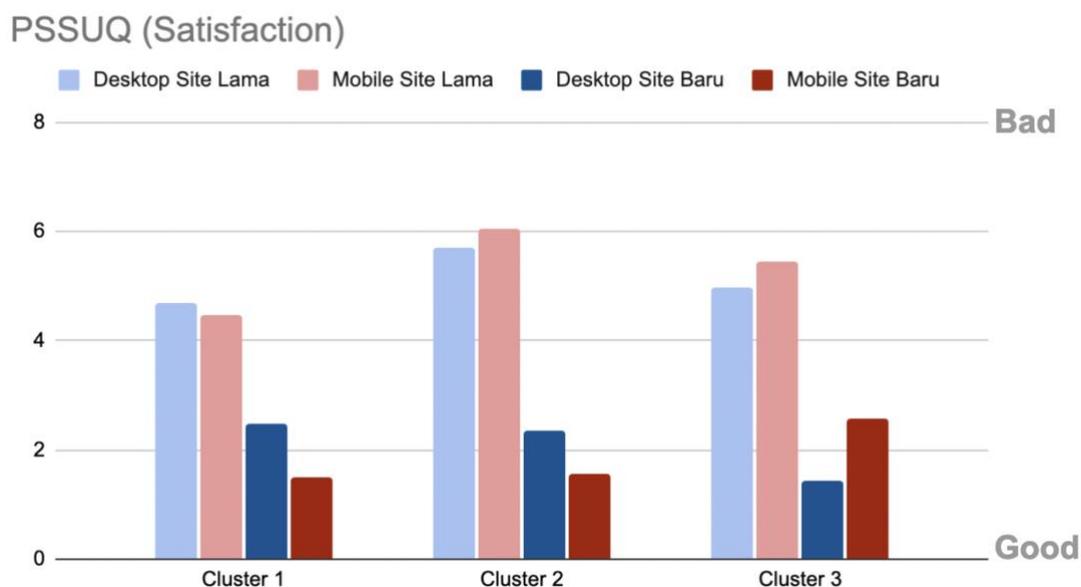

**Gambar 4.31 Perbandingan Metrik: PSSUQ**

## BAB 5

## KESIMPULAN DAN SARAN

Dalam bab 5, akan diuraikan kesimpulan dari bab-bab sebelumnya yang meliputi studi pustaka, pengumpulan data, pengolahan data, dan analisis yang telah dilakukan. Tujuan dari bab ini adalah untuk merangkum hasil penelitian ini secara komprehensif. Selain itu, bab ini juga akan menyajikan saran-saran untuk penelitian selanjutnya, sehingga dapat memberikan arahan bagi peneliti dalam melanjutkan studi ini di masa depan.

### 5.1 Kesimpulan

Penelitian ini bertujuan untuk memberikan wawasan dari sudut pandang perbaikan *user interface* dengan merancang *user interface* website SIAK-NG yang sedang dalam tahap awal pembaharuan agar dapat meningkatkan kepuasan pengguna.

Dalam penelitian ini, digunakan pendekatan *design thinking* dan metode *design thinking* yang melibatkan beberapa *tools* seperti *user persona, empathy map, storyboarding,* dan *usability testing.* Tiga persona berhasil dibangun pada tahap *empathize* menggunakan *hierarchical clustering* dan K-modes. *Empathy map* pada tahap define berhasil digunakna untuk mencari tahu permasalahan dari sudut pandang setiap *persona*.

Pada tahap ideate, dilakukan wawancara mendalam dan pembangunan *storyboard* untuk mendapatkan ide-ide baru dalam mengatasi masalah dan kesulitan yang dihadapi oleh persona saat mengakses halaman-halaman website. Selanjutnya, dilakukan perancangan ulang website SIAK-NG menggunakan wireframe *low fidelity* dan *clickable prototype* dengan Figma.

*Usability testing* dan kuesioner PSSUQ dilakukan untuk mengukur *efektivitas, efisiensi,* dan kepuasan terhadap *user interface* baru. *Usability testing* menguji performa penggunaan website, sementara kuesioner PSSUQ menunjukkan peningkatan kepuasan pengguna terhadap *user interface* baru. Hasil pengukuran secara keseluruhan menunjukkan peningkatan dalam metrik *direct success, mission unfinished, misclick rate, duration,* dan *usability score*. Skor *desktop site* secara keseluruhan lebih baik daripada situs *mobile*, tetapi keduanya tetap menunjukkan peningkatan performa dan kepuasan yang signifikan.

Penelitian ini membuktikan bahwa pendekatan *design thinking* dalam merancang ulang *user interface* dapat memberikan solusi yang sesuai dengan kebutuhan pengguna dan mencapai tujuan perbaikan website.

## 5.2 Saran

Saran yang dapat direkomendasikan oleh penulis untuk penelitian berikutnya yang berhubungan dengan topik ini adalah:

1. Melakukan survei dengan menggunakan kuesioner yang disebar kepada jumlah responden yang lebih besar dan mewakili seluruh fakultas secara proporsional.
2. Melibatkan proses desain kolaboratif dengan mendapatkan umpan balik dari profesional yang memiliki pengalaman di bidang ini.
3. Mengembangkan *prototype* menjadi lebih imersif, bukan hanya berupa *clickable mockup*, untuk mendapatkan nilai yang lebih akurat.
4. Melakukan *usability testing* dan penyebaran kuesioner PSSUQ kepada jumlah responden yang lebih banyak, sehingga dapat melihat hasilnya ketika dilakukan pada sampel populasi yang lebih besar.
5. Melakukan penelitian lebih lanjut dengan fokus pada fitur-fitur yang belum dibahas dalam penelitian ini.
6. Melakukan siklus iterasi dan umpan balik yang lebih intens dengan para pengujian.
7. Melakukan studi untuk *user role* lainnya pada SIAK-NG, seperti dosen, SBA, dll. serta membuat varian *lightmode*.
8. Menggunakan versi premium dari Maze agar bisa mengakses insight lebih banyak lagi.
9. Menyeimbangkan antara keterbatasan simplisitas dan fungsionalitas untuk performa dan kesan terbaik.

# REFERENSI

Adams, A. and Cox, A.L. (2008) 'Questionnaires, in-depth interviews and focus groups', Research Methods for Human-Computer Interaction, pp. 17–34. doi:10.1017/cbo9780511814570.003.

Borrellia, L., & Perrella, S. (2021). User Interface Design for E-Learning Platform and Institutional  Portal of University of Foggia.

Bratsberg, H.M. (2012). Empathy Maps of the FourSight Preferences.

Brickey, Jon; Walczak, Steven; and Burgess, Tony, "A Comparative Analysis of Persona Clustering Methods" (2010). AMCIS 2010 Proceedings. 217.https://aisel.aisnet.org/amcis2010/217

Brown, T., & Katz, B. (2019). Change by design: how design thinking transforms organizations and inspires innovation (Vol. 20091). HarperBusiness.

Chaturvedi, A., Green, P.E. and Caroll, J.D. (2001) 'K-modes clustering', Journal of Classification, 18(1), pp. 35–55. doi:10.1007/s00357-001-0004-3.

Cho, V, T.C.E. Cheng, W.M.J. Laia, The role of perceived user-interface design in continued usage intention of self-paced e-learning tools. In Computers & Education, 53, 2, 2009, 216-227.

Cooper, A., Reimann, R. and Cronin, D. (2007) About face 3: The Essentials of Interaction Design. Indianapolis, IN: Wiley.

Cunha, L. (2023). How your usability score is calculated. Maze. https://help.maze.co/hc/en-us/articles/360052723353-How-your-Usability-Score-is-calculated

Dillon, A. (1988). Reading from paper versus reading from Screen. The Computer Journal, 31(5), 457–464. https://doi.org/10.1093/comjnl/31.5.457

Doggett, A.M. (2005) 'Root Cause Analysis: A framework for tool selection', Quality Management Journal, 12(4), pp. 34–45. doi:10.1080/10686967.2005.11919269.

Ferrier, M. (2016, June 8). Stylewatch: Is Pantone 448C really the ugliest colour in the world?. The Guardian. https://www.theguardian.com/fashion/2016/jun/08/stylewatch-pantone-448c-ugliest-colour-world-opaque-couche-australian-smokers-fashion

Frost, B. (2016). Atomic Design.

Fu, L., Lin, J., Liu, W., Tang, R., Zhang, W., Zhang, R., & Yu, Y. (2023). An F-shape click model for information retrieval on multi-block mobile pages. Proceedings of the Sixteenth ACM International Conference on Web Search and Data Mining. https://doi.org/10.1145/3539597.3570365

Gibbons, S. (2018). Empathy mapping: The first step in design thinking. Nielsen Norman Group. https://www.nngroup.com/articles/empathy-mapping/

Gordon, K. (2020) 5 principles of visual design in UX, Nielsen Norman Group. Available at: https://www.nngroup.com/articles/principles-visual-design/ (Accessed: 02 June 2023).

Gray, D. (2017, July 16). Updated empathy map canvas. Medium. https://medium.com/@davegray/updated-empathy-map-canvas-46df22df3c8a

Gurusamy, K., Srinivasaraghavan, N., & Adikari, S. (2016). An integrated framework for design thinking and agile methods for digital transformation. Design, User Experience, and Usability: Design Thinking and Methods, 34–42.

Harley, A. (2015) Personas make users memorable for product team members, Nielsen Norman Group. Available at: https://www.nngroup.com/articles/persona/ (Accessed: 10 July 2023).

Henriksen, D., Richardson, C. and Mehta, R. (2017) 'Design thinking: A creative approach to educational problems of Practice', Thinking Skills and Creativity, 26, pp. 140–153. doi:10.1016/j.tsc.2017.10.001.

Interaction Design Foundation. (2023, April 14). What is User Interface (UI) design?. The Interaction Design Foundation. https://www.interaction-design.org/literature/topics/ui-design

Interaction Design Foundation. (2021, June 15). What is usability testing?. The Interaction Design Foundation. https://www.interaction-design.org/literature/topics/usability-testing

Interaction Design Foundation. (2022, July 12). What is design thinking?. The Interaction Design Foundation. https://www.interaction- design.org/literature/topics/design-thinking

Interaction Design Foundation. (2022, August 11). What is user experience (UX) design?. The Interaction Design Foundation. https://www.interaction-design.org/literature/topics/ux-design

Kim, K., Erickson, A., Lambert, A., Bruder, G., & Welch, G. (2019). Effects of dark mode on visual fatigue and acuity in optical see-through head-mounted displays. Symposium on Spatial User Interaction. https://doi.org/10.1145/3357251.3357584

Krause, R. (2018) Storyboards help visualize UX ideas, Nielsen Norman Group. Available at: https://www.nngroup.com/articles/storyboards-visualize-ideas/ (Accessed: 03 June 2023).

Landau, S. and Chis Ster, I. (2010) 'Cluster analysis: Overview', International Encyclopedia of Education, pp. 72–83. doi:10.1016/b978-0-08-044894-7.01315-4.

McLeod, S. (2018) Questionnaire: Definition, Examples, Design and Types. https://www.simplypsychology.org/questionnaires.html 16). Updated empathy map canvas. Medium. https://medium.com/@davegray/updated-empathy-map-canvas-46df22df3c8a

Ndukwu, D. (2020). Questionnaire: Types, Definition, Examples and How to Design your Questionnaires. KyLeads.Retrieved from www.Kyleads.com/blog/questionnaires 29.07.2020

Nessler, D. (2016, June 28). To hamburger or not to hamburger. Medium. https://medium.com/digital-experience-design/to-hamburger-or-not-to-hamburger-aad8b4a07576

Ofosu-Asare, Y. H. B. Essel, F. Mensah Bonsu, E-learning graphical user interface development using ADDIE instruction design model and developmental research: the need to establish validity and reliability, I.K. Press (2019), Volume 13, 2019.

Pengembangan dan Pelayanan Sistem Informasi Universitas Indonesia (2008) MANUAL UNTUK MAHASISWA. Depok.

Pernice, K., & Budiu, R. (2016). Hamburger menus and hidden navigation hurt UX metrics. Nielsen Norman Group. https://www.nngroup.com/articles/hamburger-menus/

Pernice, K., Gibbons, S., & Moran, K. (2021). The 6 levels of UX Maturity. Nielsen Norman Group. https://www.nngroup.com/articles/ux-maturity-model/

Pernice, K., Whitenton, K., Nielsen, J., & Group, N. N. (2014). How People Read on t he Web: The Eyetracking Evidence.

Plattner, H., Meinel, C. and Leifer, L. (2013) Design thinking understand - improve - apply. Berlin: Springer Berlin.

PRUITT, J.S. and ADLIN, T. (2006) 'Overview of the persona lifecycle', The Persona Lifecycle, pp. 46–65. doi:10.1016/b978-012566251-2/50003-4.

Pulido, Mar. (2021). Environment Guardians: strengthen the educational processes of the community in Tolima (Colombia). 10.13140/RG.2.2.30660.04485.

Quesenbery, W. and Brooks, K. (2010) Storytelling for user experience: Crafting Stories for Better Design. Brooklyn, NY: Rosenfeld Media.

Raj, A. (2022, November 10). Data-ink ratio explained with example. Code Conquest. https://www.codeconquest.com/blog/data-ink-ratio-explained-with-example/

Signal, M. (2020, February 7). What are atomic design principles, and why do we use them?. EB Pearls. https://ebpearls.com.au/atomic-design-principles/

Suleiman, A. (2023). UX Hamburger Menu Abuse & Why It's So Popular. https://ux4sight.com/blog/hamburger-menu-abuse

Tateo, L. Web accessibility and usability: limits and perspectives. In Proceedings of the First Workshop on Technology Enhanced Learning Environments for Blended Education, 2021.

Tschimmel, K. (2012). Design Thinking as an effective Toolkit for Innovation. In ISPIM Conference Proceedings (p. 1). The International Society for Professional Innovation Management (ISPIM).

Tufte, E. R. (2018). The visual display of quantitative information. Graphics Press.

WCAG. (2018). How to meet WCAG (quick reference). How to Meet WCAG (Quickref Reference). https://www.w3.org/WAI/WCAG21/quickref/#target-size

Yablonski, J. (2023) Laws of UX: Using psychology to Design Better Products & Services. Findaway World.

# APPENDIX

**1. DATA DIRI**

Nama *

Your answer

Jenis Kelamin *

○ Perempuan

○ Laki-Laki

Umur *

Your answer

Fakultas *

Choose ▾

Jurusan *("-" jika fakultas hanya ada 1 jurusan)* *

Your answer

Angkatan *

○ 2022

○ 2021

○ 2020

○ 2019

**Universitas Indonesia**

Appendix 1. Pertanyaan Pilot Survey (Lanjutan)

**2. KESAN**
DENGAN MENGISI FORM INI, SAYA **MENYATAKAN PERNAH MENGAKSES DAN MENGGUNAKAN SIAK-NG**

Jika desain SIAK-NG diperbaharui, apa pendapat saudara? *

◯ Tidak boleh, sudah sempurna

◯ Sedikit keberatan

◯ Tidak berpengaruh bagi saya

◯ Sedikit mendukung

◯ Mendukung

Jika setuju terhadap pembaharuan SIAK-NG, kapan? *

◯ Secepatnya

◯ Nanti saja (>4 tahun)

◯ Tidak setuju

Sebutkan elemen/bentuk/area **yang paling pertama menarik perhatian saudara** *
saat melihat website SIAK-NG

Your answer

Apa **perasaan** yang saudara asosiasikan saat memakai desain SIAK-NG saat ini? *
(Dapat memilih >1)

☐ Kesal

☐ Malas

☐ Bingung

☐ Apatis

☐ Bosan

☐ Capek

☐ Mengganggu

☐ Kerepotan

☐ Memandang sebelah mata

☐ Terpaksa

☐ Kecewa

☐ Malu

☐ Sulit dipelajari

☐ Netral

☐ Nikmat

☐ Menyenangkan

☐ Bersahabat

☐ Tidak bersahabat

☐ Other:

Berikan **pendapat singkat** saudara tentang desain SIAK-NG saat ini *

Your answer

Appendix 1. Pertanyaan Pilot Survey (Lanjutan)

**4. PENILAIAN**
Untuk pertanyaan selanjutnya, saudara dapat juga mempertimbangkan baik untuk
halaman login SIAK-NG maupun halaman utama aspek desain SIAK-NG dari sisi:
1. Posisi
2. Warna
3. Ukuran
4. Grouping
dan lain lain

Sebagai pengguna SIAK-NG, apa saja **pengalaman tidak menyenangkan yang** *
**saudara alami** akibat desain yang kurang baik?

Your answer

Menurut saudara apa **penyebab** dari pengalaman tersebut? *

Your answer

Menurut saudara, apa **manfaat yang dapat saudara rasakan** jika masalah *
tersebut diperbaiki?

Your answer

Sebutkan **kekurangan SIAK-NG** lainnya dari aspek desain dan layout **yang ingin** *
**saudara laporkan** selain server *("-" jika tidak ada)*

Your answer

Berikan **saran perbaikan, fitur tambahan, dan harapan** saudara untuk SIAK-NG *
yang baru *(selain server)*

Your answer

Apakah ada hal yang **tidak ingin saudara ubah** dari desain SIAK-NG? *("-" jika tidak* *
*ada)*

Your answer

Apakah di antara kekurangan-kekurangan SIAK-NG ada yang **berpengaruh** *
**terhadap kenyamanan saudara dalam melakukan pekerjaan?**

◯ Ya

◯ Tidak

**Kontak email/WA:**
Kami membutuhkan relawan sebagai narasumber kritik tentang performa SIAK-NG.
Mohon isi **jika saudara tertarik turut serta menjadi beta-tester SIAK-NG versi baru**
*(opsional)*

Your answer

**Appendix 2. Desain Prototype Desktop Site**
Coded website dapat dicoba di https://slcm.algvrithm.com/login/

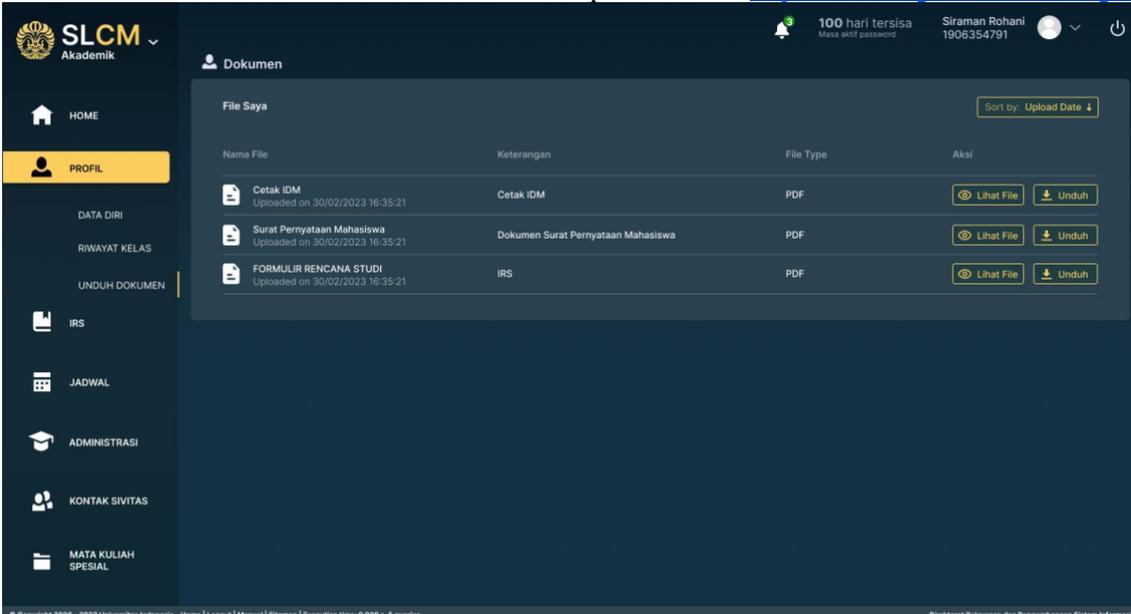

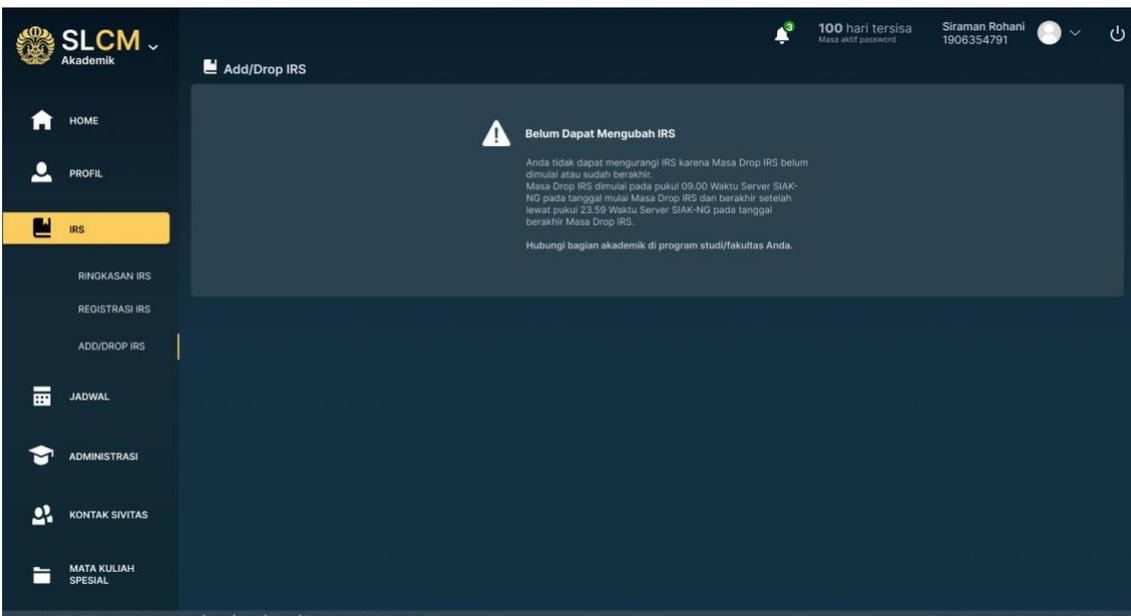

**Appendix 2. Desain Prototype Desktop Site (Lanjutan)**

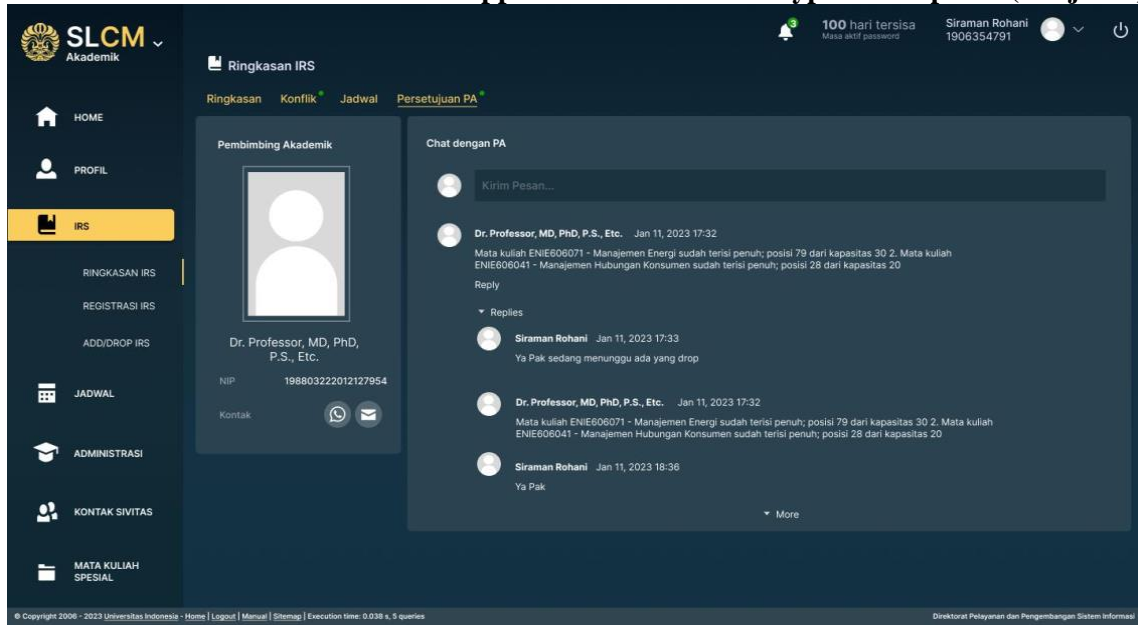

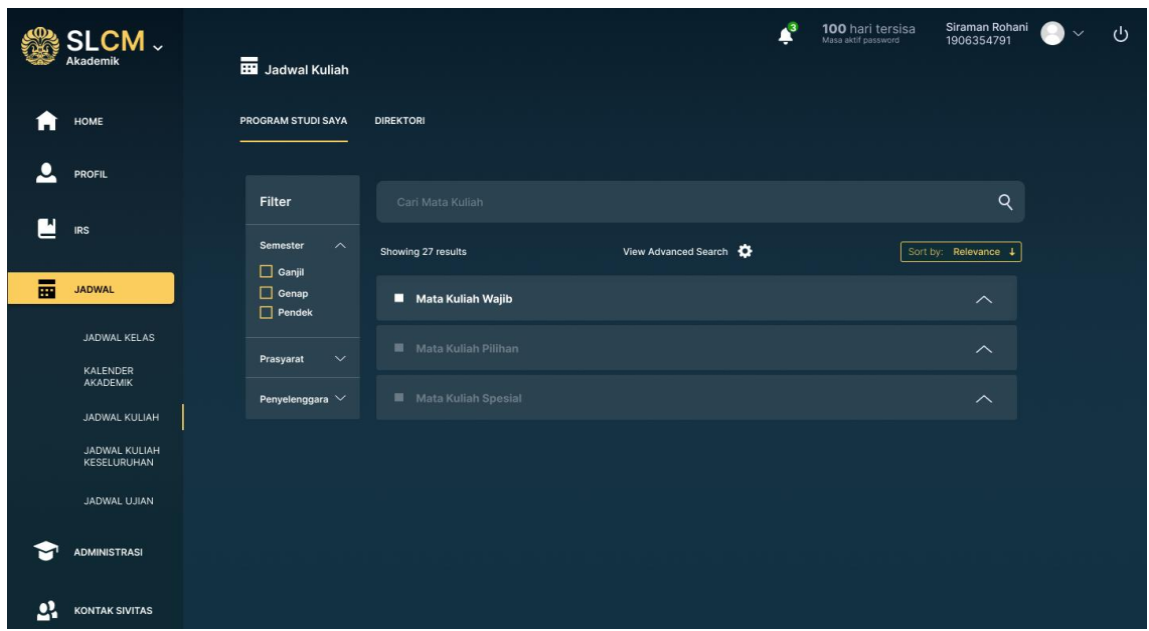

**Appendix 2. Desain Prototype Desktop Site (Lanjutan)**

**Appendix 3. Desain Prototype Mobile Site**

Coded website dapat dicoba di https://slcm.algvrithm.com/mobile-login-screen

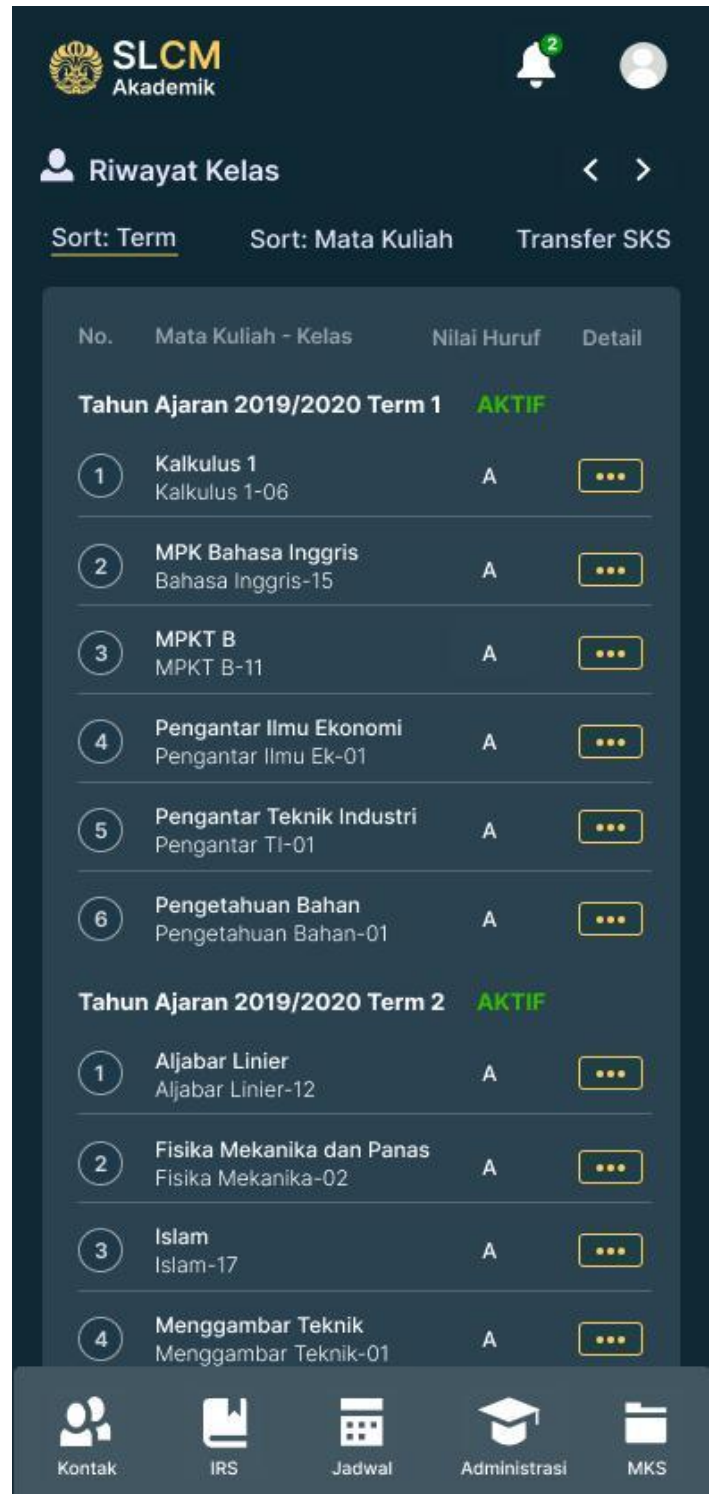

**Appendix 3. Desain Prototype Mobile Site (Lanjutan)**

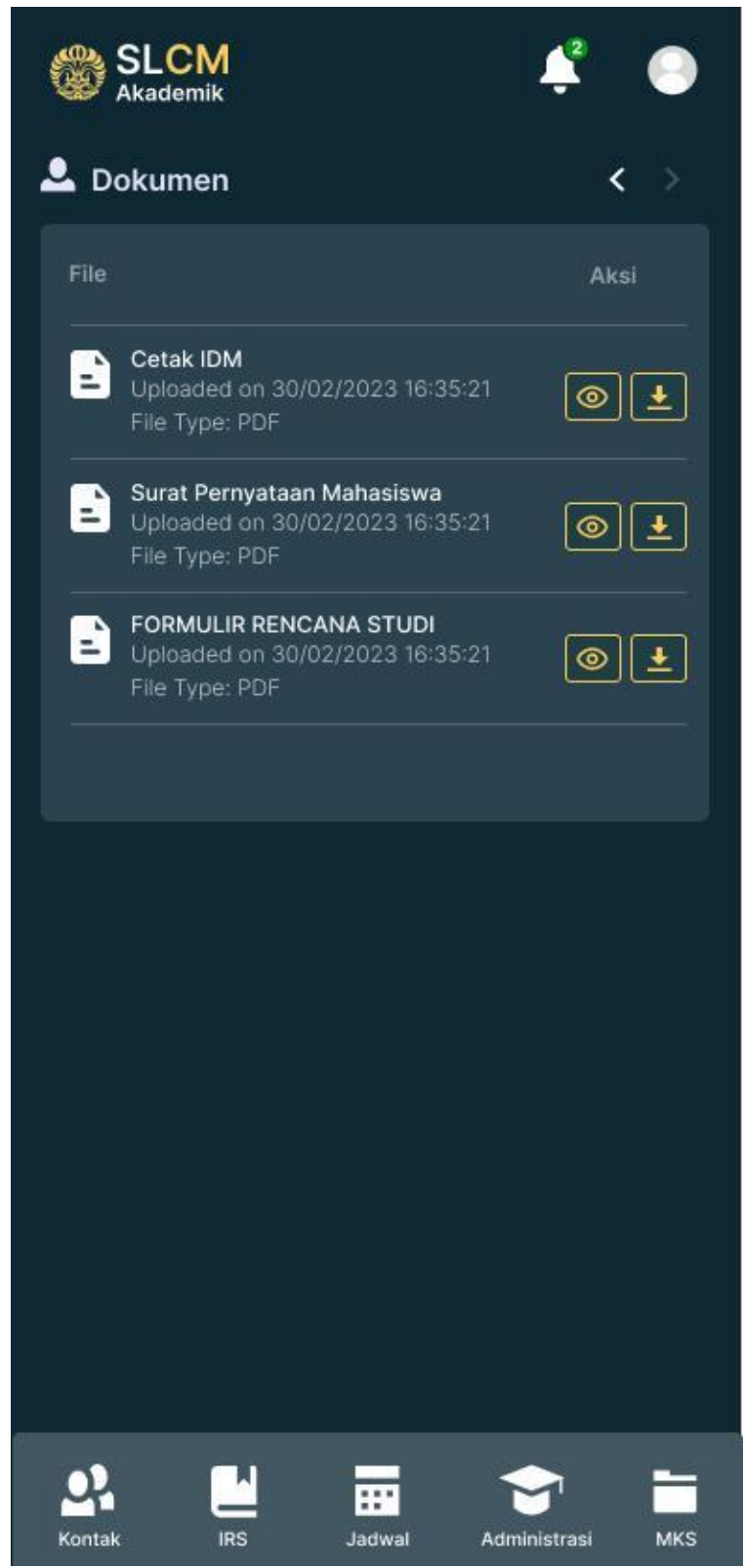